\pdfoutput=1
\documentclass[11pt,a4paper]{article}
\usepackage{jheppub}
\usepackage[T1]{fontenc}
\usepackage{booktabs}
\usepackage{amsmath,amssymb}
\usepackage[normalem]{ulem}
\usepackage{color}
\usepackage{placeins}

\usepackage[caption=false]{subfig}

\definecolor{pur}{rgb}{0.6,0.,1.}
\definecolor{diffblue}{rgb}{0.,0.,0.7}   

\usepackage{tikz}
\usetikzlibrary{arrows.meta, shadows.blur, positioning, calc}
 
\definecolor{priorBg}{HTML}{E8EDF2}
\definecolor{priorBorder}{HTML}{7B8FA1}
\definecolor{postBg}{HTML}{E3F0E8}
\definecolor{postBorder}{HTML}{5A9E6F}
\definecolor{truthBg}{HTML}{FFF5E0}
\definecolor{truthBorder}{HTML}{D4A843}
\definecolor{inputBg}{HTML}{FFF0E8}
\definecolor{inputBorder}{HTML}{E07B3C}
\definecolor{arrowOrange}{HTML}{E07B3C}
\definecolor{flowArrow}{HTML}{3B6EA5}
\definecolor{qualArrow}{HTML}{888888}

\DeclareRobustCommand{\Sec}[1]{Sec.~\ref{sec:#1}}

\DeclareRobustCommand{\App}[1]{App.~\ref{app:#1}}

\DeclareRobustCommand{\Fig}[1]{Fig.~\ref{fig:#1}}
\DeclareRobustCommand{\sec}[1]{Sec.~\ref{sec:#1}}

\DeclareRobustCommand{\Eq}[1]{Eq.~(\ref{eq:#1})}

\DeclareRobustCommand{\Reference}[1]{Ref.~\cite{#1}}
\DeclareRobustCommand{\References}[1]{Refs.~\cite{#1}}

\newcommand{\eq}[1]{Eq.~\eqref{eq:#1}}
\newcommand{\eqs}[2]{Eqs.~\eqref{eq:#1} and \eqref{eq:#2}}

\renewcommand\beforetochook{\pagestyle{myplain}\pagenumbering{roman}\begingroup\small}
\renewcommand\afterTocSpace{\endgroup\bigskip\medskip}

\newcommand{\nn}{\nonumber} 

\title{Improving parton shower predictions via precision moments of energy flow polynomials}

\author[a]{Beno\^\i t Assi,}
\author[b,c]{Kyle Lee,}
\author[d,e,f,g]{and Jesse Thaler}

\affiliation[a]{Department of Physics, University of Cincinnati, Cincinnati, OH 45221, USA}
\affiliation[b]{Department of Physics, Yale University, New Haven, CT 06511, USA}
\affiliation[c]{High Energy Physics Division, Argonne National Laboratory, Lemont, IL, USA}
\affiliation[d]{Center for Theoretical Physics -- a Leinweber Institute, Massachusetts Institute of Technology,
Cambridge, MA 02139, USA}
\affiliation[e]{Institut des Hautes \'Etudes Scientifiques, 91440 Bures-sur-Yvette, France}
\affiliation[f]{Institut de Physique Th\'eorique, CEA Paris-Saclay, 91191 Gif-sur-Yvette, France}
\affiliation[g]{The NSF Institute for Artificial Intelligence and Fundamental Interactions}

\emailAdd{assibt@ucmail.uc.edu,kyle@anl.gov,jthaler@mit.edu}

\abstract{
In this paper, we study various conceptual and practical aspects of using maximum-entropy reweighting to upgrade parton-shower event samples based on higher-accuracy theoretical constraints.
Our approach produces strictly positive per-event weights that improve parton-shower predictions while preserving full event-level exclusivity, allowing any observable to be computed on the reweighted sample without rebinning or regeneration.
On the conceptual side, we explain how theoretical principles can help determine which constraints to use and which kinds of priors lead to efficient reweighting.
On the practical side, we perform a proof-of-concept study with hemisphere observables in $e^+e^-\!\to$ hadrons, and show that even when the parton-shower prior is purposefully degraded by removing the non-singular parts of the QCD splitting functions, a small set of precision calculations can nevertheless restore the desired physical behavior.
We use energy flow polynomials (EFPs) as a systematic basis to organize infrared- and collinear-safe constraints, and study how information transfers from constrained observables to unconstrained ones.
We find rapid information saturation, where constraints from a compact set of EFP moments achieve broad improvements across observable space, including for standard hemisphere observables never used in training.
By construction, the imposed moments of the posterior are formally as accurate as the precision inputs, while the improvement of any other observable is an empirical statement about information transfer that we quantify numerically.
Physics-motivated basis reductions guided by collinear power counting achieve comparable performance to complete bases, and mixed moments combining polynomial and logarithmic terms outperform pure alternatives.
These results suggest a systematic approach to improving parton-shower event generators, where theoretical constraints of highest accuracy can be translated into full phase-space predictions of experimental relevance.}

\preprint{MIT-CTP/6020}

\begin{document}
\maketitle
\flushbottom

\section{Introduction}
 
The description of nature within quantum field theory (QFT) is, at its core, intrinsically stochastic.
From the initial production of particles in a hard scattering process to the recursive fragmentation that populates phase space, the final state of a high-energy  interaction is not a single deterministic outcome, but a probabilistic  distribution over particle multiplicities and momenta.
In quantum chromodynamics (QCD), this stochasticity is a structural feature. Infrared and collinear (IRC) dynamics generate cascades of emissions whose dominant patterns are universal, yet whose detailed realizations fluctuate event by  event~\cite{Buckley:2011ms,EuropeanStrategyforParticlePhysicsPreparatoryGroup:2019qin,
Narain:2022qud,Butler:2023glv}.

A tension currently exists between the two primary approaches to modeling these probabilistic QCD processes: exclusivity versus precision.
General-purpose parton-shower event generators~\cite{Sjostrand:2014zea,Bellm:2015jjp,Sherpa:2024mfk} 
provide a practical, fully exclusive representation of QCD radiation, 
allowing for event-level analysis.
However, they often lack the systematic precision of analytic perturbative 
methods~\cite{Maltoni:2022bqs,Craig:2022cef,Huss:2022ful} or the non-perturbative constraints from lattice QCD~\cite{Shanahan:2020zxr,Shanahan:2021tst,LatticeParton:2023xdl,Ebert:2019okf}.
On the other hand, specialized high-accuracy calculations can pin down QCD behaviors in specific regions of phase space, but they rarely produce a flexible, event-level ensemble that can be subjected to arbitrary experimental cuts.
The central question  motivating this work is how to transfer high-accuracy theoretical information into an exclusive event sample without sacrificing the flexibility of parton-shower methods.

In \Reference{Assi:2025ibi}, we proposed a maximum-entropy reweighting 
framework, which identifies the distribution closest to a parton-shower prior that is consistent with known constraints.
This framework upgrades parton-shower event samples by enforcing precision theoretical constraints in the form of observable moments, yielding strictly positive per-event weights.
We introduced logarithmic moments of event-shape observables -- motivated by the structure of Sudakov resummation -- as a new class of theoretically well-defined inputs.
We demonstrated the effectiveness of this approach using thrust moments $\langle \tau^m \ln^n \tau \rangle$ computed at NNLL$^\prime + \mathcal{O}(\alpha_s^2)$ accuracy in $e^+e^- \to \text{hadrons}$.
The resulting reweighting improved not only the thrust distribution but also correlated event shapes such as jet broadening, while enabling systematic propagation of perturbative uncertainties.

In this paper, we address the question of how to systematically choose and organize the constraints that enter the reweighting.
We employ energy flow polynomials (EFPs)~\cite{Komiske:2017aww,Komiske:2019asc,Cal:2022fnm} as a complete, IRC-safe basis for multi-observable constraints, because they admit systematic truncations of their graph basis by degree (i.e.~edge count) that target different aspects of QCD radiation.
We then perform a proof-of-concept study with hemisphere observables in $e^+e^-\!\to$ hadrons using a deliberately degraded parton shower as the prior unweighted event sample.
A key measure of reweighting performance is information saturation, which quantifies how rapidly a finite set of EFP moment constraints captures the physically relevant information, as measured by information transfer to observables not included in training.
We find that a compact set of low-degree EFP constraints achieves broad improvements across observable space, including for standard hemisphere observables never used in training, and that mixed moments combining polynomial and logarithmic terms outperform pure alternatives.
In this proof-of-concept study, both the degraded prior and the high-fidelity target are generated with \textsc{Sherpa}~\cite{Sherpa:2024mfk}.
To test the robustness of our conclusions beyond this single-generator setup, we also reweight a prior produced by the angular-ordered \textsc{Herwig} shower~\cite{Bahr:2008pv,Bellm:2015jjp} to the same \textsc{Sherpa} target, and we extend the transfer studies to jet substructure and fragmentation observables that are more directly sensitive to the shower evolution and to hadron-level dynamics.

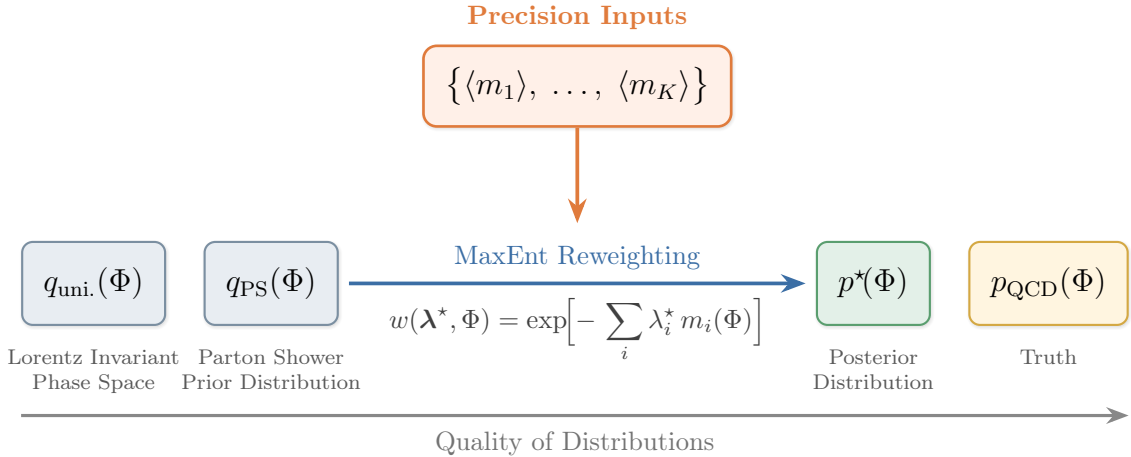
\begin{figure}[t]
\centering
\begin{tikzpicture}[
    >=Stealth,
    priorbox/.style={
        rectangle, rounded corners=4pt,
        draw=priorBorder, line width=0.8pt,
        fill=priorBg,
        blur shadow={shadow blur steps=6, shadow xshift=0.4pt, shadow yshift=-0.6pt,
                     shadow blur radius=1.5pt, shadow opacity=20},
        inner sep=8pt, font=\large, minimum height=1.1cm,
    },
    postbox/.style={
        rectangle, rounded corners=4pt,
        draw=postBorder, line width=0.8pt,
        fill=postBg,
        blur shadow={shadow blur steps=6, shadow xshift=0.4pt, shadow yshift=-0.6pt,
                     shadow blur radius=1.5pt, shadow opacity=20},
        inner sep=8pt, font=\large, minimum height=1.1cm,
    },
    truthbox/.style={
        rectangle, rounded corners=4pt,
        draw=truthBorder, line width=0.8pt,
        fill=truthBg,
        blur shadow={shadow blur steps=6, shadow xshift=0.4pt, shadow yshift=-0.6pt,
                     shadow blur radius=1.5pt, shadow opacity=20},
        inner sep=8pt, font=\large, minimum height=1.1cm,
    },
    inputbox/.style={
        rectangle, rounded corners=5pt,
        draw=inputBorder, line width=1pt,
        fill=inputBg,
        blur shadow={shadow blur steps=6, shadow xshift=0.4pt, shadow yshift=-0.6pt,
                     shadow blur radius=1.5pt, shadow opacity=20},
        inner sep=8pt, font=\large,
    },
    sublabel/.style={font=\scriptsize, text=black!65, align=center},
    scale=0.95
]
 
\node[priorbox] (quni) at (0,0) {$q_{\mathrm{uni.}}(\Phi)$};
\node[priorbox, right=14pt of quni] (qps) {$q_{\mathrm{PS}}(\Phi)$};
 
\node[postbox] (pstar) at (10.8,0) {$p^\star\!(\Phi)$};
\node[truthbox, right=14pt of pstar] (pqcd) {$p_{\mathrm{QCD}}(\Phi)$};
 
\node[sublabel, below=5pt of quni]  {Lorentz Invariant\\[-1pt] Phase Space};
\node[sublabel, below=5pt of qps]   {Parton Shower\\[-1pt] Prior Distribution};
\node[sublabel, below=5pt of pstar]  {Posterior\\[-1pt] Distribution};
\node[sublabel, below=5pt of pqcd]   {Truth};
 
\draw[->, line width=1.4pt, color=flowArrow]
    ([xshift=4pt]qps.east) --
    node[above, font=\small, text=flowArrow, yshift=1pt]
         {MaxEnt Reweighting}
    node[below, font=\small, text=black!75, yshift=-1pt]
         {$w(\boldsymbol{\lambda}^\star,\Phi)
           = \exp\!\Bigl[-\,\displaystyle\sum_i \lambda_i^\star\, m_i(\Phi)\Bigr]$}
    ([xshift=-4pt]pstar.west);
 
\coordinate (midtop) at ($(qps.north east)!0.5!(pstar.north west)$);
\node[inputbox, above=1.5cm of midtop]
    (inputs) {$\bigl\{\langle m_1\rangle,\;\ldots,\;\langle m_K\rangle\bigr\}$};
\node[above=2pt of inputs, font=\small\bfseries, text=inputBorder]
    {Precision Inputs};
 
\draw[->, line width=1.4pt, color=arrowOrange]
    (inputs.south) -- ([yshift=7pt]midtop);
 
\coordinate (botL) at ($(quni.south west) + (0, -1.25cm)$);
\coordinate (botR) at ($(pqcd.south east) + (0, -1.25cm)$);
 
\draw[->, line width=1.2pt, color=qualArrow]
    (botL) -- (botR)
    node[midway, below=2pt, font=\small, text=black!55] {Quality of Distributions};
 
\end{tikzpicture}
\caption{Schematic of information-theoretic reweighting.
A parton-shower prior $q_{\rm PS}(\Phi)$ is updated to a posterior $p^\star(\Phi)$ by a maximum-entropy projection that enforces precision constraints $\langle m_i \rangle = c_i$ via strictly positive per-event weights.
A minimal prior $q_{\mathrm{uni.}}(\Phi)$ sets $|\mathcal{M}_N|^2 = \text{const.}$ for each particle multiplicity $N$, while more realistic priors incorporate additional physical effects.
The resulting $p^\star(\Phi)$ is closer to the truth distribution $p_{\rm QCD}(\Phi)$.
}
\label{fig:map}
\end{figure}

The conceptual basis for our approach is the maximum entropy principle~\cite{Shannon:1948dpw,Jaynes:1957zza,Jaynes:1957zz}, which extends Boltzmann's reasoning in statistical mechanics~\cite{e17041971} 
to any probabilistic inference problem: among all distributions consistent 
with known constraints, the one that maximizes Shannon entropy makes the 
fewest assumptions beyond the constraints themselves.
For continuous distributions, the coordinate-invariant analogue of Shannon entropy is instead the relative entropy, also known as the (negative) Kullback–Leibler (KL) divergence, defined with respect to a reference prior distribution.
Applied to event generation, this means starting from a parton-shower prior and deriving a posterior via strictly positive, per-event reweighting that enforces precision constraints while remaining fully exclusive, as illustrated schematically in \Fig{map}.

Our maximum-entropy reweighting procedure complements ongoing efforts to systematically improve event generators.
First, improving the resummation accuracy of general-purpose 
showers~\cite{Hoche:2017iem,Dulat:2018vuy,Dasgupta:2020fwr,Bewick:2019rbu,Forshaw:2020wrq,Nagy:2020rmk,Nagy:2020dvz,Bewick:2021nhc,Gellersen:2021eci,vanBeekveld:2022zhl,vanBeekveld:2022ukn,Herren:2022jej,Assi:2023rbu,FerrarioRavasio:2023kyg,Preuss:2024vyu,Hoche:2024dee,vanBeekveld:2024wws}
and extending them beyond the leading-color 
approximation~\cite{Gustafson:1992uh,Giele:2011cb,Nagy:2012bt,Platzer:2012np,Nagy:2014mqa,Nagy:2015hwa,Platzer:2018pmd,Isaacson:2018zdi,Nagy:2019rwb,Nagy:2019pjp,Forshaw:2019ver,Hoche:2020pxj,DeAngelis:2020rvq,Holguin:2020joq,Hamilton:2020rcu}
typically requires introducing more structure into the evolution 
and more correlations between emissions.
These developments are 
essential -- a more accurate prior leads to a more stable and 
predictive posterior. Our framework can further complement them 
by enforcing precision information that may not yet be incorporated 
into the generator. 
Second, matching showers to precision fixed-order 
calculations~\cite{Campbell:2022qmc,Frixione:2002ik,Nason:2004rx},
including at 
NNLO~\cite{Hamilton:2012rf,Hamilton:2013fea,Hoche:2014uhw,Karlberg:2014qua,Hamilton:2015nsa,Alioli:2015toa,Astill:2018ivh,Re:2018vac,Monni:2019whf,Mazzitelli:2020jio,Alioli:2020qrd,Mazzitelli:2021mmm,Alioli:2021qbf,Lindert:2022qdd,Gavardi:2022ixt,Alioli:2023har,Gavardi:2023aco},
is well established for individual observables but difficult to 
propagate simultaneously to correlated, multi-differential 
distributions.
Because our reweighting produces event-level weights, 
every observable computed on the reweighted sample is automatically 
corrected, with the degree of improvement determined by how strongly 
each observable correlates with the training constraints.
Third, systematic uncertainties in shower predictions are usually 
estimated by ad hoc variations of shower and hadronization 
parameters~\cite{Mrenna:2016sih,Bothmann:2016nao,Cacciari:2011ze,Tackmann:2024kci,Lim:2024nsk}, although recent efforts leveraging ideas from information theory and machine learning offer a more systematic path toward data-driven hadronization models~\cite{Bierlich:2023zzd,Bierlich:2024xzg,Assi:2025gog,Assi:2025avy,Butter:2025wxn,Assi:2026wyo}.
In our framework, varying a set of different priors tests whether the constraint 
set is sufficient to determine the posterior independently of 
priors. Observables that are stable across prior variations are 
determined by the constraints, while those that are not indicate 
directions where further constraints are needed.
Unlike ad hoc 
parameter variations, this diagnostic has a well-defined 
information-theoretic interpretation, since it tests whether the 
constraints alone determine the posterior.
Fourth, theoretical insights such as nonperturbative power 
corrections~\cite{Dokshitzer:1995zt,Gardi:2000yh,Lee:2006nr,Mateu:2012nk,
Moult:2018jjd,Ebert:2018lzn} and multi-hadron fragmentation 
functions~\cite{Metz:2016swz,Chang:2013iba,Lee:2023tkr,Boussarie:2023izj,vonKuk:2025kbv} have yet to be fully incorporated 
into existing parton-shower frameworks; provided one can identify observables 
sensitive to such effects, our constraint-based approach offers a 
possible route to encoding them as precision inputs.
Finally, higher-precision calculations 
typically target inclusive observables or specific kinematic limits, 
and extending them to the fiducial regions defined by experimental 
cuts remains a significant computational challenge.
Because our 
posterior is a reweighted event sample with the same phase-space 
coverage as the original parton-shower generator, any fiducial cut or analysis  selection can be applied directly, and the precision improvements  from the constraints propagate automatically into the restricted 
region.

There are several important conceptual questions about our approach that deserve attention.
How much does the quality of the prior matter, and when does an ``improved'' shower 
prior make the reweighting more stable or more predictive?
What does 
it mean for a prior to be consistent with a chosen set of constraints, 
in the sense that the posterior is achievable without pathological 
weights?
When constraints are imposed through specific moments or
measurement functions, what can be said about the accuracy of the
resulting posterior, and how should one choose which constraints to
compute in the first place?
We address these questions analytically in \Sec{it_meets_qcd}, and then perform a numerical study in \Sec{results} to highlight some of the practical aspects of reweighting.
In this proof-of-concept study, we use a high-fidelity parton shower as a stand-in for precision calculations, and assume that the target constraints have zero theoretical uncertainties.
Though unrealistic, this allows us to use the same high-fidelity shower to validate the posterior and quantify information saturation without the complications associated with realistic precision calculations and their uncertainties.

The remainder of the paper is organized as follows.
In 
\Sec{it_meets_qcd}, we set up the maximum-entropy 
construction for event ensembles, explaining how a posterior 
distribution on full final-state phase space can be obtained from a 
chosen prior by imposing expectation-value constraints, and we discuss 
how these ideas map onto standard QCD language for IRC-safe observables 
and their logarithmic structure.
In \Sec{efps}, we introduce 
EFPs and motivate their use as a practical 
way to specify and systematically enlarge multi-observable constraint 
sets.
In \Sec{shower_setups}, we describe the shower setups, optimization procedure, and evaluation metrics used in our proof-of-concept study.
In 
\Sec{results}, we present results, focusing on information 
saturation as the constraint set is enlarged, transfer to non-training 
observables, and comparisons of 
reduced EFP bases, ending with a cross-generator test in which a \textsc{Herwig} prior is reweighted to the \textsc{Sherpa} target.
We summarize and outline directions for future work
in \Sec{conclusion}.
Alternatives to information-theoretic reweighting are discussed in \App{alternative_measures},
a discussion of incorporating theoretical uncertainties appears in \App{uncertainties}, and a study of two-dimensional joint distributions appears in \App{2d_transfer}.
In \App{substructure}, we study transfer to jet substructure and fragmentation observables, both for the degraded \textsc{Sherpa} priors and for the cross-generator \textsc{Herwig} prior.

\section{QCD theory meets information theory}
\label{sec:it_meets_qcd}

In this section, we present the theoretical framework for maximum-entropy event reweighting and address various conceptual questions about our methodology.
We first review how the maximum entropy principle yields a systematic method to incorporate constraints into a prior distribution, and then describe the kinds of precision constraints available from QCD perturbation theory.
We discuss the choice of observable moments used as constraints, with attention to the role of logarithmic moments motivated by the Sudakov structure of QCD, and address how the choice of prior affects reweighting efficiency.
We conclude by examining the formal accuracy of the resulting posterior.

\subsection{Review of maximum-entropy reweighting}
\label{sec:maximum_entropy}

The goal of maximum-entropy reweighting is to take a prior distribution $q(\Phi)$, where $\Phi$ denotes the full final-state phase space of an event, and deform it into a posterior distribution $p^\star(\Phi)$ that satisfies a set of physical constraints.
In the language of information theory, we scan through distributions $p(\Phi)$ to find the one that maximizes the relative entropy functional $S[p,q]$ subject to these constraints~\cite{Jaynes:1957zza,Jaynes:1957zz}. This maximal distribution is the desired posterior distribution $p^\star(\Phi)$.
In the language of statistics, maximizing the relative entropy between $p(\Phi)$ and $q(\Phi)$ is equivalent to minimizing the KL divergence between them:
\begin{align}
\label{eq:KL_divergence}
\mathcal{L}_{\mathrm{KL}}[p \| q] = - S[p,q] = \int d \Phi \,p(\Phi) \ln \frac{p(\Phi)}{q(\Phi)}\,.
\end{align}
While this is not the only measure of statistical similarity one could use, the KL divergence guarantees positive per-event weights, unlike alternative $f$-divergences discussed in \App{alternative_measures}.

The constraints take the form of expectation values over some measurement functions $m_i(\Phi)$.
For a generic distribution $p(\Phi)$, this expectation value is:
\begin{equation}
\label{eq:expmi}
\langle m_i\rangle_p = \int d\Phi\, p(\Phi)\, m_i(\Phi)\,.
\end{equation}
As discussed in \Sec{qcd_inputs}, this measurement function is often defined in terms of observable distribution moments, though more general structures are possible, so we use the generic $m_i(\Phi)$ notation here.
We denote the precision constraints as $c_i$, such that our goal is to find a $p^\star(\Phi)$ minimizing~\eq{KL_divergence} while satisfying:
\begin{align}
\label{eq:constraints}
    \langle m_i\rangle_{p^\star} = c_i\,.
\end{align}
The most basic constraint is that the distribution $p(\Phi)$ is normalized:  
\begin{align}
    m_0(\Phi) = 1\,, \quad c_0 = 1 \qquad \Rightarrow \qquad \int d\Phi\, p(\Phi) = 1\,.
\end{align}

The constrained minimization of~\eq{KL_divergence} subject to~\Eq{constraints} is equivalent to finding a stationary point of a 
new loss function:
\begin{align}
\label{eq:totalLoss}
\mathcal{L}[p, q]=
\mathcal{L}_{\mathrm{KL}}[p \| q]
+
(\lambda_0-1) \big(\langle m_0 \rangle_p - 1 \big)
+
\sum_i \lambda_i \, \big(\langle m_i\rangle_p - c_i\big)\,,
\end{align}
with respect to $p(\Phi)$ and the Lagrange multipliers $\lambda_i$.
Note that we have separated out the normalization requirement with its $\lambda_0-1$ Lagrange multiplier for later convenience. 
By construction, solving the stationary equations resulting from this loss yields the posterior distribution $p^\star(\Phi)$ that maximizes uncertainty (i.e.~entropy) subject to constraints $c_i$.
This is the same principle used in statistical mechanics to derive ensembles, where you can predict the state of a gas by maximizing your \emph{uncertainty in information} (i.e.~Shannon entropy) while imposing the \emph{information you do know} (i.e.~average energy).
In that case, one derives the Boltzmann distribution as the posterior distribution describing the canonical ensemble.

The result of this optimization is easiest to express in terms of the weight function, which is also the object we need for reweighting:
\begin{equation}
    \label{eq:weight_definition}
    w(\Phi) = \frac{p(\Phi)}{q(\Phi)}\,.
\end{equation}
With this notation, the loss in \Eq{totalLoss} can be written as:
\begin{align}
\mathcal{L}[p, q] = \int d\Phi \, q(\Phi) \Big[w(\Phi) \ln w(\Phi)
+  
(\lambda_0-1) \left(w(\Phi) - 1 \right)
+
\sum_i \lambda_i \, \big(w(\Phi) \, m_i(\Phi) - c_i \big)
\Big]\,,
\end{align}
where we are assuming that the prior $q(\Phi)$ is normalized as $\int d\Phi \, q(\Phi) = 1$.
Performing a functional variation of the loss with respect to $w(\Phi)$ to find a stationary solution with respect to $p(\Phi)$, the solution satisfies:
\begin{align}
\label{eq:weight_sol}
    \frac{\delta \mathcal{L}}{\delta w} = 0 \qquad \Rightarrow \qquad w(\Phi; \lambda_0,\boldsymbol{\lambda}) = \exp \left[-\lambda_0 - \sum_i \lambda_i \, m_i(\Phi)\right],
\end{align}
where we have made explicit that the weight function depends on the vector of Lagrange multipliers $\boldsymbol{\lambda} = \{\lambda_i\}$.

The posterior distribution also needs to be stationary with respect to $\boldsymbol{\lambda}$.
To find the optimal $\boldsymbol{\lambda}^\star$, we can substitute \Eq{weight_sol} back into the loss function in \Eq{totalLoss}.
This yields a ``dual'' objective that only depends on the Lagrange multipliers to minimize:
\begin{equation}
    \mathcal{J}(\boldsymbol{\lambda}) \equiv -\mathcal{L}|_{w=w(\Phi; \lambda_0,\boldsymbol{\lambda})}=e^{-\lambda_0} \, Z(\boldsymbol{\lambda}) - 1 + \lambda_0 + \sum_i \lambda_i \, c_i\,,
\end{equation}
where we have defined the partition function:
\begin{equation}
    Z(\boldsymbol{\lambda}) \equiv \int d\Phi \, q(\Phi) \exp \left[- \sum_i \lambda_i \, m_i(\Phi)\right].
    \label{eq:partition_function}
\end{equation}
Solving for $\lambda_0$ is straightforward:
\begin{equation}
\frac{\partial \mathcal{J}}{\partial \lambda_0} = 0 \qquad \Rightarrow \qquad \lambda_0 = \ln Z(\boldsymbol{\lambda})\,,
\end{equation}
which yields a simpler form for the dual objective:
\begin{equation}
    \mathcal{J}(\boldsymbol{\lambda}) \equiv \ln Z(\boldsymbol{\lambda}) + \sum_i \lambda_i \, c_i\,.
\label{eq:dual_objective}
\end{equation}

While the dual objective cannot be minimized in closed form, it is straightforward to minimize numerically because it is a convex optimization problem.%
\footnote{By contrast, a mean-squared error objective $\sum_i (c_i - \langle m_i \rangle_p)^2$ built from the same moment residuals shares the same global minimum but develops approximately flat directions that cause premature convergence when the number of constraints is large.}
Using the fact that $q(\Phi) \, w(\Phi) = p(\Phi)$ by \Eq{weight_definition}, the gradient of the dual objective is:
\begin{equation}
\label{eq:gradient_of_dual}
    \frac{\partial \mathcal{J}}{\partial \lambda_i}
= c_i - \int d\Phi\, p(\Phi) \, m_i(\Phi) = c_i - \langle m_i\rangle_{p}\,,
\end{equation}
so minimizing $\mathcal{J}$ is equivalent to finding a solution to the constraints from \Eq{constraints}.
The Hessian of the dual objective is:
\begin{align}
\frac{\partial^2 \mathcal{J}}{\partial \lambda_i\partial \lambda_j}
&= \int d\Phi\, p(\Phi) \, m_i(\Phi) \, m_j(\Phi) - \left(\int d\Phi\, p(\Phi) \, m_i(\Phi) \right)\left(\int d\Phi\, p(\Phi) \, m_j(\Phi) \right) \nonumber \\
&= {\rm Cov}_{p}(m_i,m_j)\,.
\label{eq:cov_hessian}
\end{align}
Because the Hessian takes the form of a covariance matrix, which is positive semi-definite by construction, the dual objective is convex.
Therefore, assuming the constraints in \Eq{constraints} are compatible and non-degenerate, there is a unique optimal $\boldsymbol{\lambda}^\star$, which yields the weight function:
\begin{align}
\label{eq:wweight}
w(\Phi; \boldsymbol{\lambda}^\star) = \frac{1}{Z(\boldsymbol{\lambda}^\star)} \exp \left[-\sum_i \lambda_i^\star \, m_i(\Phi)\right].
\end{align}

In practice, we cannot perform the integral in \Eq{partition_function} analytically, so we have to estimate it numerically.
A Monte Carlo (MC) generator sampling from $q(\Phi)$ produces a discrete set of $N_q$ events $\{\Phi_a\}_{a=1}^{N_q}$, yielding the estimate:
\begin{equation}
    Z(\boldsymbol{\lambda}) \approx \sum_{a=1}^{N_q} \exp \left[- \sum_i \lambda_i \, m_i(\Phi_a)\right].
    \label{eq:sampled_Z}
\end{equation}
Similarly, the per-event weights are:
\begin{equation}
\label{eq:wa}
    w_a \equiv w(\Phi_a; \boldsymbol{\lambda}^\star).
\end{equation}
We discuss more about the practical strategy to identify $\boldsymbol{\lambda}^\star$ in \Sec{optimization}.
In this paper, we assume that the precision inputs $c_i$ are perfectly known with no uncertainties.
In \App{uncertainties}, we discuss some of the issues involved when accounting for theoretical uncertainties on the $c_i$, including covariances.

\subsection{Precision inputs from QCD theory}
\label{sec:qcd_inputs}

In order to leverage this maximum-entropy reweighting strategy, we have to provide constraints $c_i$ to input into \Eq{constraints}.
If we had complete first-principles knowledge of QCD, then we could simply compute these constraints via:
\begin{equation}
    c_i = \int d\Phi \, p_{\rm QCD}(\Phi) \, m_i(\Phi)\,,
\end{equation}
where $p_{\rm QCD}(\Phi)$ has complete information about all of phase space.
Of course, if we already knew $p_{\rm QCD}(\Phi)$, then we could just use it as the prior directly and avoid the need to reweight entirely. 
In practice, we are nowhere near having $p_{\rm QCD}(\Phi)$ that can be used for arbitrary measurement functions $m_i(\Phi)$, but there are some measurement functions for which we can derive excellent approximations for $c_i$.

Consider a scattering process governed by a fixed hard scale $Q$, such
as the center-of-mass energy in $e^+e^-\!\to$ hadrons.
This system evolves into a final state defined by a variable number of particles $N$ that occupy a state in the full phase
space:
\begin{equation}
    \Phi \equiv \bigcup_N \Phi_N.
\end{equation}
Here, $N$-particle
Lorentz-invariant phase space (LIPS) in $d$ dimensions is defined as
\begin{align}
d\Phi_N=\delta^{(d)}\left(q-\sum_{n=1}^N p_n\right)
\prod_{n=1}^N \frac{d^{d-1} \mathbf{p}_n}{(2 \pi)^{d-1} 2 E_n}\,.
\end{align}
The probability density for finding the system in a specific $N$-particle
configuration $\Phi_N$ is given by the squared matrix element:
\begin{align}
\label{eq:Pphi}
p_{\rm QCD}(\Phi_N)\equiv \frac{1}{\sigma_{\rm tot}}
\frac{1}{2 Q^2} \left|\mathcal{M}_N (\Phi_N)\right|^2\,,
\end{align}
where $1/(2Q^2)$ is the flux factor and
$\left|\mathcal{M}_N\right|^2$ is the squared matrix element for the $N$-particle configuration.
The normalization is given by
the total cross section:
\begin{equation}
    \sigma_{\rm tot} = \sum_{N=2}^{\infty} \int d\Phi_N
\frac{1}{2 Q^2} \left|\mathcal{M}_N(\Phi_N)\right|^2,
\end{equation}
which ensures that $\int d\Phi\, p_{\rm QCD}(\Phi) = 1$.

Estimating $p_{\rm QCD}(\Phi)$ for an arbitrary phase-space
configuration is the central task of parton-shower event generators,
whether at parton level (before hadronization) or at hadron level
(after nonperturbative fragmentation).
Since the calculation of
$\left|\mathcal{M}_N\right|^2$ for high multiplicities and at
arbitrary loop orders is a monumental task, parton-shower generators
approximate these amplitudes by exploiting the universal
factorization of QCD in singular limits of $|\mathcal{M}_N|^2$.
Most parton-shower generators also interface with hadronization models to convert the phase space distribution for partons into a phase space distribution for hadrons.

Analytic approaches, by contrast, focus on computing specific observables to high (typically perturbative) accuracy.
Most commonly, one is interested in computing the multi-differential distribution for a set of event-by-event observables:
\begin{equation}
    \vec{v}(\Phi) = \{v_1(\Phi),v_2(\Phi),  \ldots, v_K(\Phi)\}.
\end{equation}
In this way, QCD calculations project the full phase-space distribution onto the space of observables $\vec{v}$:
\begin{equation}
\label{eq:r}
r(\vec{v}) \equiv \int d\Phi\, p_{\rm QCD}(\Phi)\,
\prod_{k=1}^K \delta\big(v_k - \hat{v}_k(\Phi)\big),
\end{equation}
where $r(\vec{v})$ is a probability density in observable space.
Note that event-by-event observables are not the only quantities that can be computed to high accuracy in perturbative QCD.
For example, energy
correlators~\cite{Basham:1978bw,Moult:2025nhu} yield an energy-weighted distribution of angular factors per event.
Since our reweighting framework operates event by event, we focus on event-by-event observables as precision inputs. As we explain in \Sec{relation_to_energy_correlators}, moments of energy correlators turn out to be closely related to the moments of EFPs that we use in this paper, providing a bridge between energy correlator calculations and our framework.

While analytic
calculations of $r(\vec{v})$ achieve significantly higher
precision than their parton-shower counterparts in many regions of phase space, they are
inherently less exclusive than the full event-level distributions
provided by parton-shower generators.
Following the information-theoretic perspective, we want to extract precision constraints from these calculations and use them to reweight parton-shower generators.
We cannot use \Eq{r} directly, though, since that would correspond to using a measurement function $m(\Phi) = \prod_k \delta\big(v_k - \hat{v}_k(\Phi)\big)$ in~\eq{expmi}, and one cannot exponentiate a Dirac delta function. More generally, each measurement function in~\eq{expmi} must map an event to a single real number.
One option would be to build a coarse-grained measurement function, e.g.~a histogram or a kernel-smoothed variant. Instead, we advocate for computing a set of moments, which are smooth and directly connected to the perturbative structure discussed in \Sec{Constraints}:
\begin{align}
\label{eq:mom}
\langle g_i \rangle \equiv \int d\vec{v} \,r(\vec{v})\,
g_i(\vec{v}) \,,
\end{align}
where the functions $g_i(\vec{v})$ determine which aspects of the precision calculation are imported as constraints.%
\footnote{The astute reader will notice that histograms can also be expressed in the form of \Eq{mom}, where $g_i(\vec{v})$ is an indicator function for each histogram bin.
There is nothing in the information-theoretic approach that forbids the use of histograms in this way, though it is less natural given the discussion in \Sec{Constraints}.}
Written as a measurement function, this corresponds simply to:
\begin{equation}
\label{eq:measg}
    m_i(\Phi) = g_i\big(\vec{v}(\Phi)\big).
\end{equation}
In this way, moments convert event-by-event observable distributions into ensemble-level constraints.

\subsection{Choice of constraints: observable moments}
\label{sec:Constraints}

At the core of the maximum-entropy principle is the systematic use of precision inputs $c_i$ as constraints.
In our framework for improving parton-shower generators, these inputs are provided by specific observable moments $\langle g_i \rangle$ via \Eq{mom}.
This leads to a critical question: what are the optimal inputs to employ for maximal information gain? 

Focusing on $e^+e^-$ event-shape observables, a wide variety of \emph{single-differential} event shape observables have been computed to high accuracy in both the fixed-order and resummation regimes.
Next-to-next-to-leading order (NNLO) fixed-order calculations can be carried out using tools such as NNLOJet~\cite{Gehrmann-DeRidder:2007vsv,Gehrmann-DeRidder:2007foh,NNLOJET:2025rno}, CoLoRFulNNLO~\cite{DelDuca:2016ily}, or EERAD3~\cite{Weinzierl:2008iv}.
Furthermore, many event-shape observables have been computed with resummation precision at next-to-next-to-leading logarithmic (NNLL) accuracy, some even reaching N$^3$LL and N$^4$LL~\cite{Benitez:2025vsp,Hoang:2025uaa,Benitez:2024nav,Jaarsma:2025tck}.
While precision calculations of \emph{multi-differential} observables are less common~\cite{Procura:2018zpn} and generally less precisely known, our framework remains agnostic about whether the precision inputs $\{\langle g_1\rangle, \langle g_2\rangle, \dots \}$ are derived from single-differential event shapes or involve correlations between multiple observables.
That is, the measurement function $m_i(\Phi)$ in~\eq{measg} can be derived from a single- or a multi-differential distribution.
The reweighting procedure remains identical, and any choice of precision constraints will yield a positive reweighting factor to upgrade the parton-shower prior.

With precision event-shape distributions at hand, we aim to design constraints that are more accurately determined than the corresponding predictions from the parton-shower prior.
Even within a specific distribution of an observable $v$, certain kinematic regions (such as the fixed order and resummation regions) may be much better known analytically, whereas other regions of the phase space might be better described by the prior due to effects such as hadronization modeling.
By tailoring the functional form of $g_i(\vec{v})$, we can weight these regions differently.
Specifically, for precision inputs, we can prioritize the regions where analytical computations are more reliable than the priors.
Conversely, one could construct $g_i(\vec{v})$ to weight the regions where the parton-shower prior is trusted more.
In this latter case, we may either include the moment evaluated from the parton-shower prior as a constraint to ensure that information is preserved during reweighting, or simply omit that particular moment from the optimization; we return to this point in \Sec{accuracy}.

To design a suitable form for  $g_i(\vec{v})$ to capture the precision analytic information, we look to the structure of the perturbative expansion.
Let us first consider a single-differential distribution.
For a generic Sudakov observable $v$ (assuming $v\to 0$ is the singular limit), the differential cross section can be organized by a logarithmic expansion as:
\begin{align}
\label{eq:sudakovobs}
r(v) = \sum_{m=1}^{\infty}\sum_{n=1}^{2m-1}k^{\rm LP}_{mn}\,a_s^m \left[\frac{\ln^{n}v}{v}\right]_+
+\cdots+\sum_{m=1}^{\infty}\sum_{n=1}^{2m-1}k^{{{\rm N}}^k{\rm LP}}_{mn}a_s^m \,\frac{\ln^{n}v}{v}\,v^{k}\,,
\end{align}
where $a_s = \alpha_s/(4\pi)$ and the label N$^k$LP indicates suppression by $v^{k}$ relative to the leading power (LP).
In the resummation region ($v \ll 1$), the LP terms exponentiate as:
\begin{align}
\label{eq:LPeq}
r(v)_{\rm LP} &= \frac{d}{dv}\Bigg[\Bigg(1+\sum_{m=1}^{\infty} C_m^{[0]}\, a_s^m\Bigg) \times\exp \Bigg[\sum_{i=1}^{\infty} \sum_{j=1}^{i+1} G_{i j} \, a_s^i \, \ln ^j v\Bigg]\Bigg]\,.
\end{align}
Here, the ``Sudakov logarithms'' in the exponent capture the universal collinear and soft dynamics.
Comparing this physical structure with the weight factor in \Eq{wweight}, this motivates the use of logarithmic moments as the natural choice to parameterize the resummation region: 
\begin{align}
\label{eq:logmom}
g_{0n}(v) \equiv \ln^n v\,.
\end{align}

In the language of~\References{Larkoski:2013paa,Larkoski:2015lea,Cal:2020flh,Cal:2021fla}, these logarithmic moments are known as Sudakov-safe observables, which require Sudakov resummation to regulate their singular structure.
Repeating the discussion from~\Reference{Assi:2025ibi}, we can illustrate the unique behavior of Sudakov-safe observables using the LL distribution for thrust~\cite{Farhi:1977sg,Catani:1992ua} at fixed coupling (f.c.):
\begin{align}
\label{eq:rtauLL}
r(\tau)_{\rm LL,f.c.} =  -8 a_s C_F\frac{\ln \tau}{\tau} \exp \left[-4 a_s C_F\ln ^2 \tau\right]\,.
\end{align} 
This distribution yields the following logarithmic moments:
\begin{align}
\label{eq:taulnmom}
\langle \ln^n\tau\rangle_{\rm LL,f.c.} &= \int_0^{\tau_{\rm max}} d\tau\, r(\tau)_{\rm LL,f.c.} \ln^n\tau \nn\\
&=(-1)^n\left(\frac{1}{4 a_s C_F}\right)^{n/2}\Gamma\left[1+\frac{n}{2}\right]\,,
\end{align}
where we take $\tau_{\rm max}=1$, instead of the physical $\frac{1}{2}$, for simplicity.
For generic $n$, $\langle \ln^n\tau \rangle_{\rm LL,f.c.}$ features fractional, negative powers of $a_s$, which is characteristic of Sudakov-safe observables.
In particular, such fractional powers cannot be obtained at any fixed order in $a_s$, but naturally appear in the context of resummation.

Beyond the resummation region, there are many precision fixed-order calculations available.
These calculations are important in the fixed-order region, which describes non-singular hard emissions where $v \sim 1$.
In this region, the logarithmic expansion breaks down, and the cross section is better described by a polynomial expansion in $v$.
This motivates using polynomial moments to effectively capture fixed-order corrections:
\begin{align}
\label{eq:linmom}
g_{m0}(v) \equiv v^m\,.
\end{align}

Finally, to bridge the gap between the deep resummation region and the hard tail, and to capture subleading power logarithmic effects (N$^k$LP), we can employ mixed moments:
\begin{align}
\label{eq:mixed_moment}
g_{mn}(v) \equiv v^m\ln^n v\,.
\end{align}
Note that this definition contains~\eqs{logmom}{linmom} as special cases.
The choice of the powers $m$ and $n$ dictates the balance between emphasizing the peak resummation region versus the tail fixed-order region of the distribution.
This logic extends naturally beyond single observables.
Given a set of observables $\{v_k(\Phi)\}_{k=1}^K$, we can construct joint constraints as products:
\begin{align}
g_{m_1n_1\cdots m_K n_K}(v_1,\cdots, v_K)=\prod_{k=1}^{K} g_{m_k n_k} \big(v_k(\Phi)\big)\,.
\label{eq:product_features_choice}
\end{align}
These cross-moments inject information about joint correlations invisible to marginal distributions.
While precise multi-differential calculations are currently rare, this framework is ready to ingest such correlated information as soon as it becomes available.

We note that not all event-shape observables are naturally described by the form in~\eq{sudakovobs}.
For example, ratios of Sudakov observables do not follow such form.
A well-motivated example is the ratio of $N$-jettiness observables~\cite{Stewart:2010tn,Thaler:2010tr}:
\begin{equation}
    \tau_{N+1;N} \equiv \frac{\tau_{N+1}}{\tau_N}\,,
\end{equation}
which is commonly used to detect additional resolved radiation relative to an $N$-jet description.
Such ratio observables can be written as projections of a correlated double-differential distribution:
\begin{align}
r(\tau_{N+1;N}) = \int d\tau_N \,d\tau_{N+1}\, r(\tau_N, \tau_{N+1})\, \delta\!\left(\tau_{N+1;N}-\frac{\tau_{N+1}}{\tau_N}\right).
\end{align}
While each of $\tau_N$ and $\tau_{N+1}$ exhibits Sudakov logarithms in its own small-$\tau$ limit, the ratio probes their \emph{joint} singular structure.
Therefore, this ratio also falls within the class of Sudakov-safe observables.
This kind of projected observable is well-captured by the correlated moments in~\eq{product_features_choice}.
In particular, large logarithms of the ratio,
$\ln \tau_{N+1;N}$,
arise when $\tau_{N+1}\ll \tau_N$.
This includes both the ``resolved'' regime $\tau_{N+1}\ll \tau_N\sim 1$, best captured by $\langle\ln^{n_1} \tau_{N+1} \, \tau_{N}^{m_2}\rangle$ moments, as well as the strongly ordered regime $\tau_{N+1}\ll \tau_N\ll 1$, best captured by $\langle\ln^{n_1} \tau_{N+1} \, \ln^{n_2} \tau_{N}\rangle$ moments.
In the fixed-order region $\tau_{N+1;N}\sim \mathcal{O}(1)$, $\langle\tau_{N+1}^{m_1} \, \tau_{N}^{m_2}\rangle$ moments would capture it best.
All these moments are consistent with our use of multi-observable constraint families in \Eq{product_features_choice}.%
\footnote{In practice, it may be more efficient to use $g_{0n}(\vec{v}) = \ln^n(\tau_{N+1}/\tau_N)$ and $g_{m0}(\vec{v}) = (\tau_{N+1}/\tau_N)^m$ directly as measurement functions, targeting the correlated singular structure without requiring a large set of individual-variable products; the general framework of \Eq{mom} accommodates such non-factorized measurement functions.
That said, moments of the ratio such as $\langle \ln^n(\tau_{N+1}/\tau_N) \rangle$ decompose via the binomial expansion into cross-moments $\langle \ln^{n-k}\tau_{N+1} \, \ln^k\tau_N \rangle$ already contained in the product family in \Eq{product_features_choice}.}

Finally, analytic control is not uniform across phase space.
In deep non-perturbative regimes, or for observables where theoretical calculations are lacking, the parton-shower generator itself may provide the most accurate description.
In such cases, we can design $g_i(\vec{v})$ to suppress sensitivity to these regions.
The simplest option is to restrict the integration domain with a cutoff to exclude the deep nonperturbative region, e.g.\ $g(v) = \ln^n v \, \Theta(v - v_0)$, where $v_0 \sim \Lambda_{\rm QCD}/Q$ is a typical nonperturbative scale.
Smoother alternatives include, for instance, replacing $\Theta(v - v_0)$ with a sigmoid-like function.
If necessary, moments computed from the parton-shower prior itself can also serve as constraints to ensure that trusted features of the prior are preserved during reweighting.

\subsection{Choice of priors: reweighting efficiency}
\label{sec:efficient_priors}

Our method constructs a posterior distribution on exclusive phase space by updating a prior $q(\Phi)$ with a finite set of precision constraints.
This raises the question of how to choose the prior $q(\Phi)$.
A simple guiding principle is that the posterior inherits \emph{all} aspects of the event ensemble that are left unconstrained: in any region of phase space (or along any direction in observable space) to which the imposed moments are only weakly sensitive, the posterior remains close to the prior.
Therefore, the schematic in~\Fig{map} holds practical weight: the closer the prior is to the truth $p_{\rm QCD}(\Phi)$, the higher the quality of the resulting posterior will be for a fixed set of constraints.

There is a clear hierarchy in the practical usefulness of candidate priors.
For example, a minimal prior $q_{\rm uni.}(\Phi)$ that is uniform with respect to Lorentz-invariant phase space (corresponding schematically to $|\mathcal{M}_N|^2=\mathrm{const.}$) lacks the fundamental IRC singularity structure of QCD.
Consequently, reproducing a realistic QCD distribution from such a flat prior would require an impractically rich set of constraints to build up the singularity structure from scratch.

This can be illustrated using a fixed-coupling LL model for thrust, given already in~\eq{rtauLL}.
Consider a prior that captures the correct leading-order singular structure (the prefactor) but misses the Sudakov exponent (the resummation):%
\footnote{Because of the delta and plus functions, this prior is not a proper probability distribution, but it suffices for illustrative purposes.}
\begin{align}
\label{eq:qLO}
q_{\rm LO}(\tau) = \delta(\tau)-8 a_s C_F \left[\frac{\ln\tau}{\tau}\right]_+\,.
\end{align}
Choosing the constraint $\langle \ln^2\tau \rangle_{\rm LL,f.c.}$ from~\eq{taulnmom} with $n=2$, the normalization and moment conditions for $q_{\rm LO}$ read (still keeping $\tau_{\rm max}=1$ for simplicity):
\begin{align}
\label{eq:constrtauex}
\int_0^1 d\tau\, q_{\rm LO}(\tau)\,e^{-\lambda_0} e^{-\lambda_2 \ln^2 \tau}&= 1\,,\nonumber\\
\int_0^1 d\tau\, q_{\rm LO}(\tau)\, e^{-\lambda_0}e^{-\lambda_2 \ln^2 \tau} \ln^2\tau&=  \frac{1}{4 a_s C_F}\,.
\end{align}
Here, $\lambda_0$ is capturing the reweighting needed for the normalization and $\lambda_2$ is capturing the reweighting needed for the double-logarithmic terms in the exponential.
Since the target distribution in \Eq{rtauLL} differs from $q_{\rm LO}$ only by the Sudakov factor $\exp[-4 a_s C_F \ln^2\tau]$, the solution is given by:
\begin{equation}
    (\lambda_0,\lambda_2)=(0,4 a_s C_F)\,.
\end{equation}
Therefore, our prior $q_{\rm LO}$ is improved to have the correct Sudakov exponent simply by minimizing the loss with constraints given by~\eq{constrtauex}.

By contrast, consider using a uniform prior $q_{\rm uni.}(\Phi)$ that carries no information about the singular structure of QCD.
In that case, no finite set of polynomial or logarithmic moments could capture the missing information to restore the $1/\tau$ behavior.
One can of course exponentiate any function, so we could consider expectation values of something like $g(\tau) = \ln(\ln\tau/\tau)$ to restore the leading-order QCD singularity.
But this iterated logarithmic structure would not be very well-motivated from the considerations in \Sec{Constraints}, nor would it be known \textit{a priori} how to model generic observables, defeating the purpose of trying to build a systematic basis of constraints.

Note that even though the prior in \Eq{qLO} led to the desired posterior in the LO case, this depended crucially on the set of constraints we imposed.
For example, if we only used the single-logarithmic moment $\langle\ln\tau\rangle_{\rm LL,f.c.}$ from~\eq{taulnmom}, but not the double-logarithmic moment $\langle\ln^2\tau\rangle_{\rm LL,f.c.}$, then the posterior would take the form:
\begin{equation}
p(\tau) = q_{\rm LO}(\tau)\, e^{-\lambda_0}\, e^{-\lambda_1 \ln\tau}\,,
\end{equation}
with Lagrange multipliers:
\begin{equation}
    (\lambda_0, \lambda_1) = \left(\ln \frac{\pi}{8}, -8 \sqrt{\frac{a_s C_F}{\pi}} \right).
\end{equation}
The resulting weight function scales as $\tau^{-\lambda_1} \sim \tau^{\sqrt{a_s}}$, which is a power-law tilt rather than the expected Sudakov double-logarithmic suppression 
$e^{-\lambda_2 \ln^2\tau}$.
Including $\langle\ln\tau\rangle_{\rm LL,f.c.}$ \emph{in addition} to $\langle\ln^2\tau\rangle_{\rm LL,f.c.}$ does yield a consistent result with $\lambda_1 = 0$.
Therefore, in addition to choosing a prior that is well-matched to the physics of interest, it is crucial to include a sufficient set of constraints to give the posterior distribution the functional flexibility needed to capture the correct information.

As a practical way to diagnose possible tensions between the prior and the constraints, we can compute the effective
sample fraction (ESF):%
\footnote{By the normalization constraint, the ESF numerator will be 1, but it is conceptually helpful to write it out this way.}
\begin{equation}
\mathrm{ESF} = \frac{\left(\frac{1}{N_{q}}\sum_{a=1}^{N_{q}} w_a \right)^2}{\frac{1}{N_{q}}\sum_{a=1}^{N_{q}} w_a^2}\,,
\label{eq:ESF}
\end{equation}
where $w_a$ are the event-by-event weights from \Eq{wa}.
The ESF equals unity for uniform weights and decreases as the weights
become concentrated on fewer events, providing a direct measure of the
effective statistical power of the reweighted sample as a fraction of
the original.
If $\mathrm{ESF} \ll 1$, this indicates that the constraint set is demanding structure the
prior cannot comfortably supply -- either because the prior lacks support
in the relevant phase-space region, or because the constraints are
sufficiently numerous that only a thin slice of the prior sample can satisfy all of them simultaneously.
In either case, the reweighting has exceeded the information capacity of the prior sample.
Ideally, we would like to choose priors where $\mathrm{ESF} \simeq 1$, such that the posterior is supported by a broad fraction of
the prior sample.

Because better priors yield more efficient reweightings, our framework works in tandem with efforts to systematically improve parton-shower priors.
This contrasts with the standard application of maximum entropy logic in the context of statistical mechanics, where the prior is typically a simple microcanonical ensemble (uniform phase space) constrained only by conserved quantities like energy.
In those systems, one rarely attempts to engineer a ``highly accurate'' prior to minimize the variance of the reweighting factors.
This is because the unconstrained degrees of freedom in statistical systems are generally ergodic and irrelevant to the few macroscopic state variables that one cares about.
In QCD, however, the situation is fundamentally different.
We are interested in the detailed structure of the full $N$-particle phase space, not just a few macroscopic parameters.
Furthermore, our ``precision'' inputs are themselves approximations with inherent uncertainties.
Therefore, we want to select priors that have as accurate $N$-particle phase space information as possible and carefully design the precision constraints to get the best possible posterior distributions.

\subsection{Formal accuracy of the resulting posterior}
\label{sec:accuracy}

Our reweighting approach can accommodate any consistent set of constraints, even if they have been computed to different perturbative accuracies.
This raises the question of what the formal accuracy of the posterior distribution $p^\star(\Phi)$ is after reweighting.
In a strict sense, our procedure guarantees formal accuracy only for the specific moments imposed as constraints: if a set of precision inputs $\{\langle g_i\rangle\}$ is computed at N$^{k}$LL + N$^{m}$LO, then the marginal distributions of the posterior are guaranteed to inherit such moments at the same level of the accuracy.

While this may appear to be a restricted claim, it is crucial to recognize that standard accuracy benchmarks for parton showers are also restricted (albeit less so).
For example, the statement that a shower is ``NLL accurate'' is never a blanket guarantee for the full $N$-particle phase space.
Rather, it indicates that the shower reproduces the correct NLL Sudakov exponent for a specific class of observables, typically global, recursive IRC-safe observables~\cite{Banfi:2004yd}.
In this sense, we can only state that the accuracy of the posterior distribution is inherited from the prior distribution at a certain order, with precision constraints on specific observables provided at higher orders.

One might worry that the imposition of higher-order constraints in one region of phase space could inadvertently degrade the accuracy of the prior in other regions.
For instance, if a prior $q(\Phi)$ possesses verified NLL accuracy, could 
reweighting it to match a specific NNLL constraint distort the underlying 
probability density such that NLL accuracy is lost for other observables?
In principle, such degradation cannot be rigorously ruled out: there can exist 
joint distributions of thrust and broadening that are consistent with a 
marginal NNLL thrust distribution but exhibit less-than-NLL accuracy in 
the broadening projection.
The maximum-entropy principle should mitigate this 
by design -- it makes the smallest possible update to the prior consistent 
with the constraints, leaving all unconstrained directions as close to the 
prior as possible -- but it does not eliminate the possibility.

To gain some intuition about when degradation can occur, it helps to visualize where the posterior lives in the space of all possible probability distributions.
One can think of the prior as defining a point in this space, and the constraints $\langle m_i \rangle_p = c_i$ as defining a manifold.
The posterior is then the point on this manifold that is closest (as measured by the KL divergence) to the prior.%
\footnote{Note that the KL divergence is not symmetric, so it does not define a proper notion of distance in this space, but the intuition still holds.}
Now imagine projecting this multi-dimensional probability space down to the one-dimensional probability space of a particular observable of interest.
If this projection is ``orthogonal'' to the constraint manifold, then the distribution for this observable is unchanged from the prior, by construction.
Degradation can therefore only occur along
projections that are partially correlated with the constraints --
correlated enough to be moved by the reweighting, but not constrained
enough to be moved correctly.
Whether this occurs in practice depends on
the relationship between the constraint set, the prior, and the multi-observable correlation structure of QCD.

As a practical safeguard against degradation, if a specific observable or kinematic region is known to be well-modeled by the prior, moments of that observable calculated from the prior itself can be included in the constraint set to 
explicitly enforce preservation during the reweighting.
Alternatively, one can perform various diagnostics to assess the degree to which one should trust the distribution for unconstrained observables.
The ESF metric in \Eq{ESF} is one diagnostic tool, since if the reweighting is efficient, that tells you that the prior has been minimally distorted by the constraints.
A complementary diagnostic is to vary the choice of prior.
If the posterior for a particular observable is robust across priors that differ in their treatment of subleading effects -- for example, through variations of the shower splitting kernels, the choice of evolution variable, or the value of $\alpha_s$ -- then the constraints are genuinely determining the 
posterior, and the result is not an artifact of the specific prior used.
Conversely, if the posterior for a given observable depends strongly on 
the prior even after reweighting, this indicates that the constraint set 
does not carry sufficient information about that observable, and the 
residual prior dependence is physical: it marks a direction in observable 
space that the current constraints do not control.
Monitoring information transfer 
quality across varied priors therefore separates constrained directions, 
where the posterior is determined by the precision inputs, from 
prior-dominated directions, where further constraints or better priors 
are needed.

In the numerical study in \Sec{results}, we adopt these diagnostic strategies.
Specifically, we systematically enlarge the constraint set while varying the prior, and 
monitor transfer to non-constrained observables alongside weight-health 
diagnostics, to verify that improved agreement reflects genuine information 
transfer rather than posterior collapse.
This diagnostic is also related to the practical question of convergence:  how effectively does information from a finite set of precision constraints propagate to the full phase space?
We find that convergence happens quickly, such that even with unphysically distorted priors and relatively few constraints, the posteriors are robust across a range of observables with efficient weights.

To summarize, it is worth distinguishing sharply which results in this framework are provable and which are empirical.
Provable is the accuracy of the constrained moments themselves.
The posterior reproduces each imposed moment by construction, so if a precision input is computed at N$^{k}$LL + N$^{m}$LO, then the corresponding moment of the posterior inherits exactly that accuracy.
This guarantee applies to a finite set of moments, not to full distributions.
Empirical is everything else.
At the level of expectation values, the dichotomy is exact, since the posterior mean of any observable lying in the linear span of the constrained measurement functions is fixed by the imposed moments, while the residual deviation from the truth is carried by the component outside that span (together with any accuracy limitations of the inputs themselves).
For full distributions, this span-based picture is a heuristic rather than a theorem, but it explains the qualitative pattern seen in our numerical studies, where observables dominated by the same soft and collinear dynamics as the constraints are corrected almost completely, while observables probing multi-particle correlations orthogonal to the constraint set retain residuals.
At present, we cannot assign a formal logarithmic accuracy to unconstrained predictions, nor predict a priori at which order a given observable will improve for a given constraint set.
The numerical studies of \Sec{results} and \App{substructure}, including the cross-generator test in \Sec{herwig}, are designed to characterize this information transfer quantitatively.
Establishing a criterion that connects the accuracy of the inputs to the accuracy of the posterior for a well-defined class of observables remains an important open problem.

\section{Energy flow polynomials}
\label{sec:efps}

Our proof-of-concept study will be based on moments of EFPs~\cite{Komiske:2017aww}.
Because any IRC-safe observable can be expressed as a linear combination of EFPs to any desired accuracy, the EFPs provide a natural set of observables from which to build precision constraints.
Currently, there is a lack of precision calculations of EFP distributions and no computations of their moments, so our proof-of-concept study will be based on EFP constraints derived from a high-fidelity shower.
That said, we anticipate that the special structure of the EFPs will enable systematic precision QCD calculations of their moments, which we plan to study in future work.
In this section, we define the EFPs, review their properties, and describe how to use them to build moment constraints.
We also mention some limitations of the EFPs when considering logarithmic moments, and show how EFPs and energy correlators share the same polynomial moment structures.

\subsection{Definition and graph-theoretic properties}

In $e^+e^-$ collisions, we define EFPs within a specific phase-space region $\mathcal{R}\subseteq \Phi$.
Here, $\mathcal{R}$ can represent the full event, an individual jet, or any other deterministically selected region.
For our proof-of-concept study, we use the thrust axis to partition the event into two hemispheres, and define $\mathcal{R}$ to be the heavy hemisphere (i.e.~one with larger mass).
By restricting to hemisphere observables, we can more readily relate EFPs to physical features of QCD, as discussed further in \Sec{EFPbasis}.
We note that hemisphere-level event shapes are known to receive non-global logarithms (NGLs)~\cite{Dasgupta:2001sh,Dasgupta:2002bw}, which we ignore for simplicity.
In general, though, the reweighting procedure can accommodate any choice of $\mathcal{R}$, including the whole event.

EFPs are constructed from the energy fractions and angular separations between final-state particles.
The normalized energy fractions of final-state massless particles in the region $\mathcal{R}$ are defined as:
\begin{equation}
z_i \;\equiv\; \frac{E_i}{\sum_{j\in \mathcal{R}} E_j}\,,
\qquad \sum_{i\in \mathcal{R}} z_i = 1\,.
\label{eq:z_def}
\end{equation}
The angular separation between particles $i$ and $j$, with 3-momenta $\vec{p}_i$ and $\vec{p}_j$, is given by
\begin{align}
\label{eq:thetaij}
\theta_{ij} = \sqrt{2\left(1-\frac{\vec{p}_i\cdot \vec{p}_j}{E_iE_j}\right)}\,, \qquad 0\leq  \theta_{ij} \leq 2\,.
\end{align}
With these kinematic building blocks, we can define the EFPs and their associated graph-theoretic properties. 

\begin{figure}[t]
\centering
\includegraphics[width=0.8\textwidth]{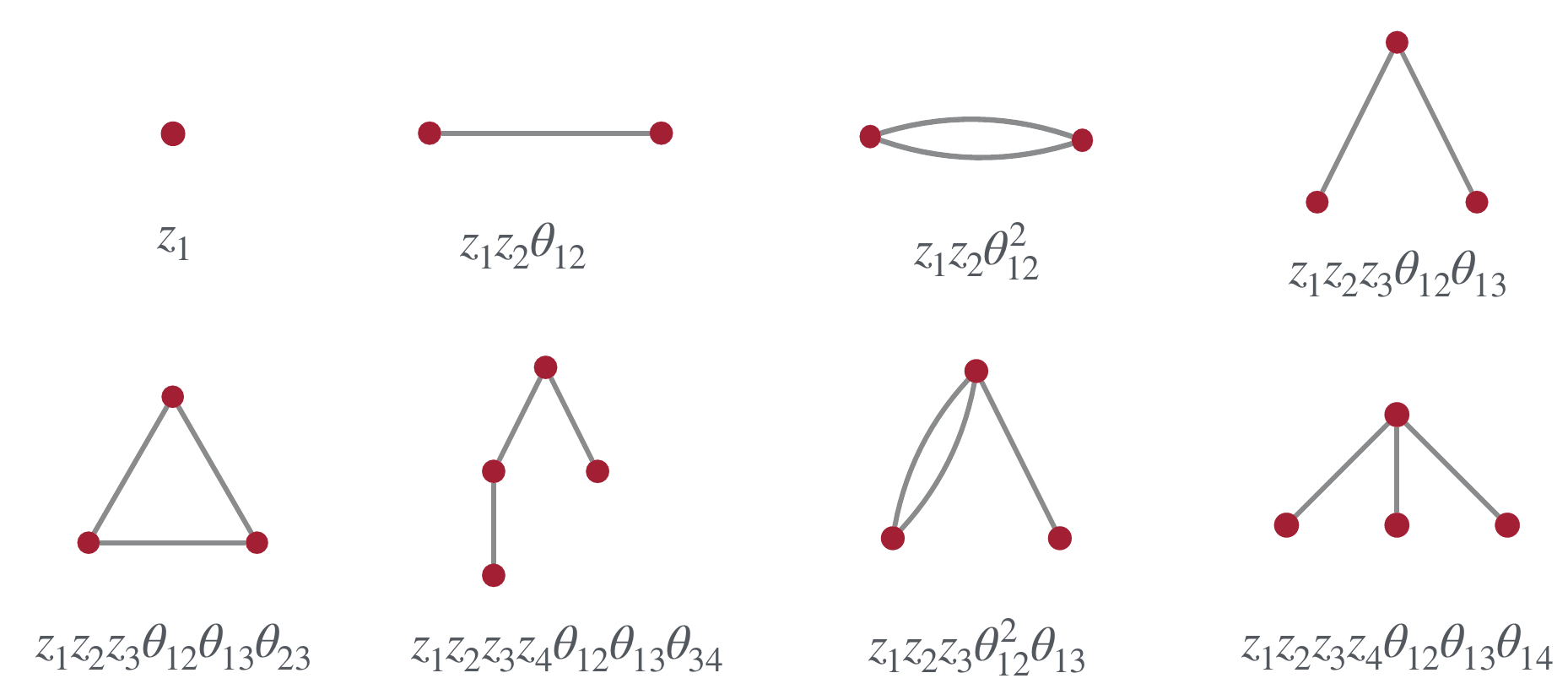}
\caption{Graphical representation of various EFPs.
Each vertex contributes an energy factor of $z_i$ and each edge contributes an angular factor of $\theta_{ij}$ when $\beta = 1$.
Multiple edges between the same pair of vertices indicate higher powers of the corresponding angle.
The degree $d$ of an EFP counts the total number of edges (with multiplicity).
The value of an EFP involves a sum over all possible assignments of particles to nodes.
}
\label{fig:efp_graphs}
\end{figure}

\begin{itemize}
\item \textbf{Graph representation and definition.}
Each EFP can be elegantly represented  using a multigraph $G=(V,E)$ with vertex set $V$ and edge set $E$, alongside a fixed angular exponent $\beta$.
The EFP associated with the graph $G$ in the $\beta$-exponent class is given by~\cite{Komiske:2017aww}:
\begin{equation}
\mathrm{EFP}_G^{(\beta)}
\;\equiv\;
\sum_{i_1,\dots,i_k\in \mathcal{R}}
\prod_{v\in V} z_{i_v}\;
\prod_{(a,b)\in E} \theta_{i_a i_b}^{\,\beta}\,.
\label{eq:EFP_def}
\end{equation}
Note that multiple edges between a pair of vertices correspond to higher powers of $\theta_{ij}^{\beta}$, and any self-loop yields a trivially vanishing EFP since $\theta_{ii}=0$.
In \Fig{efp_graphs}, we illustrate several graphical representations and their corresponding EFPs.
Throughout this work, we use $\beta=1$ and thus suppress the superscript, simply writing an EFP associated with graph $G$ as $\mathrm{EFP}_G$.
\item \textbf{Prime and composite EFPs.}
A prime EFP is one whose associated multigraph is connected. Conversely, a composite EFP corresponds to a disconnected graph and can be naturally factorized as a product of prime EFPs.
For EFPs to form a complete basis for IRC-safe observables, both prime and composite EFPs must be considered.
\item \textbf{Degree of an EFP.}
The degree of an EFP is defined as the total number of edges in its multigraph:
\begin{equation}
d(G)\equiv |E|\,.
\label{eq:degree_def}
\end{equation}
This provides a natural metric for organizing increasingly complex IRC-safe information.
The degree of a composite graph is the product of the degrees of its prime components.
\begin{figure}[t]
\centering
\includegraphics[width=0.8\textwidth]{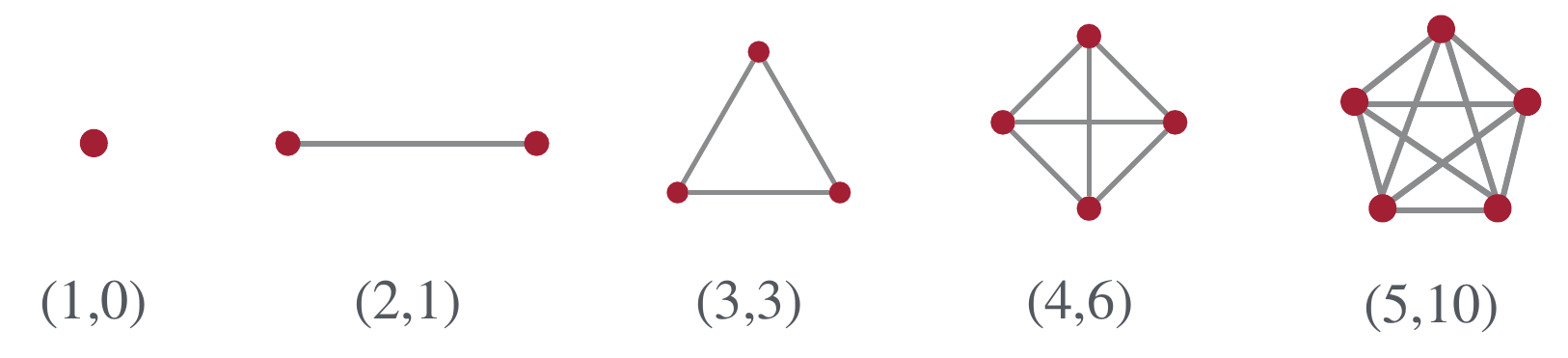}
\caption{Minimal EFP graphs for each chromatic number $\chi = 1, \ldots, 5$, labeled by $(\chi, d)$ where $d$ is the graph degree. These are the complete graphs $K_\chi$, which achieve chromatic number $\chi$ at the minimum possible degree $d = \binom{\chi}{2}$. Higher-degree graphs with the same chromatic number are obtained by adding pendant edges or additional internal edges to these minimal structures.}
\label{fig:chromatic_efps}
\end{figure}
\item \textbf{Chromatic number of an EFP.}
The chromatic number of a graph $G$, denoted $\chi(G)$, is the minimum number of colors needed to color the vertices such that no two adjacent vertices share the same color.
In the context of EFPs, this graph-theoretic property dictates the minimum number of particles required for the EFP to have a non-vanishing value.
If a final state has fewer particles than $\chi(G)$, at least two connected vertices must be assigned to the same particle, resulting in an angular factor of $\theta_{ij}=0$ and a vanishing EFP.
Consequently, much like the degree of a graph, the chromatic number provides a natural way to systematically organize increasingly complex information.
In particular, higher chromatic numbers directly probe higher multi-particle correlations. In \Fig{chromatic_efps}, we show the complete connected EFP graphs organized by chromatic number and degree up to $\chi=5$.
\end{itemize}
All EFPs evaluated in this work are computed using the \texttt{EnergyFlow} package~\cite{Komiske:2017aww,Komiske:2018cqr,Komiske:2019fks}.

\subsection{EFPs as a basis for IRC-safe observables}
\label{sec:EFPbasis}

Under the mild assumption of continuity with respect to energy flow, any IRC-safe observable defined within a phase-space region $\mathcal{R}\subseteq\Phi$ can be approximated arbitrarily well by a linear combination of EFPs~\cite{Komiske:2017aww}.
Depending on whether the basis is organized by degree $d$ or chromatic number $\chi$ (or really any systematic grading), an arbitrary IRC-safe observable $\mathcal{O}$ can be expanded as:
\begin{align}
\mathcal{O}(\mathcal{R})\;=&\; \sum_{G\in \mathcal{G}_{\le d}} c_G^{\mathcal{O}}\,\mathrm{EFP}_G(\mathcal{R}) + \delta\mathcal{O}_d(\mathcal{R})\,,\nn\\
=&\; \sum_{G\in \mathcal{G}_{\le \chi}} k_G^{\mathcal{O}}\,\mathrm{EFP}_G(\mathcal{R}) + \delta\mathcal{O}_\chi(\mathcal{R})\,,
\label{eq:O_expand_EFP}
\end{align}
where $\mathcal{G}_{\le d}$ and $\mathcal{G}_{\le \chi}$ denote the EFP basis truncated at degree $d$ and chromatic number $\chi$, respectively.
The constant coefficients $c_G^{\mathcal{O}}$ and $k_G^{\mathcal{O}}$ are observable dependent, while the truncation remainders $\delta\mathcal{O}_d$ and $\delta\mathcal{O}_\chi$ strictly vanish in the limits $d\to\infty$ and $\chi\to\infty$.
Note that the EFPs appearing on the right-hand side of~\eq{O_expand_EFP} will in general involve both prime and composite EFPs.
Also, the convergence is only guaranteed in the Stone–Weierstrass sense of uniform convergence, \emph{not} in the sense of a Taylor expansion.
Because composite EFPs are products of prime EFPs, the full set of EFPs (primes plus composites) at a given truncation degree or chromatic number forms an over-complete basis: there are more EFPs than linearly independent observables.
This over-completeness is not a problem for our reweighting framework, which does not require linear independence of the constraint set and instead benefits from the redundancy by allowing the optimization to absorb information from multiple overlapping projections of phase space.

When EFPs are defined on a highly collimated phase-space region $\mathcal{R}$, such as a narrow jet-like region where all $M$ particles satisfy $\theta_{ij}\sim \theta \ll 1$, the exact EFP expressions can be systematically simplified~\cite{Cal:2022fnm}.
Because hemispheres can be related to such a collimated region via a Lorentz boost, the analysis below holds for our proof-of-concept study.%
\footnote{This analysis would \emph{not} hold if $\mathcal{R}$ were the entire $e^+e^-$ event, which is why we focus on hemispheres for our analysis. \label{fn:Rhem}}
The strongest approximation is the strongly-ordered limit, which imposes a strict hierarchy on both the energy fractions and the emission angles:
\begin{align}
\textbf{strongly-ordered:}\qquad z_{i+1} \ll z_i, \quad \theta_{1, i+1} \ll \theta_{1, i} \text { for } i>1 .
\end{align}
A more general approximation is the $1$-collinear expansion. Here, a single hard particle dominates the energy ($z_1 \sim 1$), while the remaining $M-1$ particles are treated as collinear-soft. Crucially, this expansion makes no assumptions about the internal kinematic hierarchy among the soft emissions, which distinguishes it from the strongly-ordered case:
\begin{align}
\textbf{1-collinear:}\qquad z_1\sim 1, \quad z_i \sim z \ll 1 \text { for } i>1, \quad \theta_{i j} \sim \theta \ll 1\,.
\end{align}
This naturally extends to an $n$-collinear expansion, where $n$ hard particles are accompanied by collinear-soft radiation:
\begin{align}
\textbf{n-collinear:}\qquad z_i\sim 1 \text { for } i\leq n, \quad z_i \sim z \ll 1 \text { for } i>n, \quad \theta_{i j} \sim \theta \ll 1\,.
\end{align}
Consequently, these expansions form a nested sequence of limits, where the strongly-ordered limit is a strict sub-case of the $1$-collinear limit, which is itself a sub-case of the $2$-collinear limit, and so on.
In such highly collimated regions, observables develop large Sudakov logarithms requiring all-orders resummation.
Compared to the approximation made to carry out leading-logarithmic (LL) resummation in \Eq{sudakovobs}, the strongly-ordered limit is more restrictive, whereas the $1$-collinear expansion is less restrictive.

\begin{figure}[t]
\centering
\includegraphics[width=0.8\textwidth]{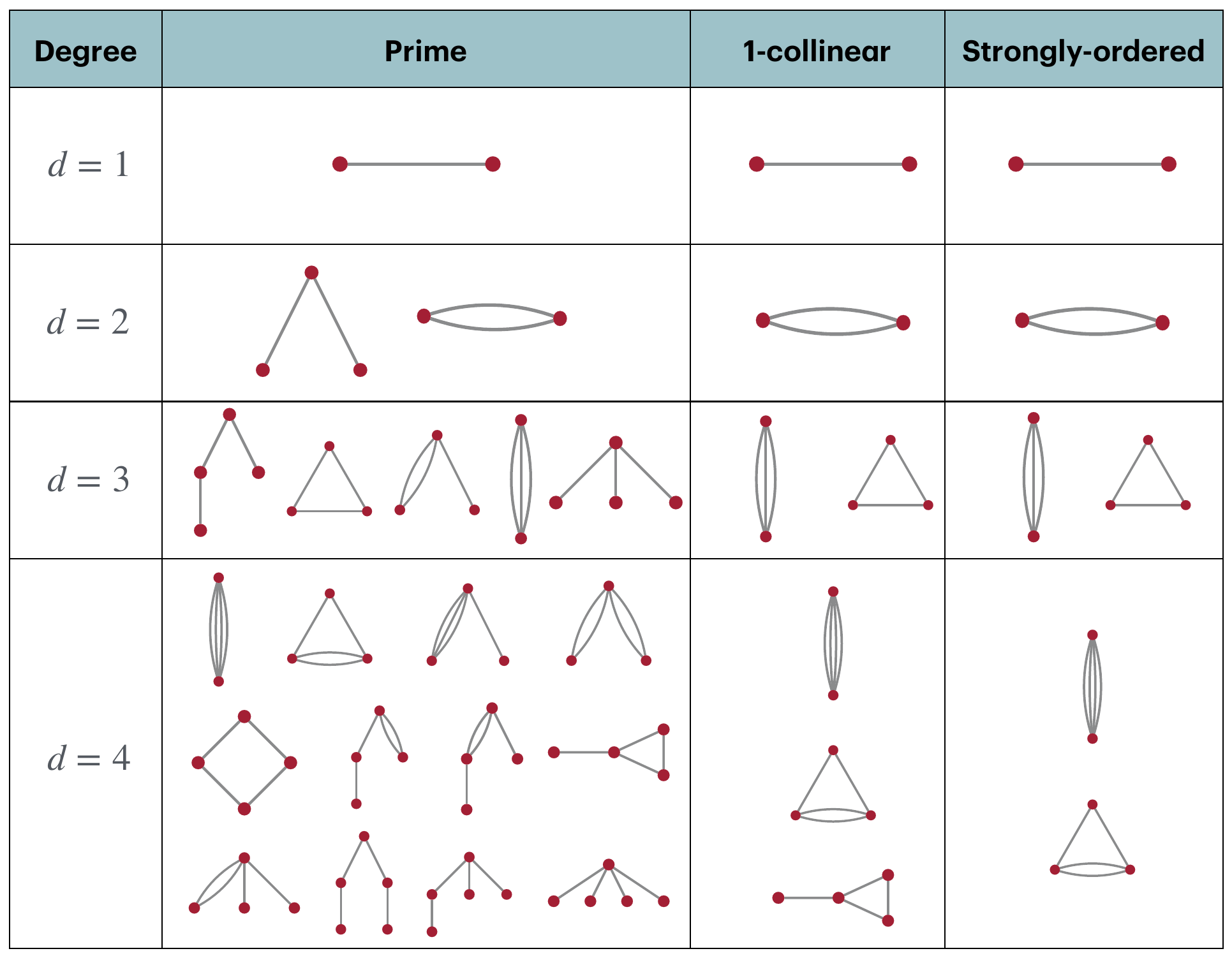}
\caption{%
Comparison of EFP basis choices for degrees $d=1$ to $d=4$.
The prime basis includes all connected multigraphs.
The 1-collinear basis retains graphs that are linearly independent when all emissions are collinear-soft relative to a single hard parton.
The strongly-ordered (SO) basis further reduces to graphs independent under hierarchical ordering of emissions in energy and angle.
The bases are nested: strongly-ordered $\subset$ 1-collinear $\subset$ prime.
In general, to have a complete linear basis, one has to also include composite EFPs built from these prime elements.
}
\label{fig:efp_bases_comparison}
\end{figure}

In~\Fig{efp_bases_comparison}, we illustrate how the complete set of prime EFPs up to degree $4$ collapses to a smaller basis of independent EFPs under these two approximations.
As expected, the $1$-collinear set retains more independent elements than the strongly-ordered set.
To form a complete basis, one has to also consider composite EFPs built from products of these prime elements.
As explained in \Reference{Cal:2022fnm}, because EFPs are related to each other in these various approximations, there is an ambiguity as to which basis elements to choose.
The particular basis sets shown here have the nice property that (to the degree shown) any composite EFPs needed for the basis can be build from products of primes already in the set.

\subsection{Precision moments of EFPs}

As explained in~\Eq{product_features_choice}, moments of the form $\langle g_{m_1 n_1 \cdots m_K n_K}(v_1, \dots, v_K) \rangle$ provide a well-motivated set of precision inputs for improving prior distributions.
For a single observable $v$, these constraints take the form of mixed moments $\langle v^m \ln^n v\rangle$.
Our goal now is to determine which EFP moments capture the desired precision information.

As shown in~\Eq{O_expand_EFP}, any IRC-safe observable $v$ can be expanded in the complete EFP basis as:
\begin{align}
v = \sum_{G} c_G^{v}\,\mathrm{EFP}_G\,,
\end{align}
where the sum runs over the full, untruncated set of multigraphs.
This means that polynomial moments of the form $\langle v^m \rangle$ can be naturally expanded in terms of EFP moments:
\begin{align}
\langle v^m \rangle = \left\langle\left(\sum_{G} c_G^{v}\,\mathrm{EFP}_G\right)^m\right\rangle &= \sum_{G_1\cdots G_m} c_{G_1}^v \cdots c_{G_m}^v \langle {\rm{EFP}}_{G_1}\cdots {\rm{EFP}}_{G_m}\rangle\,.
\end{align}
Here, the term $\langle \mathrm{EFP}_{G_1}\cdots \mathrm{EFP}_{G_m}\rangle$ represents the expectation value of a composite EFP formed by the product of $m$ individual EFPs.
In general, since multiple EFPs in the product can be identical, this involves higher moments as well.
For instance, if all $m$ graphs in a given term are identical ($G_i=G$), the expectation value reduces to the $m$-th moment of a single EFP $\langle\mathrm{EFP}_{G}^m\rangle$.

The above analysis tells us that the polynomial moments of any arbitrary IRC-safe observable $v$ can be captured by the polynomial moments of EFPs.
Of course, we have to pick a finite set of EFPs for numerical studies, so this statement is only true up to truncation remainders.
Nevertheless, there is a sense in which instead of needing to compute polynomial moments of each IRC-safe observable $v$ of interest, one can instead focus just on polynomial moments of EFPs.
This conclusion extends straightforwardly to multi-observable polynomial moments.
That is, moments of the form $\langle v_1^{m_1} v_2^{m_2} \cdots v_K^{m_K} \rangle $ can similarly be expanded into linear combinations of EFP moments.

On the other hand, the logarithmic moments of arbitrary IRC-safe observables cannot be expressed as simple linear combinations of (logarithmic) moments of EFPs.
This is intuitive, since logarithmic moments are Sudakov-safe observables, as discussed in~\sec{Constraints}, and EFPs are a natural basis for IRC-safe observables, \emph{not} for Sudakov-safe observables.
However, because the EFP basis largely collapses in highly collimated limits (see~\sec{EFPbasis}), the logarithmic moments of generic observables, which are dominated by these singular phase-space regions, should be well-approximated by the singular behavior of this much smaller, collapsed set of EFPs.
Therefore, in practice, we expect logarithmic EFP moments to serve as highly efficient constraints for capturing the resummation structure of a wide variety of observables.
Ultimately, a more rigorous and better alternative would be to construct a complete, natural basis for Sudakov-safe observables, though identifying such a basis remains elusive.%
\footnote{Part of the challenge is that there isn't even a rigorous way to determine whether an observable is Sudakov safe, though \Reference{Komiske:2020qhg} made an attempt.}

\subsection{Connection to energy correlators}
\label{sec:relation_to_energy_correlators}

In recent years, there has been a renewed theoretical interest in correlations of the energy flow operator~\cite{Hofman:2008ar,Kravchuk:2018htv,Dixon:2019uzg,Chen:2020vvp}:
\begin{align}
\mathcal{E}(\vec{n})=\int_0^{\infty} d t \lim _{r \rightarrow \infty} r^2 n^i T_{0 i}(t, r \vec{n})\,,
\end{align}
which relates the energy-momentum tensor $T_{\mu\nu}$ to the energy captured by an asymptotic detector in the direction $\vec{n}$.%
\footnote{Despite the naming similarity, these energy correlators differ from energy correlation functions~\cite{Larkoski:2013eya,Moult:2016cvt} which behave more like EFPs.}
In our notation, the standard two-point energy-energy correlator (EEC)~\cite{Basham:1979gh,Basham:1978zq,Basham:1978bw,Basham:1977iq} can be expressed as an angular-differential expectation value:\footnote{Our angle $\theta_{ij}$ defined in~\eq{thetaij} is related to the standard EEC variable $z=(1-\cos\chi_{ij})/2$ by $z=\theta_{ij}^2/4$, where $\chi_{ij}=\arccos(\hat{p}_i\cdot\hat{p}_j)$ is the geometric opening angle; both $\theta_{ij}$ and $z$ are chord-length-based and coincide with $\chi_{ij}$ at small angles.}
\begin{equation}
\mathrm{EEC}(\theta) \;\equiv\; \left\langle \sum_{ij \in R} z_i z_j\, \delta(\theta - \theta_{ij}) \right\rangle\,,
\label{eq:EEC_def}
\end{equation}
where the expectation value corresponds to a cross-section weighted average over phase space.
While traditionally defined over the entire event ($\mathcal{R} = \Phi$), this EEC definition naturally generalizes to a restricted phase-space region $\mathcal{R} \subset \Phi$.

The polynomial moments of EFPs are intimately connected to the moments of energy correlators. For instance, the $k$-th angular moment of the EEC yields:
\begin{align}
\int d\theta\, \theta^k\,\mathrm{EEC}(\theta) = \left\langle \sum_{ij\in R} z_i z_j\, \theta_{ij}^k \right\rangle\,.
\end{align}
This is exactly the expectation value of an EFP corresponding to a two-vertex multigraph with $k$ edges. This correspondence generalizes straightforwardly: the polynomial moments of higher-point energy correlators map directly onto the expectation values of EFPs with correspondingly more vertices.

In contrast, logarithmic moments highlight the differences between EFPs and EECs.
Evaluating the logarithmic moment of the EEC gives:
\begin{align}
\int d\theta\, \ln^n\theta\,\mathrm{EEC}(\theta) = \left\langle \sum_{ij\in R} z_i z_j\, \ln^n\theta_{ij} \right\rangle \neq \left\langle \ln^n\left(\sum_{ij\in R} z_i z_j\, \theta_{ij}^k\right) \right\rangle\,,
\end{align}
where the rightmost term represents the logarithmic moment of the two-vertex EFP with $k$ edges. Physically, $\big\langle \sum_{ij\in R} z_i z_j\, \ln^n\theta_{ij} \big\rangle $ is dominated by events containing at least one sufficiently energetic collinear pair in $\mathcal{R}$.%
\footnote{In the small-angle limit, the EEC($\theta$) distribution scales schematically as $\theta^{\omega-1}$ within the perturbative regime. The exponent $\omega$ is determined by a competition between the positive contribution from the twist-2, spin-3 anomalous dimension and the negative contribution from the QCD $\beta$-function~\cite{Dixon:2019uzg}. A net negative $\omega$ would render logarithmic moments formally non-integrable. However, this divergence is physically regulated by the confinement transition \cite{Lee:2025okn}: as $\theta$ approaches the non-perturbative scale ($\theta \lesssim \Lambda_{\rm QCD}/Q$), the scaling shifts from $\theta^{\omega-1}$ to a linear $\theta$ dependence, ensuring integrability of log moments.} By contrast, $\big\langle \ln^n\big(\sum_{ij\in R} z_i z_j\, \theta_{ij}^k\big) \big\rangle$ becomes large only when the entire sum is small, which requires a collective singular configuration, i.e.\ collinear or soft.
In this way, the former signals the presence of \emph{any} highly collinear pair within $R$, while the latter probes whether the particles are \emph{globally} in a singular configuration.

It would be interesting to explore how complementary information from EEC logarithmic moments could be incorporated into our reweighting procedure.
Of course, one could simply include EEC logarithmic moments as a separate set of constraints.
Alternatively, there is an intriguing identity:
\begin{equation}
\lim_{\beta \to 0} \frac{\theta^\beta  - 1} {\beta} = \ln \theta,
\end{equation}
which suggests that it may be possible to extract logarithmic information about EECs by studying the $\beta$-class of EFPs in the $\beta \to 0$ limit.
Nevertheless, even without this logarithmic information, the direct connection between their polynomial moments provides a clear pathway for leveraging precision EEC calculations~\cite{Dixon:2018qgp,Tulipant:2017ybb,Dixon:2019uzg,Duhr:2022yyp,Gao:2020vyx,Jaarsma:2025tck} to evaluate the EFP constraints required by our reweighting framework.


\section{Two-shower proof-of-concept setup}
\label{sec:shower_setups}

In this section, we describe the two-shower setup used for our proof-of-concept study.
To rigorously test our reweighting framework, we employ two distinct parton shower settings.
First, we use a purposely degraded shower that serves as our baseline prior.
Second, we use a high-fidelity shower that acts as the ``truth'' baseline.
Taking the high-fidelity shower as the ``truth'' means we can use it to numerically compute arbitrarily accurate precision constraints for our reweighting procedure.
While unrealistic, it is convenient for this proof-of-concept study, as calculating and collecting state-of-the-art precision computations for multiple observables, especially for  multi-differential distributions with correlated uncertainties, is a challenging task.
Furthermore, having access to the full event-level ``truth'' shower provides a complete phase-space baseline, allowing us to test the improvements in our posterior distribution after the reweighting procedure.

We first describe the nature of the high-fidelity and degraded showers, then describe our numerical optimization strategy, and finally define the performance metrics used to test the level of improvements in our posterior distribution.
Common to both sets of showers, we consider $e^+e^-\to{\rm hadrons}$ at center-of-mass energy $Q=91.2~{\rm GeV}$, and we generate samples with $N_{\rm ev}=10^6$ events with \textsc{Sherpa}~3.1~\cite{Sherpa:2024mfk}.
The beams are $e^\pm$ with energies $E_{e^\pm}=45.6~{\rm GeV}$ and hadronization is performed with the \texttt{AHADIC} cluster model~\cite{Chahal:2022rid} with its default settings: a shower cutoff of $1$~GeV for timelike splittings and $1.414$~GeV for spacelike splittings.
At the hard-process level, we generate the inclusive partonic final state at Born level.
Unless stated otherwise, we set $\alpha_s(m_Z)=0.118$, and use two-loop running for $\alpha_s$ in the CMW scheme~\cite{Catani:1990rr}.

\subsection{Degraded and high-fidelity showers}

Both the degraded and high-fidelity showers are defined by the standard \texttt{CSSHOWER} in \textsc{Sherpa}, which is based on Catani--Seymour dipole factorization~\cite{Sherpa:2024mfk}.
Emissions are generated from color-connected emitter--spectator dipoles using spin-averaged dipole kernels and exact momentum mappings.
The shower is ordered by its default transverse-momentum-like evolution variable $t$ and is implemented as a Markov chain with unitary evolution: the probability for an emission at scale $t$ is accompanied by a Sudakov form factor $\Delta(t,t_{\rm max})$ that encodes the corresponding no-emission probability.
Schematically, the probability for an emission at scale $t$ from a given dipole is given by:
\begin{equation}
\label{eq:dP}
  {\rm d}\mathcal{P}_{i\to jk}(t) =\Delta(t,t_{\rm max})\;
  \frac{\alpha_s(\mu_R(t))}{2\pi}\,\mathcal{K}_{i\to jk}(z,\phi;\,t)\,
  {\rm d}\Phi_{\rm emit}\,,
\end{equation}
where $\mathcal{K}_{i\to jk}$ is the relevant dipole kernel, $z$ is an energy-sharing variable, $\phi$ is an azimuth, and ${\rm d}\Phi_{\rm emit}$ is the one-emission phase space. The Sudakov form factor $\Delta(t,t_{\rm max})$ encodes the probability of evolving from the starting scale $t_{\rm max}$ down to $t$ without any resolvable emissions.

In the massless collinear limit, the dipole kernels reduce to the universal Altarelli--Parisi splitting functions~\cite{Altarelli:1977zs}.
At leading order, they are given by:
\begin{align}
P_{q\to qg}(z) &= C_F\,\frac{1+z^2}{1-z}
\;\sim\; C_F\,\frac{2}{1-z}
\qquad [z\to 1]\,,
\nonumber\\
P_{g\to gg}(z) &= 2C_A \!\left(\frac{z}{1-z}+\frac{1-z}{z}+z(1-z)\right)
\;\sim\; 2C_A\!\left(\frac{1}{z}+\frac{1}{1-z}\right)
\qquad [z\to 0,1]\,,
\nonumber\\
P_{g\to q\bar q}(z) &= T_R\Big(z^2+(1-z)^2\Big)\,,
\label{eq:standard_kernels}
\end{align}
where $C_F=4/3$, $C_A=3$, and $T_R=1/2$.
Here, $z$ denotes the momentum fraction carried by parton $j$ in an $i\to jk$ splitting.
Because momentum is conserved, the parton $k$ carries the remaining fraction $1-z$.
Therefore, the endpoint singularities at $z\to 0$ and $z\to 1$ physically correspond to the soft limit of the parton $j$ and $k$, respectively.
To see how these singularities drive the logarithmic structure of the shower, we note that the one-emission phase space ${\rm d}\Phi_{\rm emit}\propto \frac{d\theta^2}{\theta^2}$ at small angles.
This collinear singularity combined with the soft singularities of the splitting functions generate the double-logarithmic structure $\alpha_s^n \ln^{2n} v$ for a typical Sudakov observable $v$.

We take this default setting to be our high-fidelity ``truth'' shower.
Since this will be the target for our degraded shower, we will also call it the ``target'' shower.
While our use of leading-order Altarelli--Parisi splitting implies that our target shower is not formally NLL accurate, it provides a well-defined and fully exclusive baseline that possesses the correct leading QCD singularity structure.

To create a deliberately degraded prior, we strip the splitting functions of their non-singular contributions -- the terms that control the rate of hard-collinear emissions -- and disable the $g\to q\bar{q}$ channel entirely:
\begin{align}
P_{q\to qg}(z) &\;\longrightarrow\;
P_{q\to qg}^{\rm broken}(z)= C_F\,\frac{2z}{1-z}\,,
\nonumber\\
P_{g\to gg}(z) &\;\longrightarrow\;
P_{g\to gg}^{\rm broken}(z)= 2C_A\left[\frac{z}{1-z}+\frac{1-z}{z}\right]\,,
\nonumber\\
P_{g\to q\bar q}(z) &\;\longrightarrow\;
P_{g\to q\bar q}^{\rm broken}(z)=0\,.
\label{eq:broken_kernels}
\end{align}
The Sudakov factors are also recomputed from these modified kernels so that the shower remains unitary within the altered evolution kernels.
In the CAESAR decomposition of the Sudakov radiator, $R = L \, g_1(\alpha_s L) + g_2(\alpha_s L) + \cdots$~\cite{Banfi:2004yd,Catani:1992ua}, the preserved soft singularities $\sim 1/(1-z)$ leave the LL function $g_1$ unchanged.
The removed pieces $C_F\, (1-z)$ and $C_A\,z(1-z)$ control emission at finite energy fraction: compared to the standard kernels, the broken shower underproduces hard collinear gluons in $q\to qg$ and suppresses symmetric energy-sharing in $g\to gg$, while leaving the soft emission rate untouched.
Disabling $g\to q\bar{q}$ removes all secondary quark production, eliminating the shower's $n_f$ dependence.
Together, these modifications shift the NLL function $g_2$; the non-singular integrals entering the radiator change as:
\begin{align}
\int_0^1\!dz\,P_{q\to qg}^{\rm reg}(z) &= -\tfrac{3}{2}C_F \;\longrightarrow\; -2C_F\,,\nonumber\\
\int_0^1\!dz\,P_{g\to gg}^{\rm reg}(z) &= -\tfrac{11}{3}C_A \;\longrightarrow\; -4C_A\,,\nonumber\\
\int_0^1\!dz\,P_{g\to q\bar{q}}(z) &= \tfrac{2}{3}n_fT_R \;\longrightarrow\; 0\,,
\label{eq:nll_shifts}
\end{align}
where $P^{\rm reg}$ denotes the remainder after subtracting the soft poles, and the energy-sharing and flavor composition of the parton cascade are correspondingly altered.%
\footnote{The broken shower retains the leading collinear singularities; by contrast, the uniform prior $q_{\rm uni.}(\Phi)$ mentioned in \sec{efficient_priors} lacks even these.}
We generate three prior variation samples using these broken kernels with different strong coupling values:
\begin{equation}
\alpha_s(m_Z) = 0.118 \,(\pm 20\%)\,.
\label{eq:alphas_variations}
\end{equation}
Varying $\alpha_s$ rescales the overall emission rate but cannot restore the $z$-dependent shape of the splitting functions or reintroduce secondary quark production.

\subsection{Choice of EFP moment constraints}
\label{sec:EFPtraining}

With the degraded and high-fidelity showers established, we implement the maximum-entropy reweighting procedure by enforcing moment constraints of the form discussed in~\sec{Constraints}.
Motivated by the fact that any IRC-safe observable can be expanded in the EFP basis as shown in~\eq{O_expand_EFP}, we construct our training sets using various subsets of EFPs.
We organize these sets systematically, such as by degree or chromatic number, and detail the specific sets as they appear in~\sec{results}.
Finally, to ensure we evaluate the EFPs in a regime where the approximations discussed in~\sec{EFPbasis} and \Reference{Cal:2022fnm} hold true, we restrict their definition to the heavy hemisphere of each event (see footnote~\ref{fn:Rhem}).

For a given training set of EFPs, $S =\{\mathrm{EFP}_{G_1},\mathrm{EFP}_{G_2},\cdots,\mathrm{EFP}_{G_N}\}$, we construct three distinct families of measurement functions to probe different regions of the phase space:
\begin{align}
\text{Logarithmic family:}\quad & \mathcal{B}^{\mathrm{log}}(S)=\{\ln^m \mathrm{EFP}_{G_i}\,|\,G_i\in S\}, \nonumber\\
\text{Polynomial family:}\quad &  \mathcal{B}^{\mathrm{poly}}(S)=\{\mathrm{EFP}_{G_i}^m\,|\,G_i\in S\}, \nonumber\\
\text{Mixed family:}\quad &  \mathcal{B}^{\mathrm{mix}}(S)= \{\mathrm{EFP}_{G_i}^m\ln^n\mathrm{EFP}_{G_i}\,|\,G_i\in S\}\,,
\label{eq:constraint_families}
\end{align}
where we take $m\in\{1,2,3,4\}$ for the logarithmic and polynomial families,
yielding four measurement functions per graph.
For the mixed family, we use the combinations:
\begin{equation}
(m,n)\in\{(1,0),(0,1),(0,2),(1,1)\},
\end{equation}
also yielding four measurement functions per graph.
Logarithms are regulated with a small cutoff $\epsilon$ as $\ln(\mathrm{EFP}_G+\epsilon)$ to ensure numerical stability for events where the EFP vanishes; the precise value of $\epsilon$ is irrelevant as long as it is negligible compared to typical EFP values.
For a given family of measurement functions, we input the corresponding precision constraint values.

To avoid a combinatoric explosion of possibilities, we only consider moments of a single EFP at a time.
To probe correlations between different EFPs, one could also consider measurement functions of the form suggested in \Eq{product_features_choice}.
For example, to study correlation between two EFPs, we could consider:
\begin{align}
 \mathcal{B}^{\mathrm{corr}}(S)= \{\mathrm{EFP}_{G_i}^m\mathrm{EFP}_{G_j}^n,\,\ln^m \mathrm{EFP}_{G_i}\ln^n\mathrm{EFP}_{G_j},\,\mathrm{EFP}_{G_i}^m\ln^n\mathrm{EFP}_{G_j}\,|\,G_i,G_j\in S,\,i\neq j\},
\label{eq:cross_constraints}
\end{align}
which allows inputting information about the double-differential joint distributions beyond the marginal moments of individual EFPs.
Generalizing to the case involving even higher numbers of EFPs is straightforward, but becomes even more computationally heavy.
Perhaps surprisingly, we find that even without these cross EFP moments, the reweighting does an excellent job modeling correlations, as studied in \App{2d_transfer}.

Note that some EFP correlations are already captured implicitly.
If the training set contains a composite graph $G = G_a \cup G_b$, the corresponding EFP factorizes as $\mathrm{EFP}_G = \mathrm{EFP}_{G_a} \mathrm{EFP}_{G_b}$.
In this case, a linear constraint on the composite EFP is equivalent to enforcing a specific correlated polynomial moment $\langle \mathrm{EFP}_G \rangle=\langle \mathrm{EFP}_{G_a} \mathrm{EFP}_{G_b} \rangle$.
Higher polynomial moments of $\langle \mathrm{EFP}^m_G \rangle$ would also give sensitivity to different polynomial moments of correlated form.
Similarly, because the logarithm of a product is the sum of the logarithms, logarithmic and mixed moments of composite operators yield some of the terms given in~\eq{cross_constraints}.
In this work, we primarily focus on marginal moments, supplemented by the correlated moments that arise naturally through polynomial moments of composite EFPs in a training set.
The effect of adding explicit cross-moment constraints is explored in \App{2d_transfer}, where we find that they provide diminishing returns as the marginal basis grows.


\subsection{Optimization strategy}
\label{sec:optimization}

As noted in~\Eq{sampled_Z}, a parton-shower generator samples from $q_{\rm PS}(\Phi)$ to produce a discrete set of $N_q$ events $\{\Phi_a\}_{a=1}^{N_q}$.
For a given event $\Phi_a$, therefore, the reweighting factor associated with a Lagrange multiplier $\boldsymbol{\lambda}$ in~\eq{weight_sol} takes the discrete form 
\begin{equation}
w_a(\boldsymbol{\lambda}) = \frac{ \exp\!\left(-\sum_i \lambda_i \, m_i(\Phi_a)\right)}{\sum_{b=1}^{N_q} \exp\!\left(-\sum_i \lambda_i \, m_i(\Phi_b)\right)}\,.
\label{eq:weights_discrete}
\end{equation}
If we now use a set $S$ of EFPs with an associated family of measurement functions $\mathcal{B}(S)$, as described in~\sec{EFPtraining}, the event weight becomes:
\begin{equation}
w_a(\boldsymbol{\lambda}) = \frac{ \exp\!\left(-\sum_{f\in \mathcal{B}(S)} \lambda_f f(\Phi_a)\right)}{\sum_{b=1}^{N_q} \exp\!\left(-\sum_{f \in \mathcal{B}(S)} \lambda_f f(\Phi_b)\right)}\,,
\label{eq:weights_exp}
\end{equation} 
where $\lambda_f$ is the Lagrange multiplier associated with the measurement function $f$. Consequently, there are $|\mathcal{B}(S)|$ Lagrange multipliers to optimize. This is achieved by minimizing the finite-sample form of the dual objective from \eq{dual_objective}
\begin{equation}
\mathcal{J}(\boldsymbol{\lambda}) = \ln \left[\sum_{a=1}^{N_q} \exp\!\left(-\sum_{f \in \mathcal{B}(S)} \lambda_f f(\Phi_a)\right)\right]
- \sum_{f\in \mathcal{B}(S)} \lambda_f c_f\,,
\label{eq:loss}
\end{equation}
where $c_f$ represents the target constraint value, i.e.\ the precision input for the expectation value of the measurement function $f$.

As explained in~\sec{maximum_entropy}, this dual objective is convex and therefore possesses a unique global minimum.
We only need to evaluate $f(\Phi_a)$ once for all $N_q$ events and all $|\mathcal{B}(S)|$ measurement functions. After this initial $\mathcal{O}(N_q \times |\mathcal{B}(S)|)$ computation, the values are cached and can be reused without re-evaluation of observables or re-generation of events.
The optimization then simply iterates over new choices of the vector $\boldsymbol{\lambda}$ to find the global minimum.
During each iteration, computing the argument of the exponent, $-\sum_{f\in \mathcal{B}(S)} \lambda_f f(\Phi_a)$, reduces to a single, highly efficient dense matrix-vector multiplication on $\mathbb{R}^{N_{q} \times |\mathcal{B}(S)|}$.

We perform the minimization using the L-BFGS algorithm~\cite{liu:1989}, leveraging automatic differentiation for exact gradients and the log-sum-exp trick to maintain numerical stability.\footnote{Specifically, the partition function $Z = \sum_k e^{z_k}$ is computed as $e^c \sum_k e^{z_k - c}$ with $c = \max_k z_k$ to avoid numerical overflow from the exponent.}
For our samples with $N_{\rm ev}=10^6$ events and constraint sets up to $|\mathcal{B}(S)|\sim 10^3$, the optimization typically converges within $\mathcal{O}(100)$ iterations.
This process takes only a few minutes on a standard multi-core desktop CPU.
For large constraint sets ($|\mathcal{B}(S)|\gtrsim 500$), we accelerate convergence via warm starts, first optimizing on a random subset of events before refining the multipliers on the full dataset.
Finally, while both the prior and target shower samples possess finite MC statistics, we treat the target moments $c_f$ as exact for our proof-of-concept study.
See \App{uncertainties} for a discussion of ways to incorporate (correlated) uncertainties on the moments into our framework.

Since the cached measurement functions can be reused for any new set of constraint values, updating the reweighting for revised precision inputs carries negligible additional cost beyond the one-time $\mathcal{O}(N_q \times |\mathcal{B}(S)|)$ evaluation described above.
The maximum-entropy reweighting code used in this work, together with the analysis scripts used to produce the figures in this paper, is publicly available at \url{https://github.com/benleo12/efp_maxent}, and the event samples and derived per-event tables are archived on Zenodo~\cite{Assi:2026data}.

\subsection{Performance metrics}
\label{sec:PerfMet}

When we present our numerical results in the next section, we evaluate our reweighting framework in two distinct ways: the fidelity of the posterior and the efficiency of the prior.
As described more below, to quantify the quality of the posterior relative to the target, we evaluate the triangular divergence for the marginal distributions of various observables, comparing both the prior and posterior distributions against the target ``truth'' baseline.
To diagnose the efficiency of our chosen prior, we employ weight-health diagnostics.

In principle, the most direct way to assess whether the posterior has converged to the target would be to measure the distance between the two distributions at the full phase-space level.
This is highly nontrivial given the high dimensionality of phase space, so we leave such explorations to future work.
Instead, we evaluate the agreement between marginal distributions of the posterior and target for a large collection of individual observables.
Each EFP projects the full phase space onto a different one-dimensional distribution, and achieving good agreement across many such projections simultaneously is highly nontrivial.
Close agreement of the marginal distributions across EFPs therefore provides strong evidence that the posterior and target are close in the full phase space as well; see \App{2d_transfer} for a study of two-dimensional projections.

As our specific measure of statistical similarity, we use the triangular divergence. 
The triangular divergence is an $f$ divergence (see \App{alternative_measures}) that quantifies the similarity between two distributions.
Consider two distributions $r$ and $s$ for an observable $v$, represented by normalized histograms $\{r_{v,\ell}\}$ and $\{s_{v,\ell}\}$ with uniform bins indexed by $\ell$.
In this binned context, the triangular divergence is defined as:
\begin{equation}
\mathrm{TD}_v(r,s)=\frac{1}{2}\sum_\ell\frac{(r_{v,\ell}-s_{v,\ell})^2}{r_{v,\ell}+s_{v,\ell}+\varepsilon}\,,
\label{eq:TD}
\end{equation}
where we include a small regularization parameter $\varepsilon=10^{-10}$ to prevent division by zero in empty bins.
The triangular divergence is 0 if the distributions are identical and 1 if they have no overlapping support.
Unlike the KL divergence, the triangular divergence is symmetric between $r$ and $s$.
In our results, we report the divergence of the prior relative to the target before reweighting, $\mathrm{TD}_v^{\rm prior} = \mathrm{TD}_v(q, p_{\rm target})$, as well as the divergence of various posterior distributions determined from different sets of EFPs and measurement families, $\mathrm{TD}_v^{\rm posterior} = \mathrm{TD}_v(p^\star, p_{\rm target})$.%
\footnote{We also tested using the Wasserstein distances between distributions as a measure of fidelity, and found qualitatively similar results.}

Weight health is monitored using the effective sample fraction (ESF) defined in~\eq{ESF}, alongside the tail concentration metric
\begin{equation}
\kappa_{99} = \frac{w_{0.99}}{\bar{w}}\,,
\label{eq:kappa99}
\end{equation}
which measures the ratio of the 99th percentile weight to the mean weight $\bar{w}$.
A robust, healthy reweighting generally maintains an $\mathrm{ESF} \gtrsim 10\%$ and a moderate $\kappa_{99}$ (typically $\lesssim 5$).
Satisfying these conditions indicates that the prior distribution is efficient, possessing sufficient natural support to capture the information imposed by the precision constraints without relying on pathological, heavy-tailed reweighting factors.

\section{Results of EFP reweighting}
\label{sec:results}

We now present results for the EFP reweighting procedure, as described in \Sec{shower_setups}.
Starting from three degraded-shower priors, the target distributions are determined by an unbroken CSS shower at $\alpha_s(m_Z)=0.118$.
All results use EFPs computed in the heavy (larger invariant mass) hemisphere as defined by the thrust axis, following the conventions described in \Sec{efps}.
We start by studying how information saturates as the EFP basis is enlarged.
We then test how well the learned reweighting transfers to hemisphere observables not included in training, and explore how physics-motivated EFP basis reductions compare with generic basis choices.
In all of these studies, both the degraded priors and the target are \textsc{Sherpa} showers.
Finally, we test whether our conclusions persist when the prior comes from a different event generator, reweighting an angular-ordered \textsc{Herwig} prior to the same \textsc{Sherpa} target.

\subsection{Information saturation}
\label{sec:saturation_degree}

Our central question is how rapidly information saturates as we expand the set of EFP observables used for training, while holding the moment structure per observable fixed.
As discussed in \Sec{efps}, EFPs form an over-complete basis for IRC-safe observables, and one can organize them by degree $d$ or chromatic number $\chi$.
As discussed in \Sec{EFPtraining}, we can summarize EFP distributions using different moment families:  polynomial, logarithmic, or mixed.
The key question is then: given a fixed number of moment constraints per EFP (four for this study), how many EFPs must be included before additional observables provide diminishing returns?

\begin{figure}[t]
\centering
\subfloat[][]{
    \includegraphics[width=0.32\textwidth]{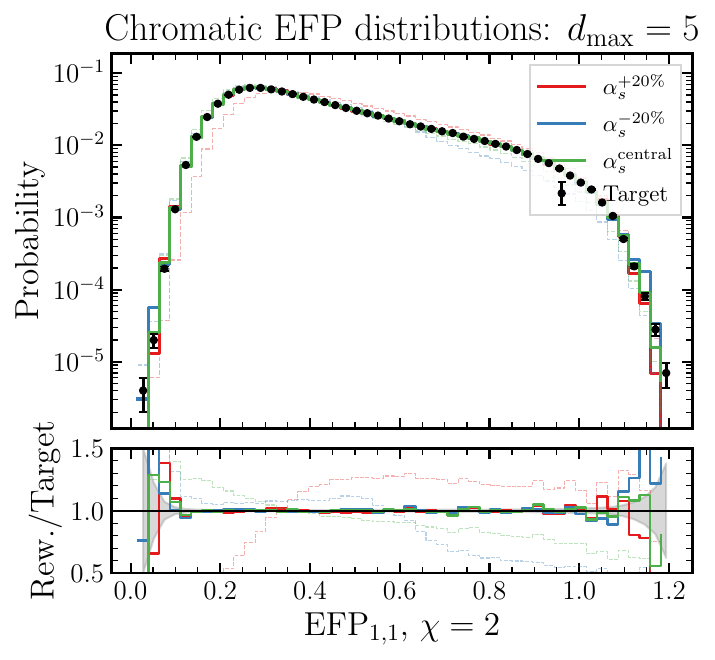}
    }
\subfloat[][]{
    \includegraphics[width=0.32\textwidth]{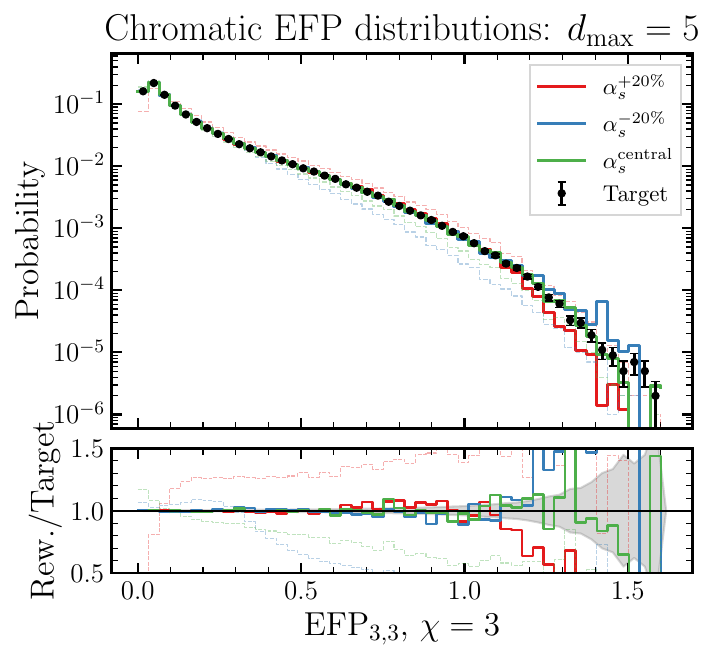}
    }
\subfloat[][]{
    \includegraphics[width=0.32\textwidth]{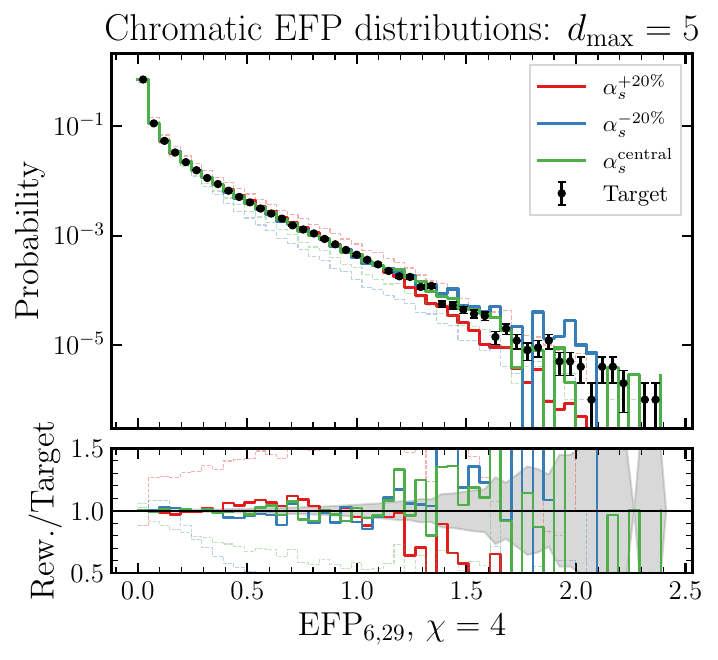}
    }
\caption{Comparison of marginal distributions with $d_{\rm max}=5$ reweighting with mixed moments for (a) $\mathrm{EFP}_{1,1}$ ($\chi=2$, trained), (b) $\mathrm{EFP}_{3,3}$ 
($\chi=3$, trained), and (c) $\mathrm{EFP}_{6,29}$ ($\chi=4$, transfer).
Black points show the target with statistical uncertainties, dashed lines show the unweighted priors, and solid lines show the reweighted distribution.
Colors indicate the $\alpha_s$ variations: 
central (green), $+20\%$ (red), $-20\%$ (blue).
Lower panels show the ratio to target with gray bands indicating the target statistical uncertainty. 
The complete graph $K_4$ (i.e.~$\mathrm{EFP}_{6,29}$) is outside the training set yet it is substantially corrected, demonstrating transfer across chromatic complexity.}
\label{fig:dist_d5}
\end{figure}

To start, we consider the largest set of EFPs used in our proof-of-concept study -- all prime and composite EFPs with degree $d\leq 5$, corresponding to 101 EFPs in total -- using the mixed moment representation.
In \Fig{dist_d5}, we show the impact of reweighting on the marginal distribution for three representative EFPs with different chromatic number:  $\mathrm{EFP}_{1,1}$ ($\chi=2$, single edge), $\mathrm{EFP}_{3,3}$ ($\chi=3$, triangle graph), and $\mathrm{EFP}_{6,29}$ ($\chi=4$, complete graph $K_4$).
The notation $\mathrm{EFP}_{d,k}$ indicates the $k$-th EFP (as ordered by the \texttt{EnergyFlow} package) with degree $d$, and the specific graphs chosen here are the lowest degree ones at a given chromatic number, as studied further in \Sec{chromatic_study}.
With $d_{\rm max}=5$ training, the $\chi = 2$ and $\chi = 3$ observables are part of the training, so it is not surprising that the reweighting corrects their distributions to near-perfect agreement with the target across all $\alpha_s$ variations. 
More surprisingly, the $K_4$ graph, which lies outside the $d_{\rm max} = 5$ training set, is substantially corrected through transfer, demonstrating that information propagates across chromatic complexity.
This is already a preliminary hint that information will saturate quickly and robustly, though with some exceptions studied in \Sec{aplanarity}.

\begin{figure}[t]
\centering
\includegraphics[width=0.9\textwidth]{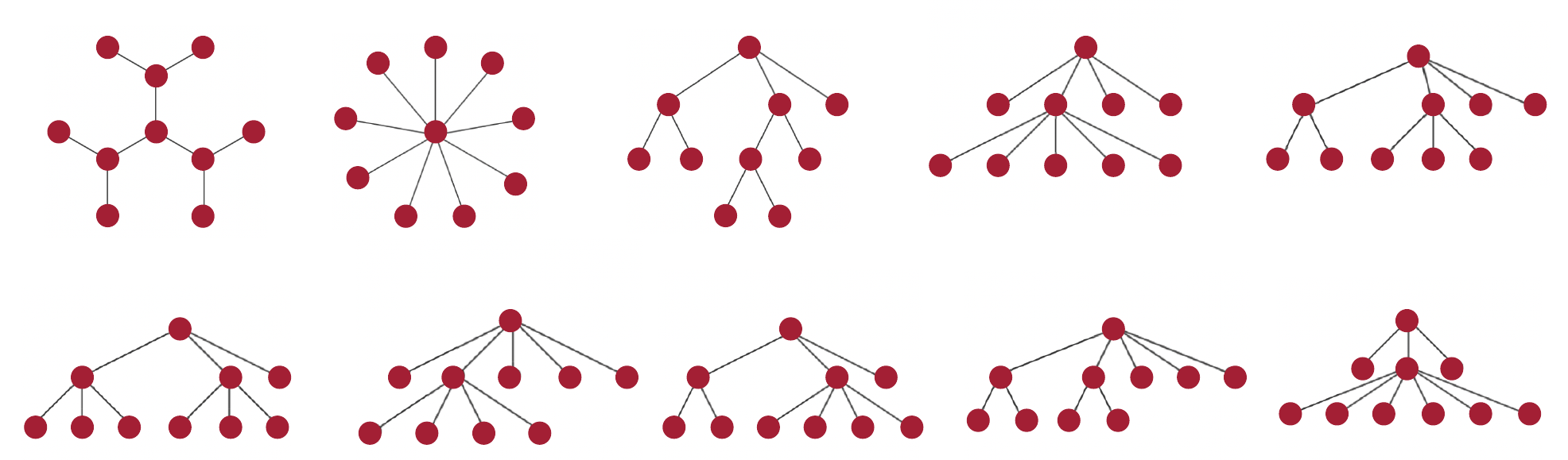}
\caption{The 10 homeomorphically irreducible trees with $n=10$ vertices, 
constituting the complete degree-9 EFP set used for validation in 
\Fig{saturation_prime_d5}. Their enumeration is a classic problem 
in combinatorial graph theory~\cite{Harary1973,Gessel2023}.}
\label{fig:deg9_trees}
\label{fig:deg9_subset}
\end{figure}

\subsubsection{Saturation organized by degree}

To study information saturation more systematically, we train on EFPs up to degree $d\leq d_{\rm max}$ and evaluate how well the reweighted posterior reproduces the target shower, both on the trained EFPs and on higher-degree EFPs not included in training.
To quantify information transfer, we compute the triangular divergence in \Eq{TD} between the target and reweighted distribution for all EFPs up to degree 6, as well as ten degree-9 homeomorphically irreducible trees shown in \Fig{deg9_subset},\footnote{The enumeration of degree-9 homeomorphically irreducible trees featured as a blackboard problem in the 1997 film \emph{Good Will Hunting}~\cite{Gessel2023}.} for a total of 166 observables.
Smaller values of the triangular divergence indicate better transfer of information, which we can track as we increase $d_{\rm max}$.

\begin{figure}[p]
\centering
\subfloat[][]{
    \includegraphics[width=0.95\textwidth]{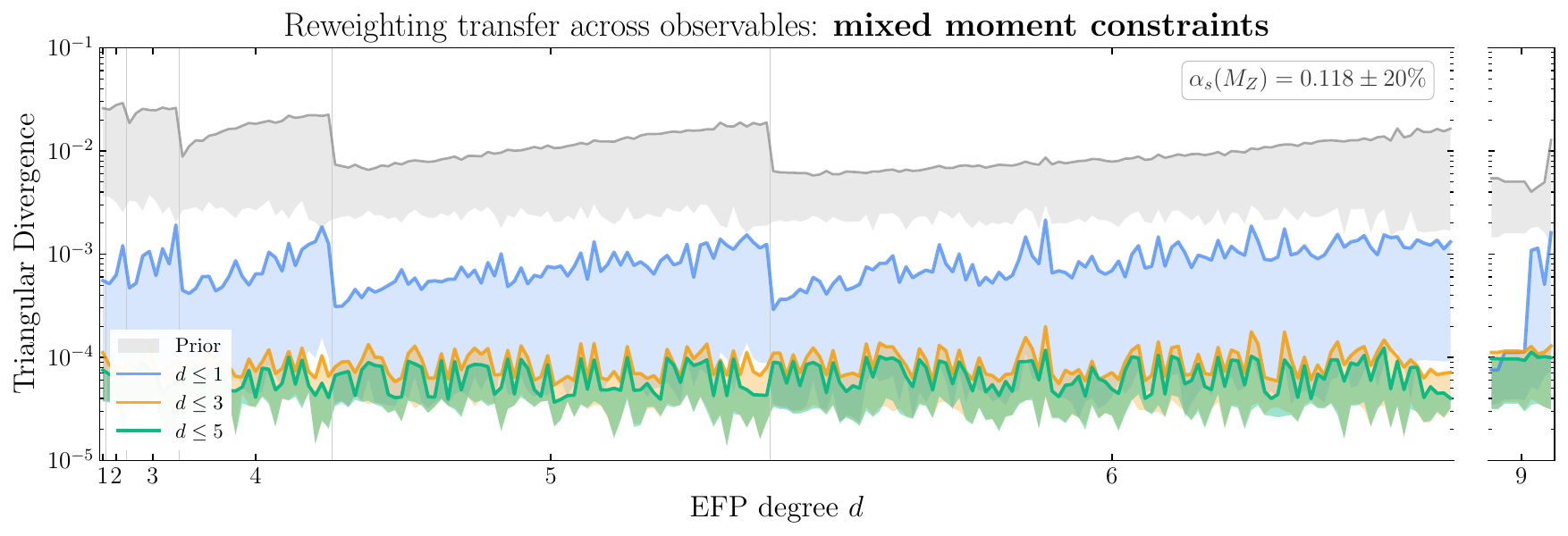}
    }\\
\subfloat[][]{
    \includegraphics[width=0.95\textwidth]{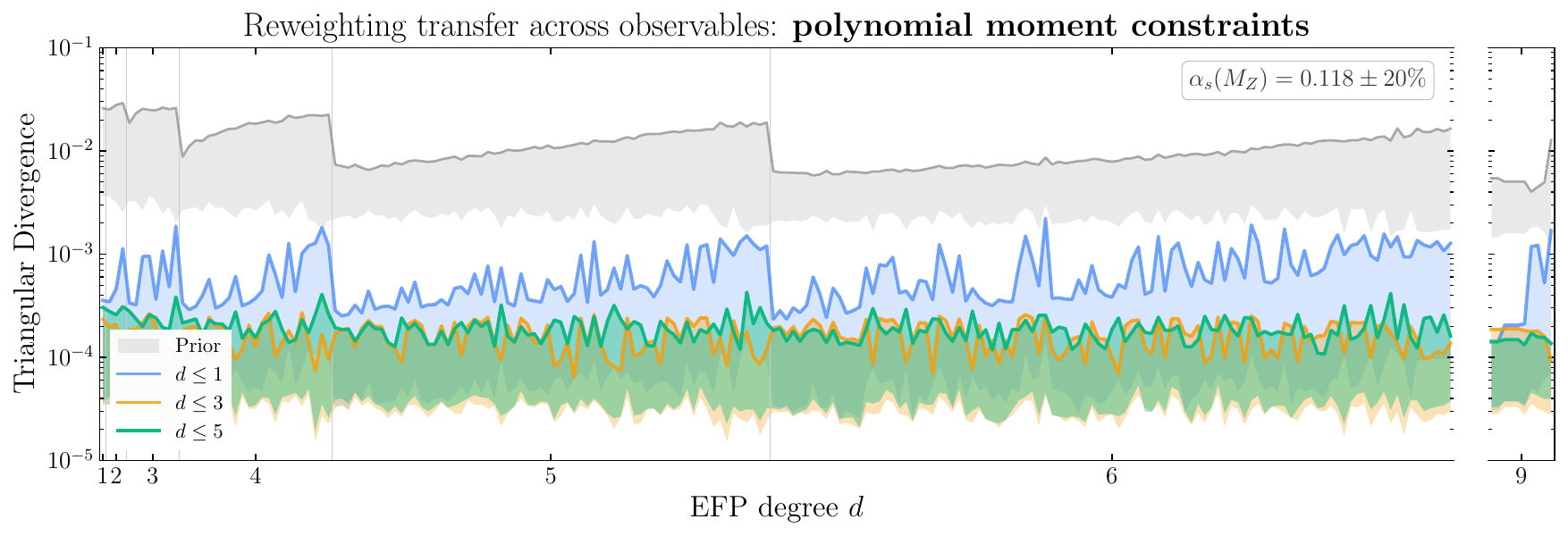}
    }\\
\subfloat[][]{
    \includegraphics[width=0.95\textwidth]{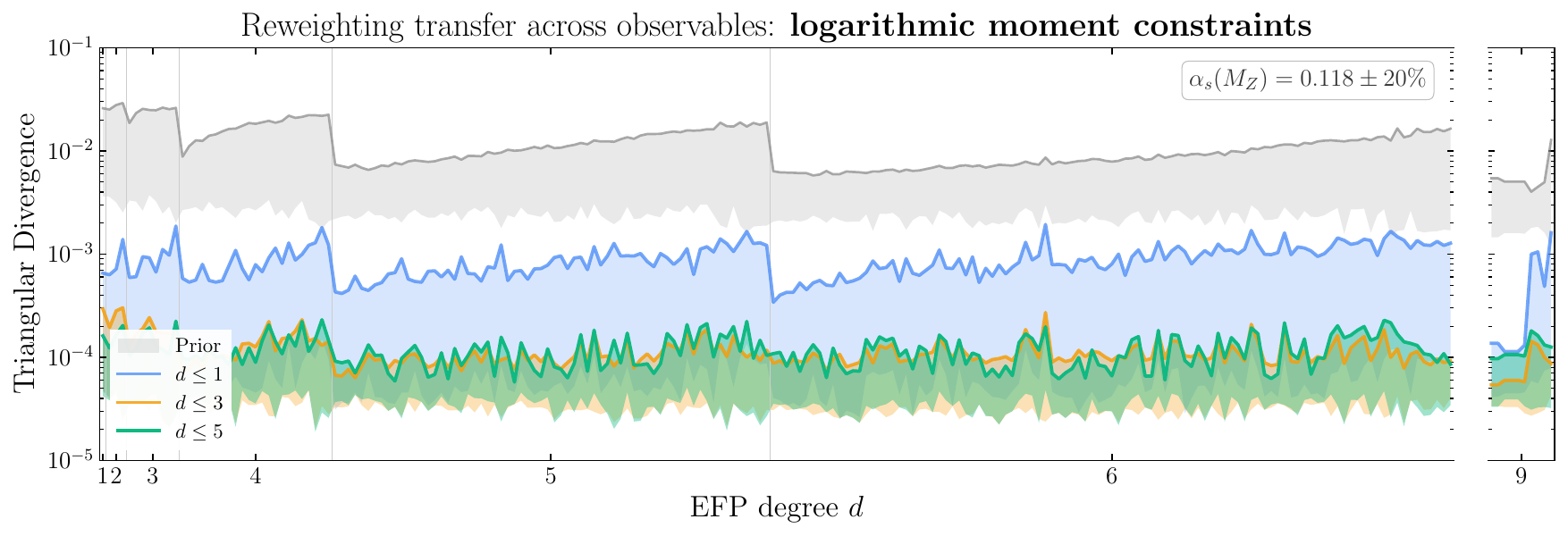}
    }
\caption{Information saturation of EFP-based reweighting organized by graph degree,
for (a) mixed, (b) polynomial, and (c) logarithmic moment features, as quantified across 156 prime EFPs.
In each panel, the triangular divergence
between reweighted and target distributions is shown for individual EFP observables, comparing
the unweighted prior (gray) to training sets including EFPs up to $d \leq 1$ (blue),
$d \leq 3$ (orange), and $d \leq 5$ (green).
Solid lines show the worst-case
$\alpha_s$ variation, while the bands span the full $\alpha_s(M_Z) = 0.118 \pm 20\%$ envelope.
Training on low-degree EFPs
alone yields substantial improvement across all test observables, with progressive
gains from including higher degrees.
The rightmost region shows an additional cinematic subset of degree-9 EFPs from \Fig{deg9_subset}.}
\label{fig:saturation_prime_d5}
\end{figure}

In \Fig{saturation_prime_d5}, we show the saturation behavior for different choices of $d_{\rm max}$, for the mixed moment (top), polynomial moment (middle), and logarithmic moment (bottom) measurement functions.
Each vertical column corresponds to a different EFP, and the shaded bands show the envelope of triangular divergences for the three different priors.
The gray band corresponds to the bare prior before reweighting, where the triangular divergence is around $10^{-2}$.
The blue, orange, and green bands show the reweighting based on training with EFPs of degree up to $d_{\rm max}=1$, $d_{\rm max}=3$, and $d_{\rm max}=5$, respectively.
We mark the top of each band with a solid curve, to emphasize that we are most interested in tracking the performance of the least accurate reweighting, though we also want to see that the spread among the priors decreases as we add more information.

We see that all three moment families exhibit rapid saturation.
Even for $d_{\rm max}=1$, which only has a single EFP, the triangular divergence decreases by an order of magnitude across all observables.
Roughly speaking, this single EFP imposes a constraint on an averaged cusp anomalous dimension across all observables, mitigating to a large extent the prior variation coming from the change in $\alpha_s$.
Training on EFPs up to degree $d_{\rm max}=3$ already achieves substantial improvement on higher-degree EFPs.
For $d_{\rm max}=5$ with mixed moments, the reweighted distributions are nearly indistinguishable from the target for the majority of observables, though this is hard to tell from the triangular divergence alone, which tends to saturate due to finite MC statistics.
Interestingly, when using only polynomial or logarithmic moments, the $d_{\rm max}=5$ reweightings can sometimes be worse than for $d_{\rm max}=3$, indicating a competition between the prior and the moment constraints.

\begin{figure}[t]
\centering
\subfloat[][]{
    \includegraphics[width=0.32\textwidth]{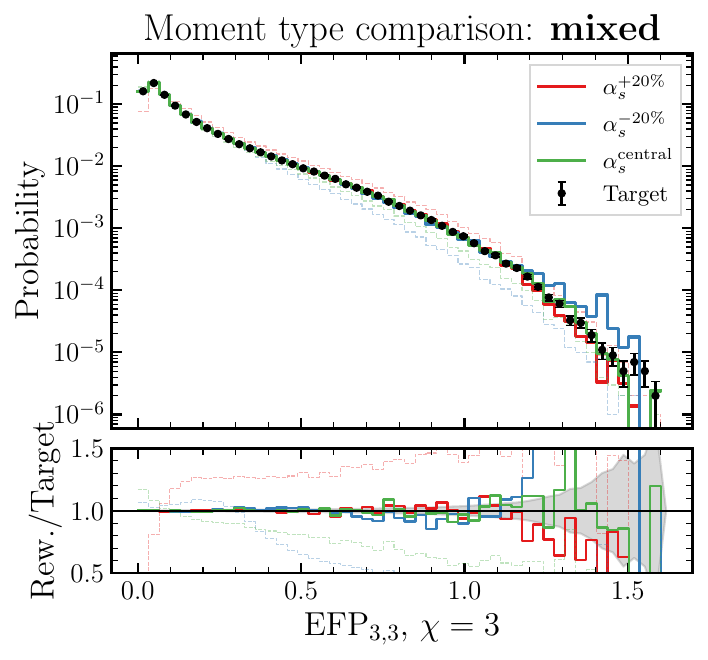}
    }
\subfloat[][]{
    \includegraphics[width=0.32\textwidth]{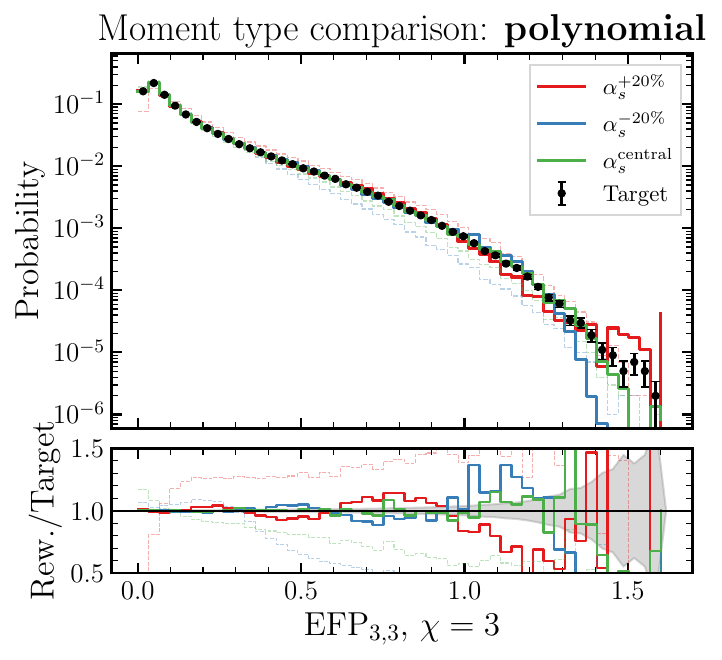}
    }
\subfloat[][]{
    \includegraphics[width=0.32\textwidth]{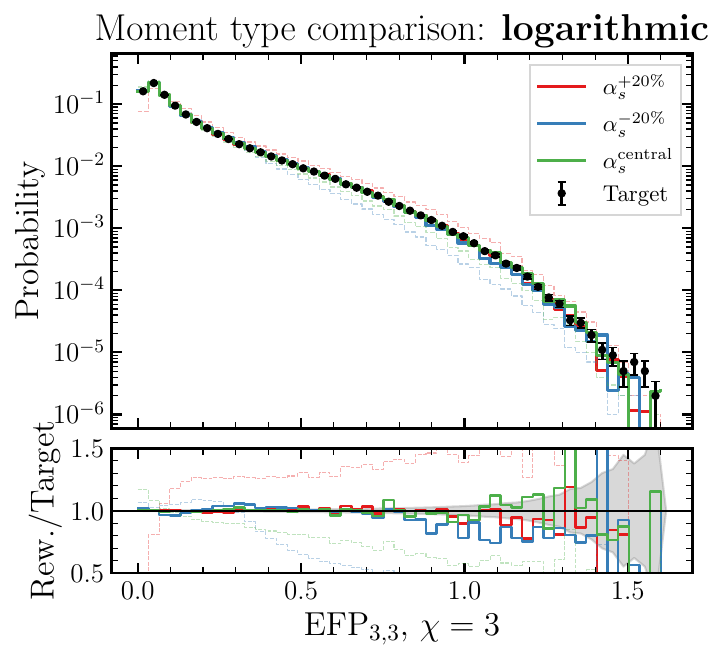}
    }
\caption{Comparison of moment families for reweighting $\mathrm{EFP}_{3,3}$ ($\chi=3$), using $d_{\rm max}=3$ training for (a) mixed moments, (b) polynomial moments only, and (c) logarithmic moments only.
The plot formatting follows \Fig{dist_d5}.
Mixed moments achieve the best corrections overall, combining
Sudakov-peak sensitivity from logarithmic moments with tail sensitivity from polynomial moments.}
\label{fig:moment_comparison}
\end{figure}

Among the three moment families, the mixed family exhibits the most efficient saturation, achieving comparable or superior improvement to pure log or pure polynomial families at lower $d_{\rm max}$.
The mixed moments $\langle\mathrm{EFP}_G\ln\mathrm{EFP}_G\rangle$ play a particularly important role, since it naturally interpolates between two regimes: the factor of $\mathrm{EFP}_G$ downweights the extreme Sudakov peak where logarithmic features would otherwise be dominated by rare fluctuations, while the logarithmic factor retains sensitivity to the exponentiated structure that characterizes the missing corrections in the degraded prior.
We illustrate this complementarity for $\mathrm{EFP}_{3,3}$ ($\chi=3$) in
\Fig{moment_comparison}: mixed moments achieve the flattest ratio across the full distribution, while polynomial moments alone show larger residuals in the intermediate region and logarithmic moments alone undershoot the tail.

\subsubsection{Discussion and weight diagnostics}

The rapid information saturation observed in \Fig{saturation_prime_d5} reflects the strong correlations among EFPs induced by the underlying soft-collinear structure of QCD.
Hemisphere energy flow is effectively a low-dimensional space, so many different EFP graphs probe overlapping kinematic configurations.
Constraining a relatively modest number of low-degree moments therefore injects sufficient information to reconstruct the distribution over a much larger set of observables.
We will return this point when discussing reduced EFP bases in \Sec{basis_selection}.

This rapid information saturation raises two natural questions.
First, why do the distributions for higher-degree EFPs improve when only lower-degree EFPs are trained?
This is natural because many higher-degree EFPs become composites of lower-degree EFPs under the strongly-ordered or collinear approximations discussed in \Sec{EFPbasis}.
As a result, some of the information in higher-degree EFPs is already captured by the lower-degree constraints.
Even beyond the strongly-ordered limit, the collinear structure of QCD ensures that low-degree EFPs constrain the dominant kinematic regions that also govern higher-degree observables.

Second, why do lower-degree EFPs continue to improve as higher-degree EFPs are added to the training set?
In principle, lower-degree information is already directly constrained, so one might not expect further improvements.
However, because we use only a small number of moments per EFP (four), the lower-degree constraints are not fully saturated.
A higher-degree EFP that contains a lower-degree EFP as a composite factor effectively provides additional moments of the lower-degree observable.
Concretely, when $\mathrm{EFP}_{G} = \mathrm{EFP}_{G_a} \cdot \mathrm{EFP}_{G_b}$, constraining $\langle \mathrm{EFP}_{G}^m \rangle$ amounts to constraining correlated higher moments of $\mathrm{EFP}_{G_a}$ and $\mathrm{EFP}_{G_b}$.
These additional constraints help pin down the lower-degree marginal distributions more precisely.

\begin{figure}[t]
\centering
\subfloat[][\label{fig:weight_diagnostics_esf}]{
    \includegraphics[width=0.48\textwidth]{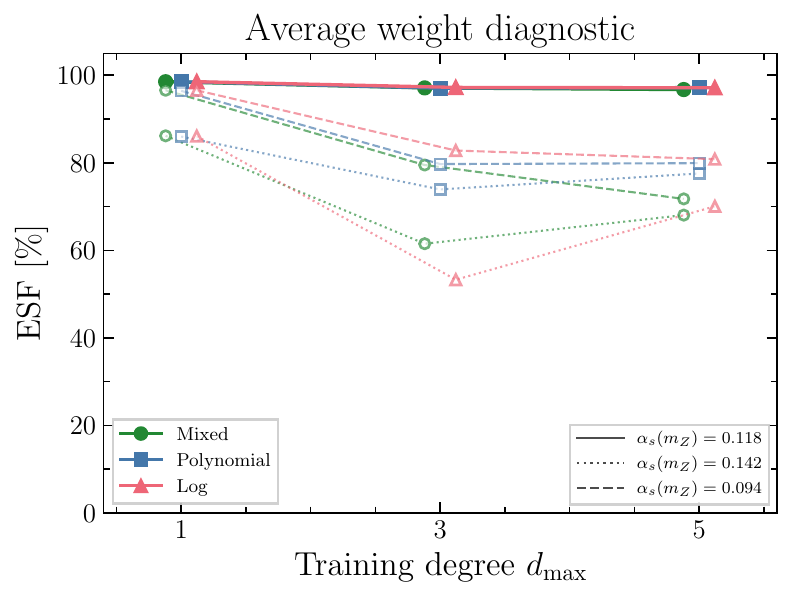}
    }
\hfill
\subfloat[][\label{fig:weight_diagnostics_kappa99}]{
    \includegraphics[width=0.48\textwidth]{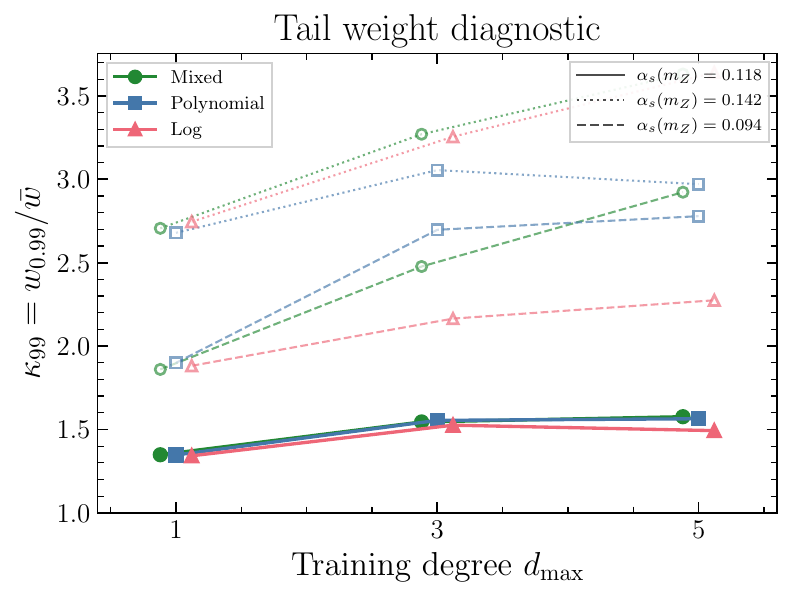}
    }
\caption{Weight diagnostics as a function of training degree $d_{\rm max}$
for mixed (green circles), polynomial (blue squares), and logarithmic (red triangles)
moment families.
Solid lines show the central $\alpha_s(m_Z) = 0.118$ prior;
dashed and dotted lines show the $\pm 20\%$ variations. All configurations remain
healthy ($\mathrm{ESF} \gtrsim 50\%$, $\kappa_{99} \lesssim 4$) at
$d_{\rm max} = 5$.}
\label{fig:weight_diagnostics}
\end{figure}

Another concern one might have is that the excellent  performance we are observing could arise from reduced statistical power of the reweighted distribution.
To guard against this, we can monitor weight-health diagnostics and robustness across the $\alpha_s$ variations in the degraded prior to ensure that saturation is not achieved at the cost of pathological weight distributions.
In \Fig{weight_diagnostics}, we use two weight diagnostics to confirm that information saturation is achieved with healthy weight distributions.
For all three moment families at $d_{\rm max} = 5$, the effective sample fraction satisfies $\mathrm{ESF} \gtrsim 68\%$ across the $\alpha_s$ variations, with tail concentration ratios $\kappa_{99} \lesssim 3.7$.
The central $\alpha_s = 0.118$ prior achieves $\mathrm{ESF} > 96\%$, reflecting its close proximity to the target.
The $+20\%$ variation, which has the highest emission rate and largest prior-target gap, requires larger corrections but remains healthy at $\mathrm{ESF} \simeq 68\%$.
These values are well above the $\mathrm{ESF} \gtrsim 10\%$ threshold discussed in \Sec{PerfMet}, indicating that the prior has sufficient phase-space support to accommodate the precision constraints without collapsing onto a small subset of events.

\subsubsection{Saturation organized by chromatic number}
\label{sec:chromatic_study}

\begin{figure}[t]
\centering
\subfloat[][]{
    \includegraphics[width=0.95\textwidth]{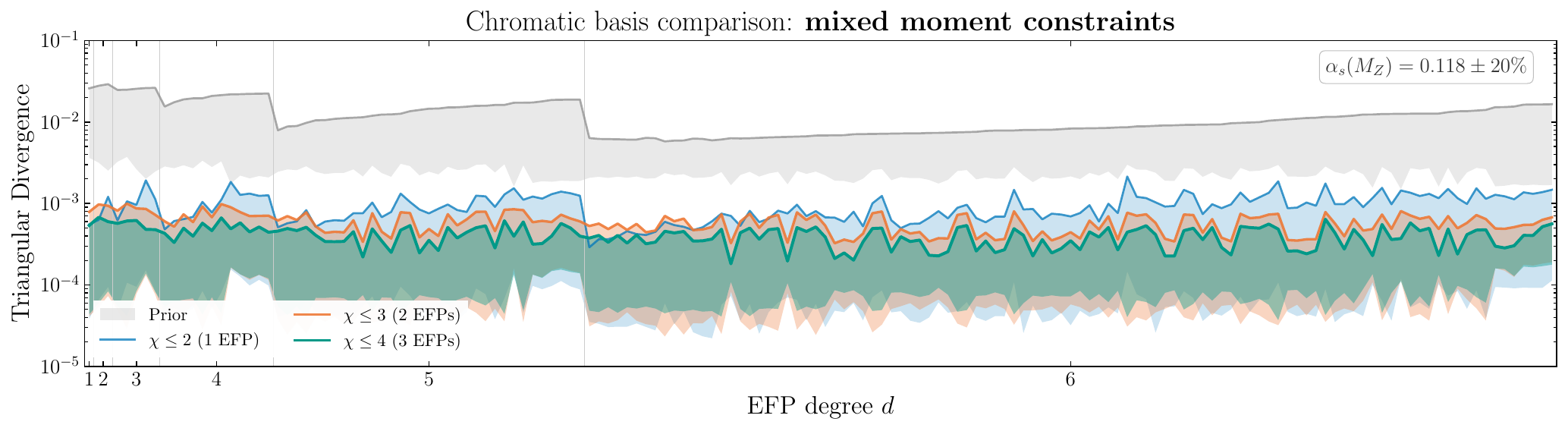}
    }\\
\subfloat[][]{
    \includegraphics[width=0.95\textwidth]{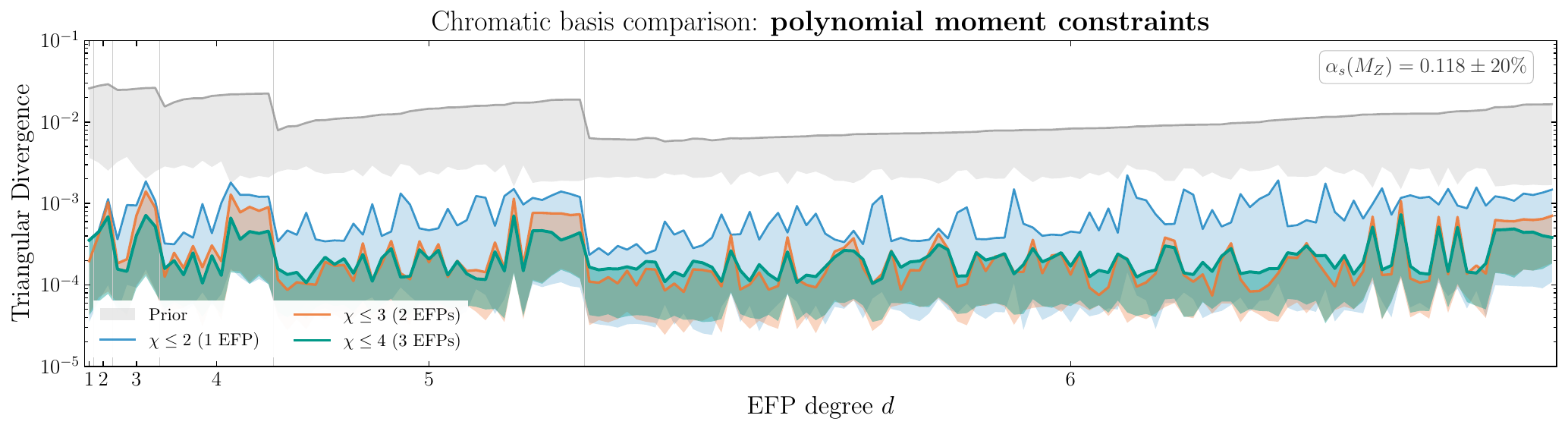}
    }\\
\subfloat[][]{
    \includegraphics[width=0.95\textwidth]{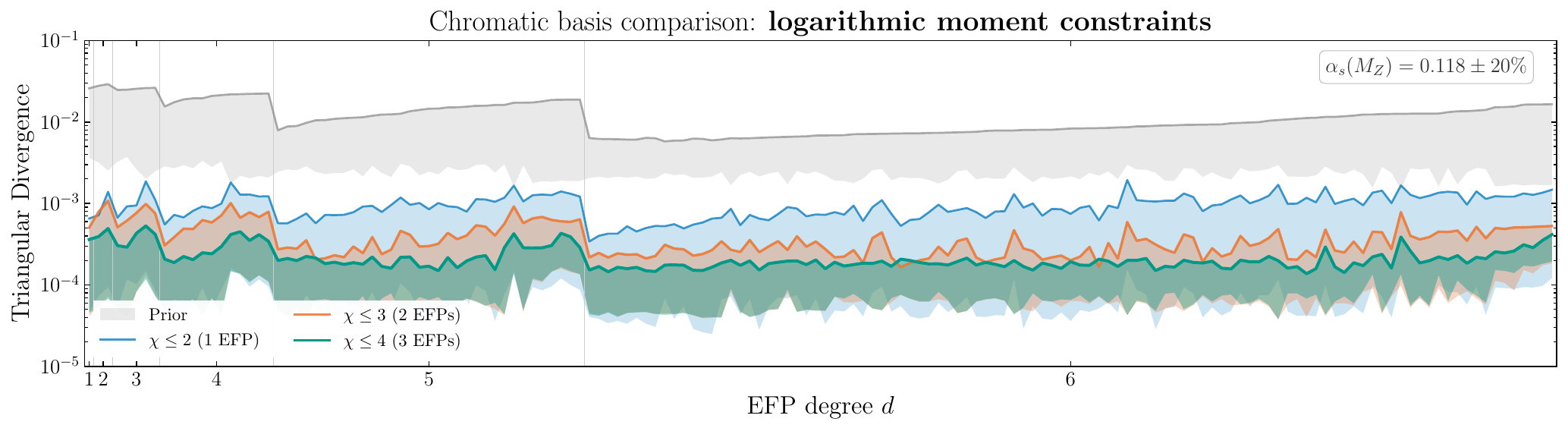}
    }
\caption{Information saturation using chromatic-number organized training bases for (a) mixed, (b) polynomial, and (c) logarithmic moment features.
The training sets consist of the lowest-degree representative
of each chromatic class: $\chi \leq 2$ (single edge, 1 EFP), $\chi \leq 3$
(adds triangle $K_3$, 2 EFPs), and $\chi \leq 4$ (adds complete graph $K_4$,
3 EFPs).
Evaluation spans the 156 prime EFPs up to degree 6.
These 1--3 EFP
training sets achieve 10--100$\times$ improvement in the triangular divergence, meaningful but below the
$\sim$100--1000$\times$ achieved by the complete degree-organized $d \leq 5$ basis in \Fig{saturation_prime_d5}.
}
\label{fig:saturation_chromatic}
\end{figure}

In addition to organizing EFPs by their degree, we can also organize EFPs by chromatic number $\chi$, the minimum number of colors needed to label graph vertices such that no edge connects same-colored vertices.
Higher chromatic number encodes increasingly complex multi-particle correlations: $\chi = 2$ (bipartite graphs) probes two-group correlations, $\chi = 3$ requires triangle substructures, and $\chi = 4$ requires the complete graph $K_4$, all shown in~\Fig{chromatic_efps}.
Since the number of EFPs at fixed chromatic number (and capped degree) grows more slowly with $\chi$ than the number at fixed degree grows with $d$, this provides a complementary test of information transfer.

For the chromatic-number expansion, we train on a small subset of EFPs within each chromatic class, rather than including all EFPs of a given chromatic number.
Indeed, there are an infinite number of graphs at a given chromatic number.
For this study, we use only the lowest-degree representative of each class: $\chi \leq 2$ includes only the single edge ($\mathrm{EFP}_{1,1}$, 1 EFP), $\chi \leq 3$ adds the triangle ($\mathrm{EFP}_{3,3}$, 2 EFPs total), and $\chi \leq 4$ adds the complete graph $K_4$ ($\mathrm{EFP}_{6,29}$, 3 EFPs total).
This contrasts with the degree-organized training, where all EFPs (primes and composites) up to $d_{\rm max}$ were included.

The results of the chromatic expansion are shown in \Fig{saturation_chromatic}.
These ultra-compact bases achieve $10$--$100\times$ improvement across all 156 prime EFPs up to degree 6, meaningful but less than the ${\sim}\,1000\times$ improvement from the complete $d \leq 5$ basis shown in~\Fig{saturation_prime_d5}.
Notably, the $\chi \leq 4$ basis provides only modest incremental gain over $\chi \leq 3$, consistent with the aplanarity results discussed in \Sec{aplanarity} below, where adding $K_4$ does not systematically improve multi-particle tails.
The gap between chromatic-minimal and complete bases reveals that graph topology within each chromatic class encodes genuinely distinct angular correlations: different graphs sharing the same $\chi$ probe different projections of multi-particle phase space and drives the full basis to ${\sim}\,1000\times$ improvement.

Comparing the three panels of \Fig{saturation_chromatic} also reveals an interesting contrast with the degree-organized results. For the chromatic bases, polynomial moments achieve the best performance at $\chi \leq 3$ and $\chi \leq 4$ (${\sim}\,47\times$), outperforming both logarithmic (${\sim}\,40\times$) and mixed (${\sim}\,36\times$) moments.
This is opposite to the degree-organized hierarchy, where mixed moments dominate.
With only 1--3 EFPs, polynomial moments extract the maximum angular information from each graph topology.
Apparently, it is only when many EFPs are included that the Sudakov sensitivity of logarithmic terms outweighs the per-graph efficiency of polynomial moments.

\subsection{Transfer to traditional observables}
\label{sec:transfer_eventshapes}

A more stringent test of the reweighting performance is information transfer to observables that were not included in the EFP training set.
Because the weights $w_i$ from \Eq{weights_exp} are event-level quantities, any observable computed on the same events can be reweighted via $\langle \mathcal{O} \rangle_{\rm rew} = \sum_a w_a \, \mathcal{O}(\Phi_a)$.
In the following, all observables are evaluated on the same heavy hemisphere and the same event sample used for the EFP computation, so the learned weights $w_a$ apply directly without any approximation.
Since any IRC-safe observable can be expanded in EFPs as in \eq{O_expand_EFP}, constraining a sufficiently rich set of EFP moments should in principle improve all such observables, with the degree of improvement determined by how strongly each observable correlates with the training set.

\subsubsection{Hemisphere shapes}

\begin{figure}[t]
\centering
\subfloat[][]{
    \includegraphics[width=0.32\textwidth]{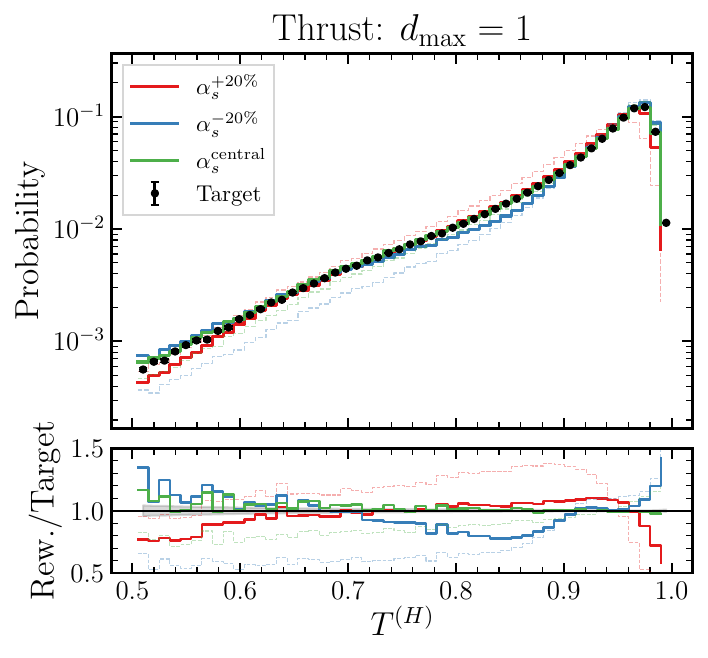}
    }
\subfloat[][]{
    \includegraphics[width=0.32\textwidth]{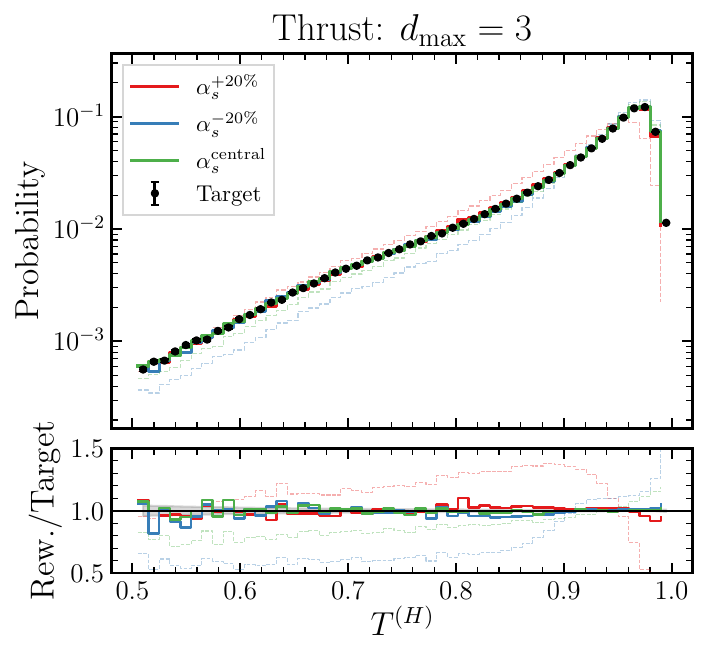}
    }
\subfloat[][]{
    \includegraphics[width=0.32\textwidth]{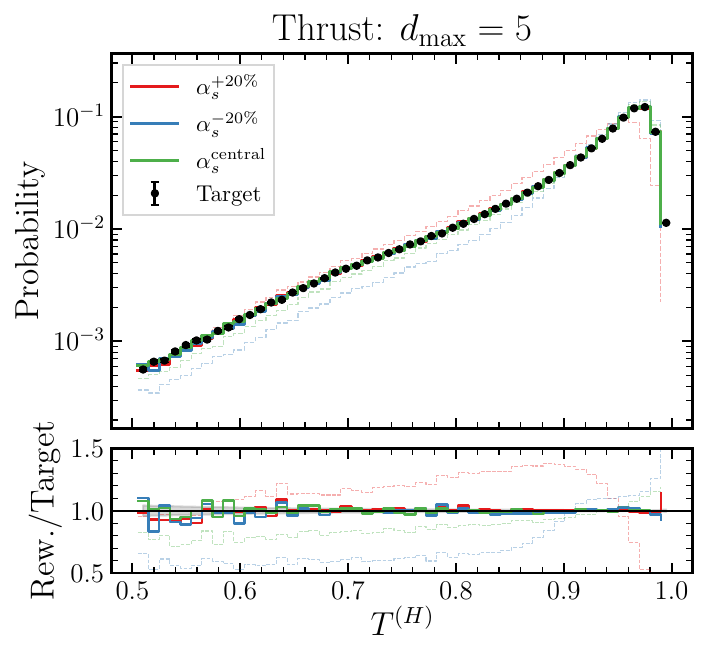}
    }
\caption{Hemisphere thrust distributions for training degrees (a) $d_{\rm max}=1$, (b) 
$d_{\rm max}=3$, and (c) $d_{\rm max}=5$.
The target distribution is shown as black points 
with statistical uncertainties, the dashed lines show unweighted prior, and solid 
lines show the reweighted result.
Colors indicate $\alpha_s$ variations:  central (green), $+20\%$ (red), $-20\%$ (blue).
Lower panels show the ratio to target with gray bands indicating the target statistical uncertainty. 
Saturation is rapid, with $d_{\rm max}=3$ achieving near-complete correction.}
\label{fig:thrust_saturation}
\end{figure}

We start with the classic event shape thrust, restricted to the heavy hemisphere.
In \Fig{thrust_saturation}, we show the hemisphere thrust distributions for training degrees $d_{\rm max}=1,3,5$, using the mixed moment family.
Even with $d_{\rm max}=1$, which involves only a single EFP and its four moments, the reweighted distributions show substantial improvement over the prior across all three $\alpha_s$ variations.
By $d_{\rm max}=3$ the correction is nearly complete, and $d_{\rm max}=5$ provides only marginal further improvement.
This rapid saturation mirrors the behavior observed for EFP observables in \Fig{saturation_prime_d5}, indicating that transfer to other Sudakov observables follows the same pattern.

\begin{figure}[t]
\centering
\subfloat[][]{
    \includegraphics[width=0.32\textwidth]{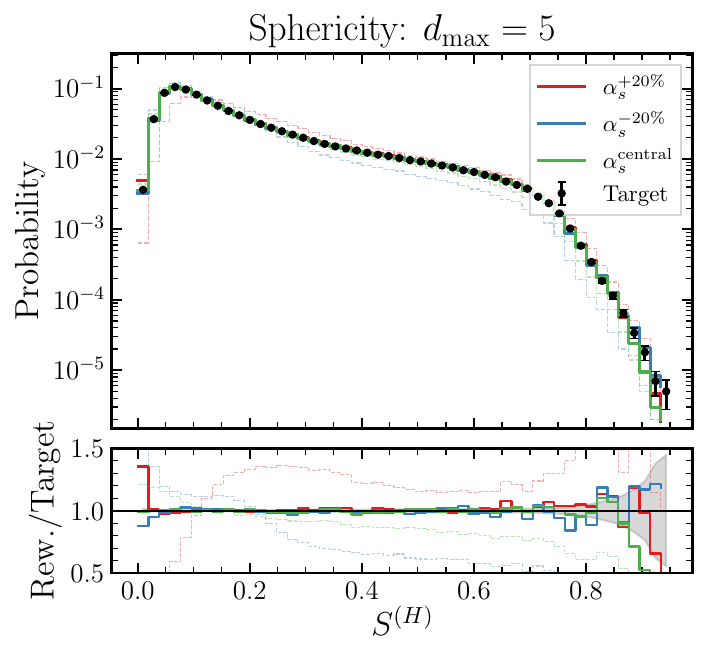}
    }
\subfloat[][]{
    \includegraphics[width=0.32\textwidth]{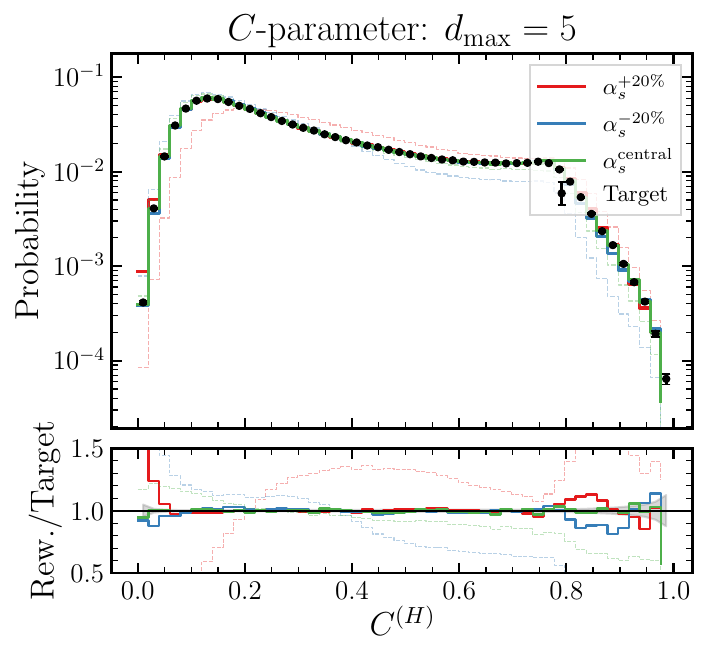}
    }
\subfloat[][]{
    \includegraphics[width=0.32\textwidth]{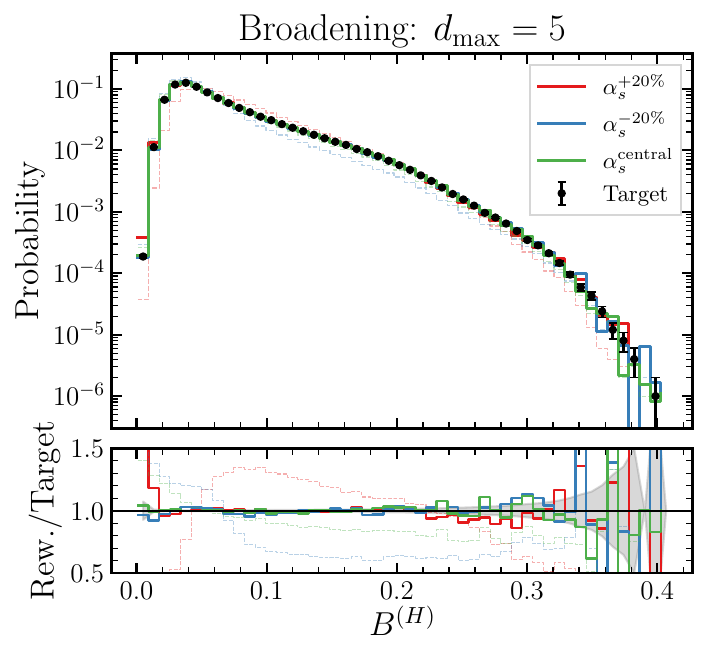}
    }
\caption{Additional hemisphere observables at $d_{\rm max}=5$ with mixed moments: (a) sphericity, (b) $C$-parameter, and (c) broadening.
The plot formatting follows \Fig{thrust_saturation}.
All three observables achieve good agreement with the target across the full $\alpha_s$ variation range.}
\label{fig:shapes_d5}
\end{figure}

Focusing on $d_{\rm max}=5$ with mixed moments, we show three additional hemisphere observables in \Fig{shapes_d5}: sphericity, $C$-parameter, and broadening.
All observables derived from the sphericity tensor use the linearized (IRC-safe) definition~\cite{Eden:1998gq}, in which each outer product is weighted by $1/|\vec{p}_i|$ rather than $1/|\vec{p}_i|^2$.
All three observables are corrected to near-target agreement across the full $\alpha_s$ variation range. This is consistent with the expectation from \eq{O_expand_EFP}: these event shapes are dominated by low-degree EFPs in the collinear limit, and the degree-5 training set constrains these contributions effectively.

\begin{figure}[t]
\centering
\subfloat[][]{
    \includegraphics[width=0.32\textwidth]{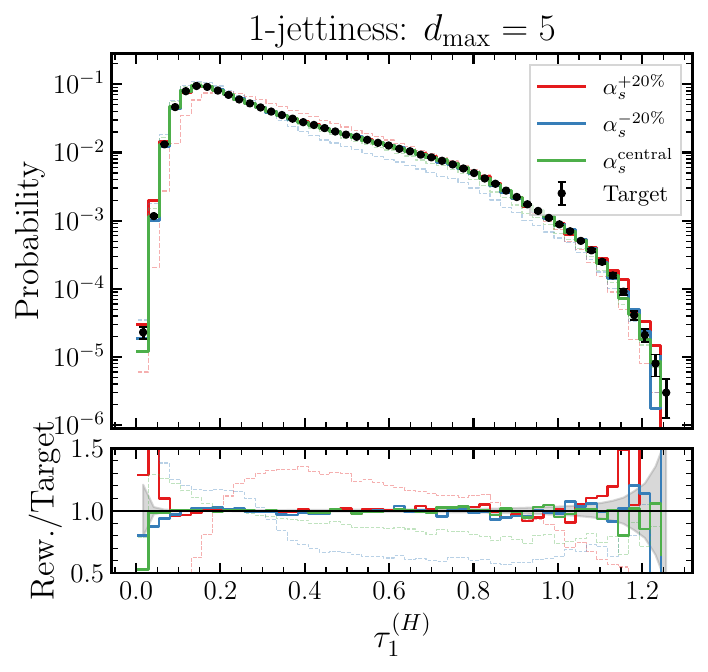}
    }
\subfloat[][]{
    \includegraphics[width=0.32\textwidth]{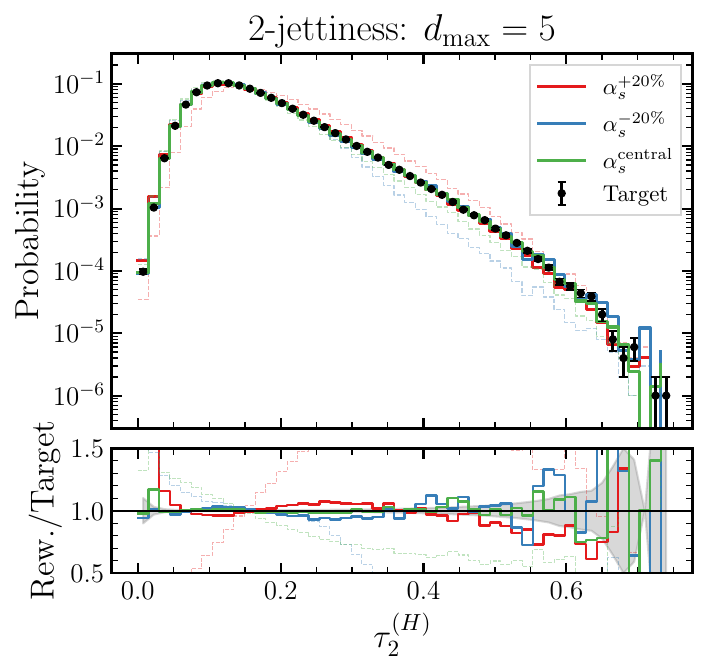}
    }
\subfloat[][]{
    \includegraphics[width=0.32\textwidth]{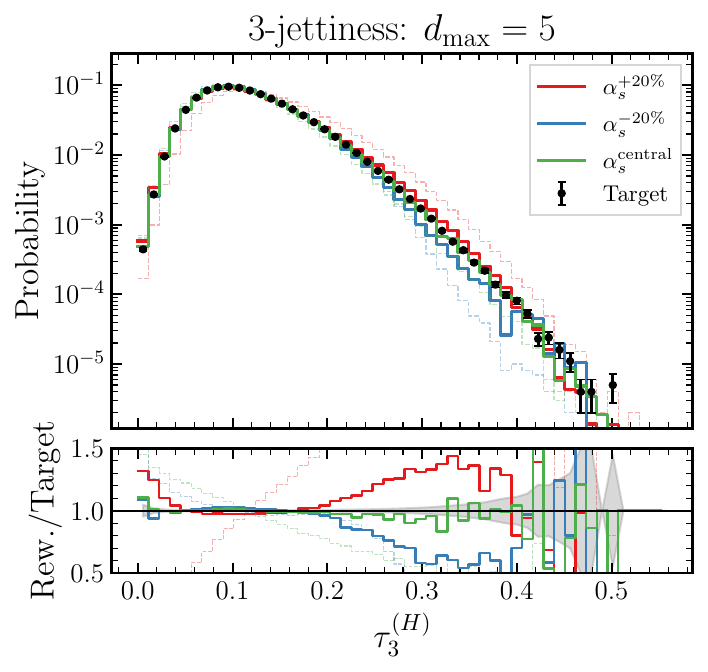}
    }
\caption{Hemisphere $N$-jettiness distributions at $d_{\rm max}=5$ with mixed moments: (a) $\tau_1$, (b) $\tau_2$, and (c) $\tau_3$.
The plot formatting follows \Fig{thrust_saturation}. Transfer succeeds for $\tau_1$ and $\tau_2$ despite the distinct kinematic definition involving minimization over jet axes.
The $\tau_3$ tail shows degradation, reflecting missing sensitivity to multi-particle angular correlations.}
\label{fig:jettiness_d5}
\end{figure}

In \Fig{jettiness_d5}, we show distributions for hemisphere $N$-jettiness~\cite{Stewart:2010tn,Thaler:2010tr} for $N=1,2,3$, again reweighted using $d_{\rm max}=5$ and mixed moments.
These observables measure the degree to which radiation in a hemisphere is clustered around $N$ axes and involve a minimization over axis directions that has no direct analog in the EFP construction.
Nevertheless, the EFP-based reweighting transfers successfully to $\tau_1$ and $\tau_2$.
The quality of the transfer degrades as $N$ increases, with $\tau_3$ showing incomplete correction at large values.
This is physically sensible: higher-$N$-jettiness probes increasingly fine-grained multi-particle configurations.
In particular, the large-$\tau_3^{(H)}$ tail signals energy flow along a fourth direction that cannot be absorbed into three jet axes, probing four-particle correlations that are not fully captured by the $d_{\rm max}=5$ training set.

\subsubsection{Multi-particle tails}
\label{sec:aplanarity}

In the $\tau_3^{(H)}$ analysis above, we saw that the $d_{\rm max}=5$ basis was not sufficient to fully correct the multi-particle tails.
As a further probe of this region, we study the hemisphere aplanarity and $D$-parameter distributions.
Defined as $A = \frac{3}{2}\lambda_3$, where $\lambda_3$ is the smallest eigenvalue of the linearized sphericity tensor, aplanarity measures the extent to which particle momenta spread out of the event plane.
In a hemisphere, where momentum conservation does not constrain the particles, three momenta can be linearly independent, so aplanarity can be nonzero already for three particles ($\chi = 3$).
However, the tails of the aplanarity and $D$-parameter distributions require substantial momentum flow out of the event plane, involving additional emissions beyond this minimum.

\begin{figure}[t]
\centering
\subfloat[][]{
    \includegraphics[width=0.45\textwidth]{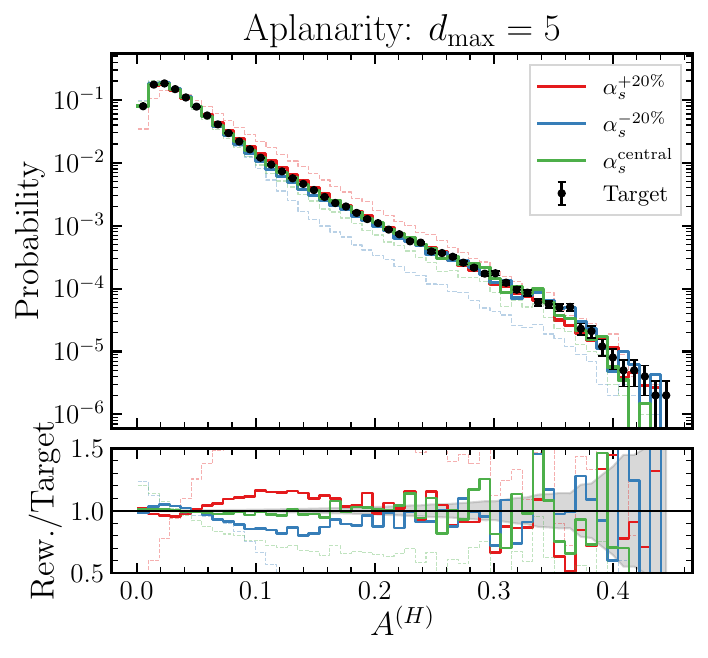}
    }$\quad$
\subfloat[][]{
    \includegraphics[width=0.45\textwidth]{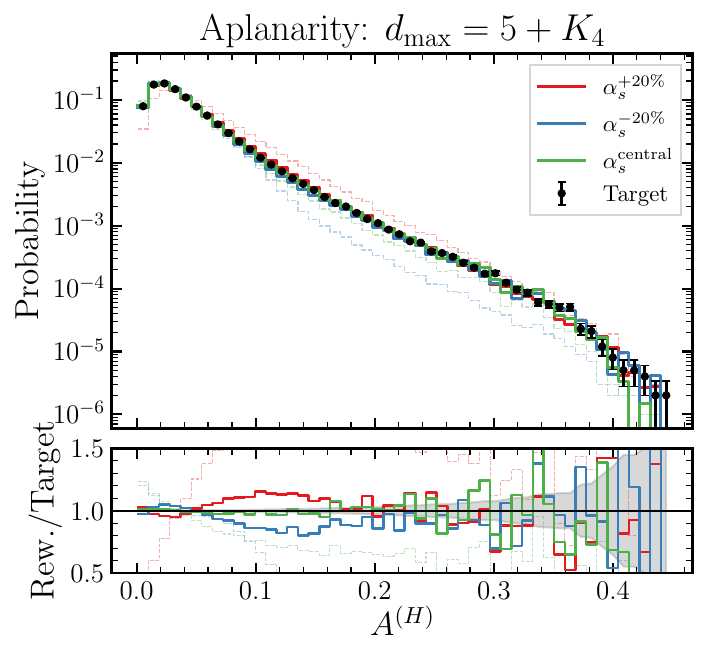}
    }
\caption{Aplanarity distribution using the mixed moment reweighting with (a) $d_{\rm max}=5$ and (b) 
$d_{\rm max}=5$ with $K_4$ ($\mathrm{EFP}_{6,29}$) added.
The 
central-$\alpha_s$ prior (green) is corrected in both cases. The $\pm 20\%$ 
priors (red, blue) show incomplete correction that is not improved by 
adding $K_4$.
Adding $K_4$ does not systematically improve the aplanarity correction.}
\label{fig:aplanarity}
\end{figure}

\begin{figure}[t]
\centering
\includegraphics[width=0.45\textwidth]{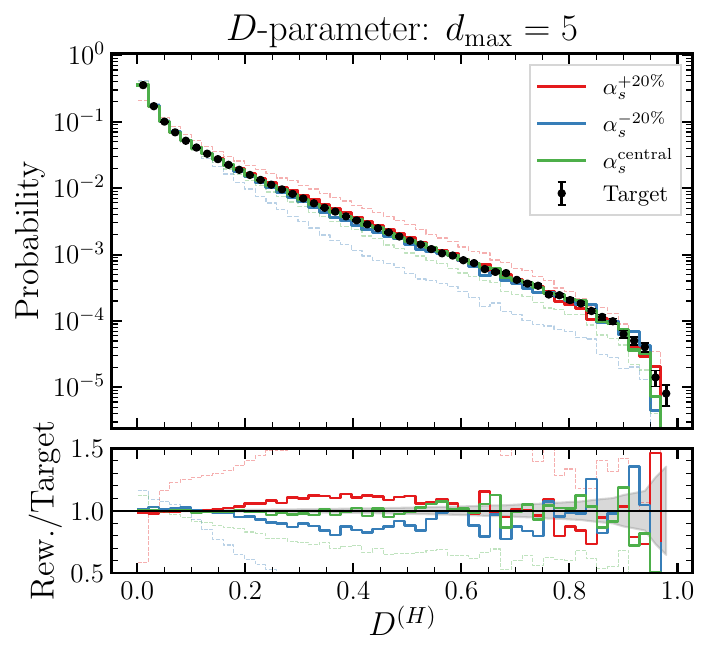}
\caption{$D$-parameter distribution using the $d_{\rm max}=5$ mixed moment reweighting. The
central-$\alpha_s$ prior (green) is corrected, while the $\pm 20\%$
priors (red, blue) show incomplete correction similar to aplanarity.}
\label{fig:d_parameter}
\end{figure}

In \Fig{aplanarity}, we show the aplanarity distribution using the $d_{\rm max}=5$ mixed moment reweighting.
We also show an augmented training with the $K_4$ graph ($\mathrm{EFP}_{6,29}$) added to the training set.
The central-$\alpha_s$ prior is corrected successfully in both cases, but the $\pm 20\%$ variations show substantially less improvement.
The $D$-parameter, shown in \Fig{d_parameter}, exhibits similar behavior.
Adding $K_4$ explicitly does change the learned weights and produces small shifts in the aplanarity distribution.
However, these shifts are not systematic improvements: for the $+20\%$ prior the triangular divergence increases slightly, while for $-20\%$ it decreases slightly.
This indicates that $K_4$ captures some of the multi-particle phase space relevant to aplanarity, but it alone is not sufficient to fully constrain it.
The $K_4$ EFP measures a specific product of pairwise angles, whereas aplanarity measures out-of-plane spread through the eigenvalue structure of the sphericity tensor.

When the prior is close to the target (central $\alpha_s$), the reweighting reaches aplanarity indirectly through correlations in EFP space.
When the prior is far from the target ($\pm 20\%$ variations), this indirect path is insufficient.
It would be interesting to find an explicit decomposition of aplanarity in terms of EFPs to understand quantitatively how information saturates, but this is beyond the scope of the present work.
At minimum, one can expect that a single additional EFP like $K_4$ is not sufficient to capture aplanarity, and a complete correction would likely require training to higher degree to better span the relevant multi-particle phase space.

\subsubsection{Energy-energy correlators}

As discussed in \Sec{relation_to_energy_correlators}, the polynomial moments of EFPs are directly connected to the angular moments of multi-point energy correlators.
Since energy correlators can be defined within the heavy hemisphere in the same way as our EFPs, it is natural to ask whether the reweighting also improves the hemisphere-level energy correlator distribution.

\begin{figure}[tb]
\centering
\includegraphics[width=0.45\textwidth]{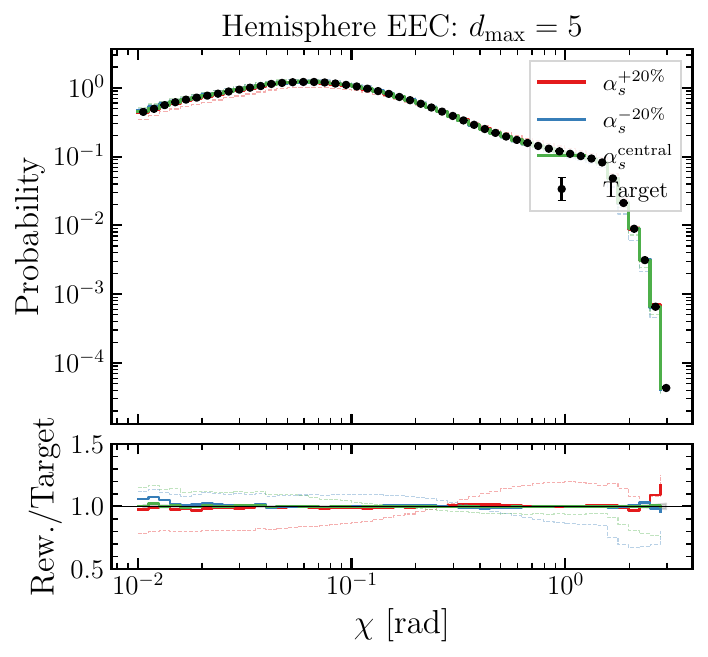}
\caption{Hemisphere-level EEC distribution as a function of the opening angle $\chi$, using the mixed moment $d_{\rm max}=5$ reweighting. The plot formatting follows
\Fig{thrust_saturation}. The reweighted distributions show good improvement, reflecting the close connection between EFP polynomial moments and energy correlator angular moments established in \Sec{relation_to_energy_correlators}.}
\label{fig:eec_transfer}
\end{figure}

In \Fig{eec_transfer}, we show the two-point EEC distribution before and after reweighting, using the $d_{\rm max} = 5$ reweighting with mixed moments.
It is instructive to contrast the hemisphere-level EEC shown here with the standard event-wide EEC, which has been measured and computed with high precision across its full angular range~\cite{Electron-PositronAlliance:2025fhk,Jaarsma:2025tck,Schindler:2023cww,Lee:2024esz,Gao:2026xuq}.
The event-wide EEC exhibits distinct physical regimes as a function of angle: a collinear region ($\chi \ll 1$) governed by DGLAP-type splitting and anomalous scaling, a back-to-back region ($\pi-\chi \ll 1$) dominated by Sudakov double logarithms from soft and collinear emissions, and confinement regions ($\chi , \pi-\chi \lesssim \Lambda_{\rm QCD}/Q\ll 1$) where hadronization sets in.
For the hemisphere-level EEC, the collinear region retains the same physical interpretation, but the back-to-back region now corresponds to correlations between soft radiations at the opposite boundary of the hemisphere.
This is also the region where non-global logarithms from the hemisphere boundary become important~\cite{Dasgupta:2001sh,Dasgupta:2002bw}.
In \Fig{eec_transfer}, we observe that the collinear and the bulk region shows marked improvement after reweighting, which may be expected given the direct connection between polynomial moments of EFPs and angular moments of the EEC established in \Sec{relation_to_energy_correlators}. The confinement regions also show substantial improvement, which is noteworthy since this regime is governed by nonperturbative dynamics.
In the back-to-back region, improvement is less pronounced, possibly due to the sensitivity of this region to NGLs. The improvements overall are encouraging, as they indicate that precision information from EEC calculations~\cite{Dixon:2018qgp,Tulipant:2017ybb,Dixon:2019uzg,Duhr:2022yyp,Gao:2020vyx,Jaarsma:2025tck} can be leveraged by our framework both as input constraints and as independent validation observables.

The event shapes and correlators considered in this subsection all probe the hemisphere energy flow through global momentum sums.
In \App{substructure}, we extend the transfer studies to a complementary class of observables that probe the shower evolution and hadron-level dynamics more directly, namely subjet rates, the primary Lund plane, the charged-particle multiplicity, and the charged fragmentation spectrum.

\subsection{QCD structure and basis selection}
\label{sec:basis_selection}

The above results demonstrate that EFP-based reweighting achieves broad improvement across observable space.
We now ask a more targeted question motivated by the discussion in \Sec{EFPbasis}: can QCD understanding help design a better choice of the training observables?
This is a natural question because the EFP basis simplifies considerably in the strongly-ordered and collinear limits, and different EFPs that appear distinct at the exact level can become degenerate under these approximations.

\begin{figure}[t]
\centering
\includegraphics[width=0.9\textwidth]{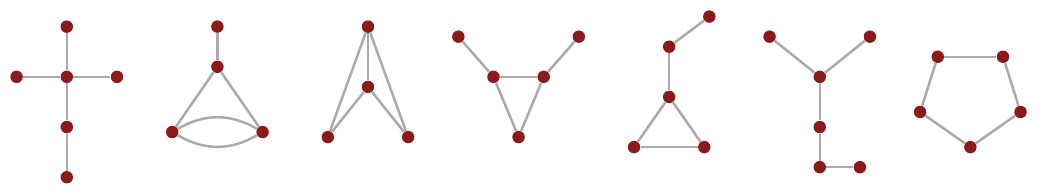}
\caption{The seven degree-5 EFP graphs constituting the maximally redundant ``worst'' training set. All are tree or near-tree graphs with mutual Pearson correlations $|r_{ij}| > 0.996$ and identical leading collinear scaling under strong ordering.}
\label{fig:worst_set_graphs}
\end{figure}

Motivated by \Fig{efp_bases_comparison} and the collinear power-counting of \Reference{Cal:2022fnm}, we consider a 7-EFP training set consisting of the four strongly-ordered (SO) prime EFPs at $d \leq 3$ plus 3 composite EFPs at $d \leq 3$.
This ``SO + composites'' basis is significantly smaller than the full set of 12 EFPs at $d \leq 3$.
To benchmark the performance of this QCD-motivated basis, we construct a maximally redundant ``worst'' set of 7 EFPs.
This ``worst'' set was constructed by computing the Pearson correlation coefficient,
\begin{equation}
r_{ij} = \frac{\sum_{k=1}^{N_{\rm ev}} \bigl(\mathrm{EFP}_i^{(k)} - \overline{\mathrm{EFP}}_i\bigr)\bigl(\mathrm{EFP}_j^{(k)} - \overline{\mathrm{EFP}}_j\bigr)}{\sqrt{\sum_{k=1}^{N_{\rm ev}} \bigl(\mathrm{EFP}_i^{(k)} - \overline{\mathrm{EFP}}_i\bigr)^2\;\sum_{k=1}^{N_{\rm ev}} \bigl(\mathrm{EFP}_j^{(k)} - \overline{\mathrm{EFP}}_j\bigr)^2}},
\label{eq:pearson}
\end{equation}
between all pairs of the 53 prime EFPs up to $d = 5$ evaluated on target events, where $\mathrm{EFP}_i^{(k)}$ denotes the value of the $i$-th EFP on event $k$ and $\overline{\mathrm{EFP}}_i$ is its sample mean.
The 7-EFP subset with the highest mean pairwise $|r_{ij}|$ was then selected greedily.
The resulting set consists entirely of degree-5 tree graphs with mutual correlations $|r_{ij}| > 0.996$. One can also formally show that under strong-ordering, all 7 of the selected ``worst'' EFPs become identical, i.e.\ they share the same leading collinear scaling.
By construction, this is the set of 7 distinct EFPs that carry the least independent information when polynomial moments are used.
The seven graphs are shown in \Fig{worst_set_graphs}.

\begin{figure}[t]
\centering
\subfloat[][]{
    \includegraphics[width=0.95\textwidth]{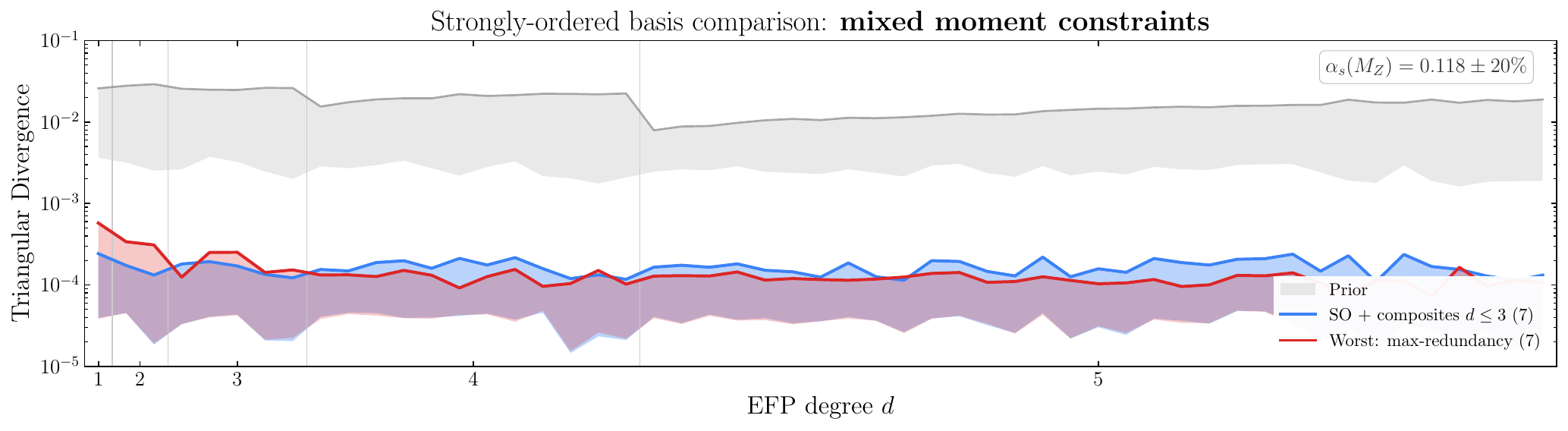}
    }\\
\subfloat[][]{
    \includegraphics[width=0.95\textwidth]{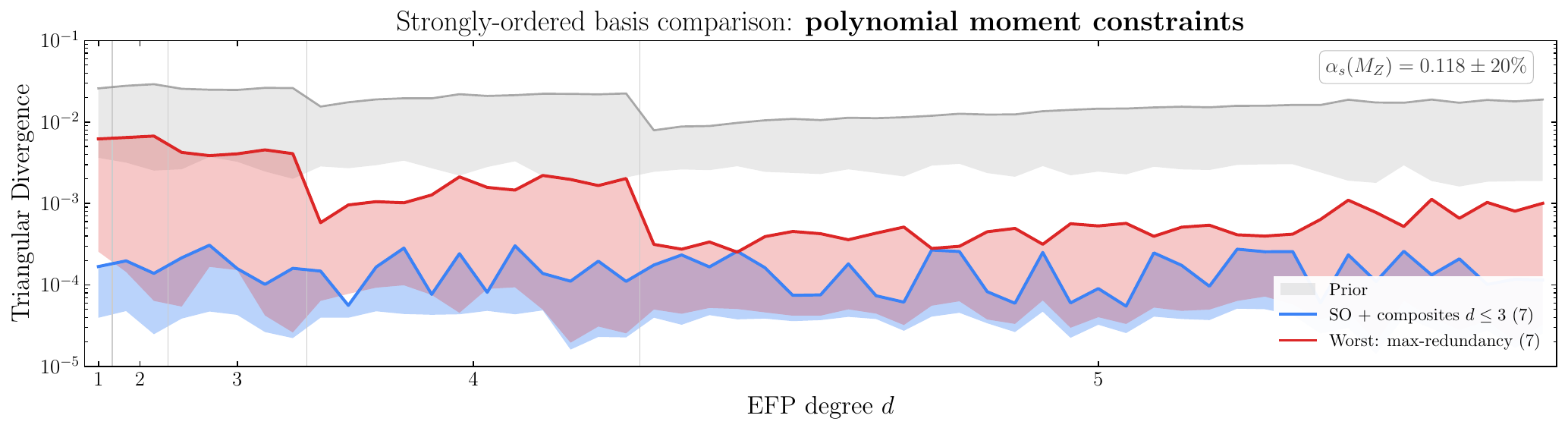}
    }\\
\subfloat[][]{
    \includegraphics[width=0.95\textwidth]{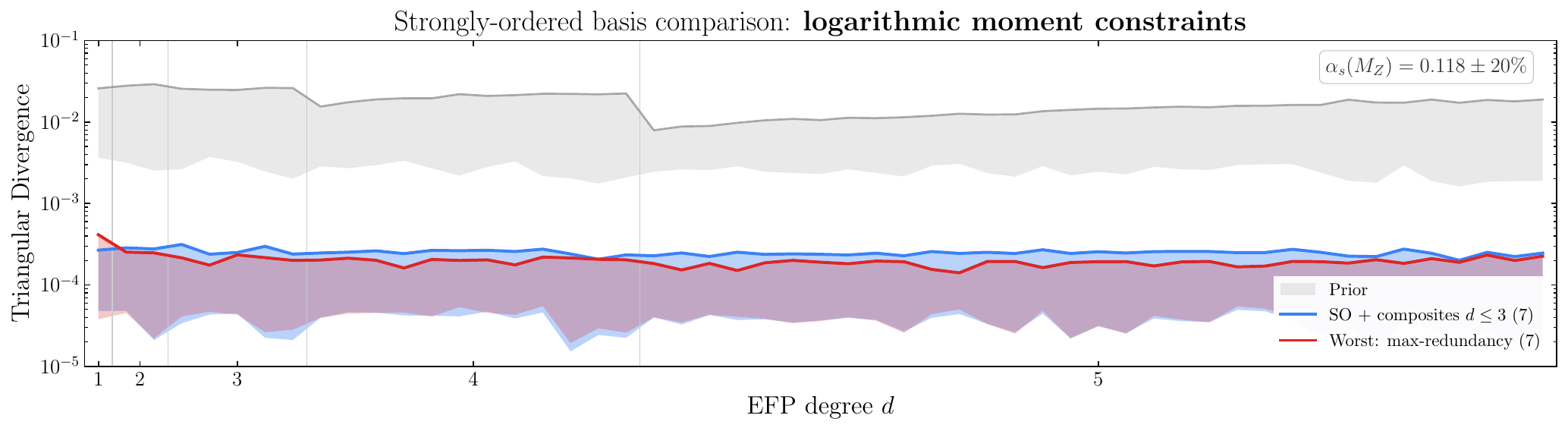}
    }
\caption{Basis comparison: SO+composites ($d \leq 3$, 7 EFPs, blue) versus
the maximally redundant worst set (7 EFPs, red), evaluated on all 53 prime
EFPs, for mixed (top), polynomial (middle), and logarithmic (bottom) moments.
For polynomial moments the physics-motivated SO basis clearly outperforms
the worst set, confirming that the collinear power-counting hierarchy
identifies genuinely distinct angular correlations. For mixed and logarithmic
moments the hierarchy collapses, with both bases achieving comparable
performance. Bands show variation across $\alpha_s(M_Z) = 0.118 \pm 20\%$.}
\label{fig:basis_SO_comparison}
\end{figure}

In \Fig{basis_SO_comparison}, we compare the SO+composites basis to the worst set for all three moment families.
For polynomial moments, the SO+composites basis achieves ${\sim}\,75\times$ median improvement on the transfer EFPs, compared to only ${\sim}\,25\times$ for the worst set.
This factor-of-3 difference is expected: polynomial moments preserve the collinear power-counting, so the degree-5 trees provide genuinely redundant constraints while the absence of low-degree EFPs leaves the leading multi-particle correlations unconstrained.
This confirms that QCD understanding of the collinear hierarchy identifies genuinely distinct angular correlations and should not be ignored when designing training sets.

The picture changes for logarithmic and mixed moments, though.
For mixed moments, both bases achieve comparable performance across observables (${\sim}\,75\times$ versus ${\sim}\,80\times$), with the worst set actually slightly outperforming the SO+composites basis.
Logarithmic moments alone reproduce this behavior: the worst set again matches SO+composites (${\sim}\,65\times$ each).
This collapse of the basis hierarchy can be understood from the fact that logarithmic moments are Sudakov-safe observables (as discussed in \Sec{Constraints}), and they probe the soft/collinear region in a fundamentally different way from polynomial moments.
Under polynomial moments, the collinear power-counting hierarchy distinguishes which graph topologies contribute independent angular information.
Under logarithmic moments, even graphs that are degenerate in the strongly-ordered limit probe distinct projections of the Sudakov logarithms, rendering the collinear hierarchy irrelevant for basis selection.

This observation motivates finding an alternative organizing principle for EFP bases when Sudakov-safe observables are involved. The strongly-ordered hierarchy studied in \Reference{Cal:2022fnm} provides clear guidance for IRC-safe (polynomial) moments but none for logarithmic or mixed measurement functions.
Finding a complete basis of observables that span the space of Sudakov-safe observables would enable more efficient capture of logarithmic-moment information and a more principled design of the training set. Identifying such a basis remains an open question.

As noted in \Sec{EFPtraining}, the constraints used throughout this section are based on marginal distribution moments of individual EFPs.
Through \Eq{product_features_choice}, our framework generalizes naturally to moments involving multi-dimensional distributions, including cross-moments between pairs of EFPs.
We leave a systematic exploration of such correlated constraints to future work, but in \App{2d_transfer} we investigate two-dimensional distributions and their correlated moments, confirming that joint distributions are also improved by marginal-only training and that cross-moment constraints provide diminishing returns as the marginal basis grows.

\subsection{Reweighting a Herwig prior to the Sherpa target}
\label{sec:herwig}

In the studies presented so far, both the degraded priors and the target are variants of the same \textsc{Sherpa} shower.
They share the dipole-shower framework, the transverse-momentum-like evolution variable, and the hadronization model, and they differ only through the controlled kernel modifications of \Eq{broken_kernels} and the $\alpha_s$ variations of \Eq{alphas_variations}.
A more demanding test is a prior produced by a different event generator, whose emissions populate phase space in a genuinely different way.
In this subsection, we therefore reweight a prior generated with the angular-ordered \textsc{Herwig} shower to the same \textsc{Sherpa} target used throughout this section.
Beyond differences in the treatment of the hard matrix element and the effective value of $\alpha_s$, \textsc{Herwig} orders its emissions in angle rather than in transverse momentum, uses different kinematic mappings, and employs its own cluster hadronization model and tune.
Unlike for the degraded \textsc{Sherpa} priors, these differences are not aligned with any simple parametric deformation of the target, so this test probes whether the EFP moment constraints carry enough information to correct a prior whose deficits were not designed into the study.

For the prior sample, we use \textsc{Herwig}~7.3~\cite{Bahr:2008pv,Bellm:2015jjp} to generate $e^+e^-\to\gamma^*/Z\to q\bar{q}$ events (five massless flavors) at $\sqrt{s}=91.2$~GeV, with the default angular-ordered $\tilde{q}$ shower, the default cluster hadronization model, and the default tune, leaving all shower and hadronization parameters unchanged.
We generate $N_{\rm ev}=10^6$ events, matching the size of the \textsc{Sherpa} samples, and analyze all final-state particles excluding neutrinos, using the same heavy-hemisphere definition as in \Sec{shower_setups}.
The reweighting setup is identical to the degraded-\textsc{Sherpa} study. We train on EFPs up to $d_{\rm max}=1,3,5$ for all three moment families of \Eq{constraint_families}, with the Lagrange multipliers optimized again for this prior using the L-BFGS procedure of \Sec{optimization}.
Since there is a single \textsc{Herwig} prior rather than a family of $\alpha_s$ variations, the results below show single curves rather than envelope bands.

\begin{figure}[t]
\centering
\subfloat[][]{
    \includegraphics[width=0.32\textwidth]{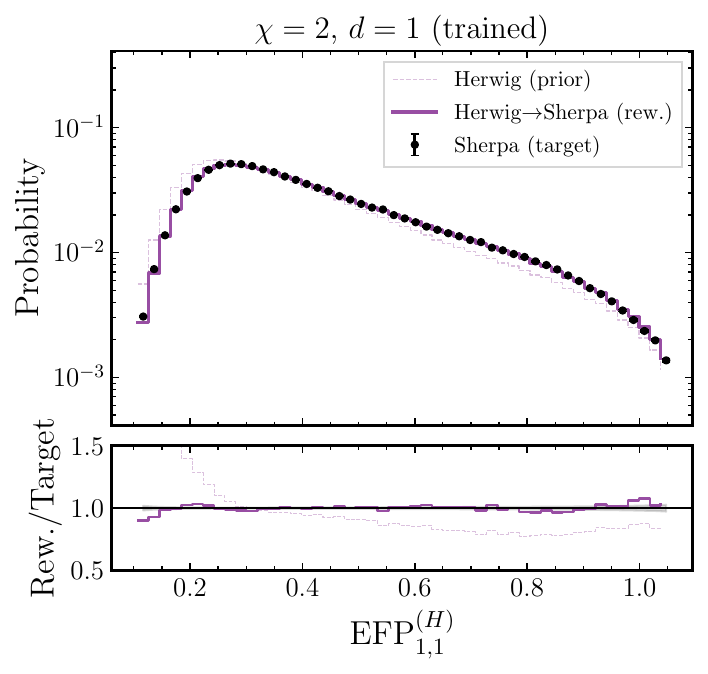}
    }
\subfloat[][]{
    \includegraphics[width=0.32\textwidth]{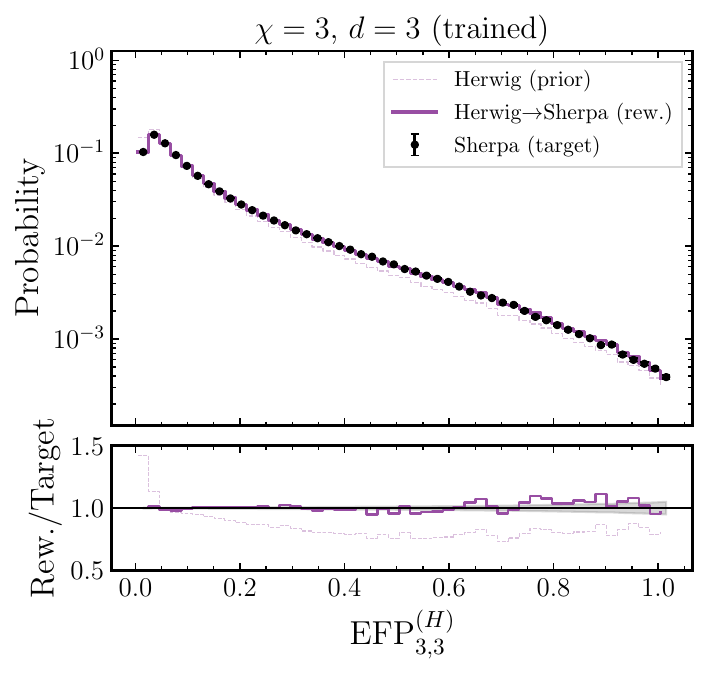}
    }
\subfloat[][]{
    \includegraphics[width=0.32\textwidth]{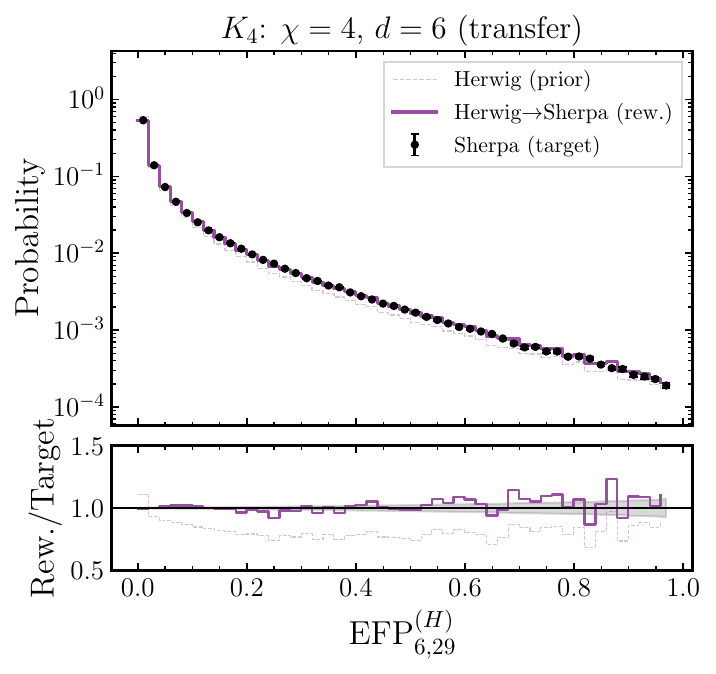}
    }
\caption{Cross-generator reweighting of an angular-ordered \textsc{Herwig} prior to the \textsc{Sherpa} target with $d_{\rm max}=5$ mixed moments, for the same three representative EFPs as in \Fig{dist_d5}: (a) $\mathrm{EFP}_{1,1}$ ($\chi=2$, trained), (b) $\mathrm{EFP}_{3,3}$ ($\chi=3$, trained), and (c) $\mathrm{EFP}_{6,29}$ (the complete graph $K_4$, $\chi=4$, transfer).
Black points show the \textsc{Sherpa} target with statistical uncertainties, dashed lines show the unweighted \textsc{Herwig} prior, and solid lines show the reweighted distribution.
Lower panels show the ratio to the target with gray bands indicating the target statistical uncertainty.
As in the degraded-\textsc{Sherpa} study, the untrained $K_4$ graph is substantially corrected through transfer.}
\label{fig:herwig_efp_dists}
\end{figure}

In \Fig{herwig_efp_dists}, we show the impact of the cross-generator reweighting on the marginal distributions of the same three representative EFPs studied in \Fig{dist_d5}, namely the trained $\mathrm{EFP}_{1,1}$ ($\chi=2$) and $\mathrm{EFP}_{3,3}$ ($\chi=3$), and the untrained complete graph $K_4$ ($\chi=4$).
The unweighted \textsc{Herwig} prior differs from the \textsc{Sherpa} target at the 10--30\% level through the bulk of these observables, undershooting the tails and overshooting the region of small EFP values, where the deviations grow larger still.
After reweighting with the $d_{\rm max}=5$ mixed moments, the trained distributions agree with the target at the few-percent level across the bulk of their range, with residual fluctuations confined to the sparsely populated tails.
The untrained $K_4$ distribution is corrected to a similar degree, demonstrating that transfer across chromatic complexity survives the change of generator.

\begin{figure}[p]
\centering
\subfloat[][]{
    \includegraphics[width=0.95\textwidth]{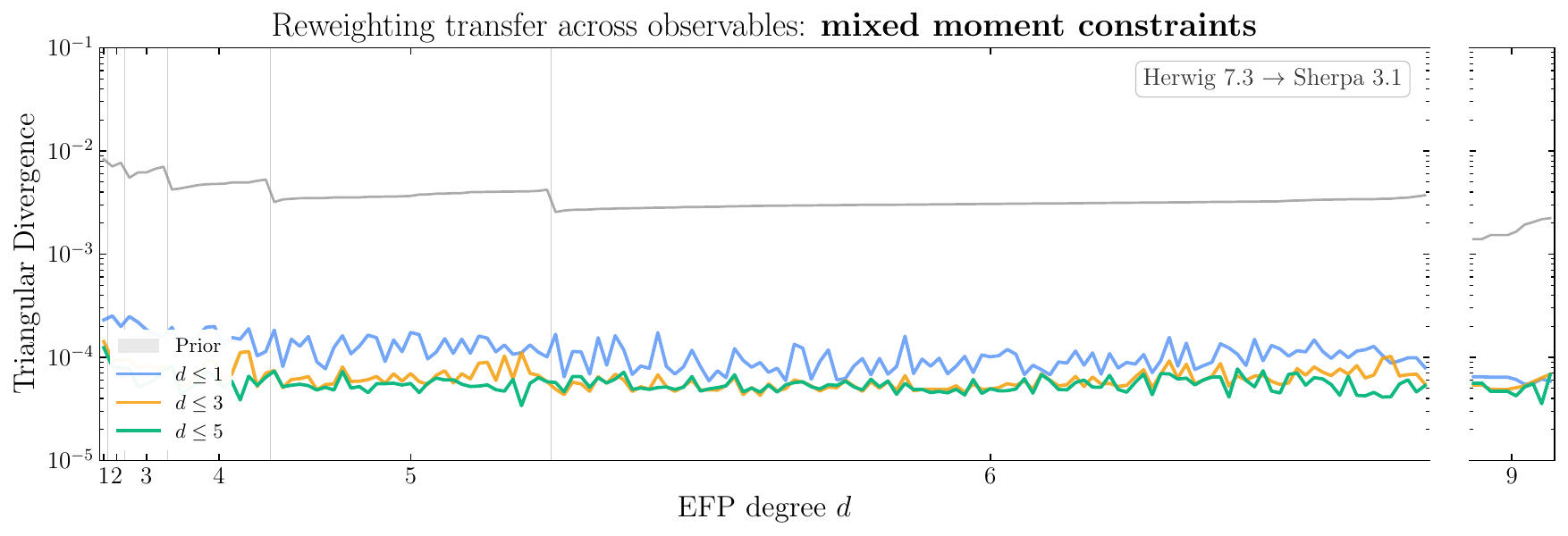}
    }\\
\subfloat[][]{
    \includegraphics[width=0.95\textwidth]{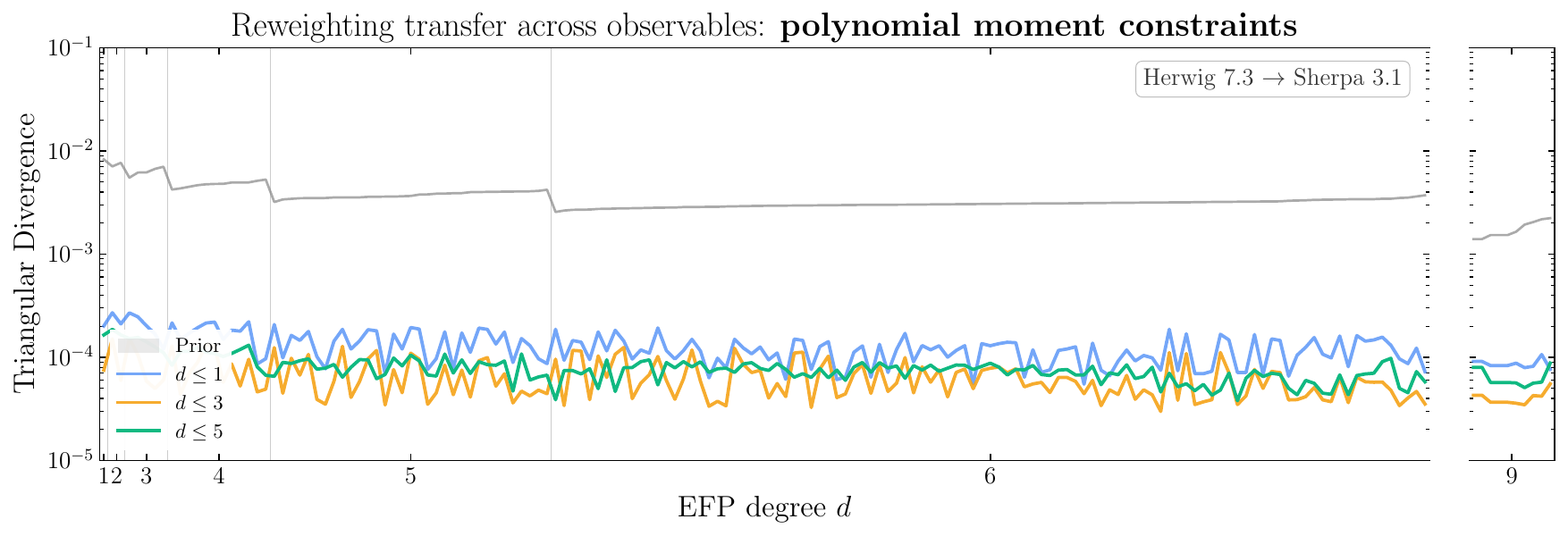}
    }\\
\subfloat[][]{
    \includegraphics[width=0.95\textwidth]{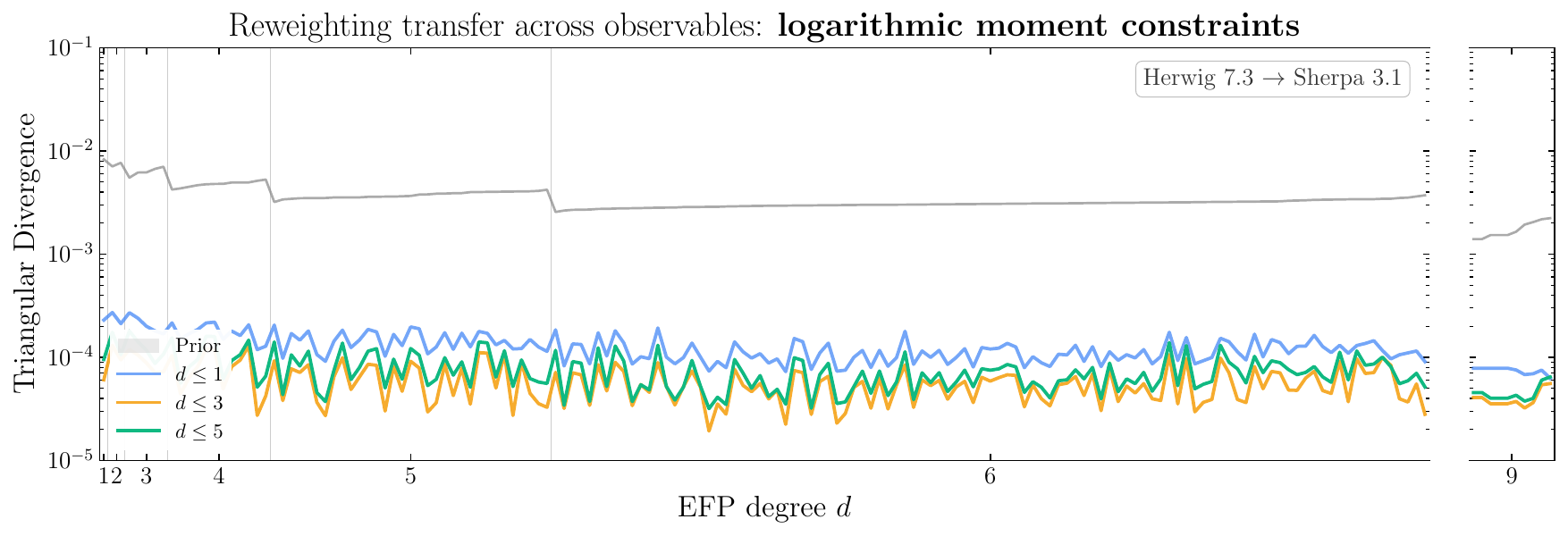}
    }
\caption{Information saturation for the cross-generator reweighting of the \textsc{Herwig} prior to the \textsc{Sherpa} target, organized by graph degree for (a) mixed, (b) polynomial, and (c) logarithmic moment features, in direct analogy to \Fig{saturation_prime_d5}.
In each panel, the triangular divergence between reweighted and target distributions is shown for the same 156 prime EFPs as in \Fig{saturation_prime_d5}, comparing the unweighted \textsc{Herwig} prior (gray) to training sets including EFPs up to $d \leq 1$ (blue), $d \leq 3$ (orange), and $d \leq 5$ (green).
The rightmost region shows the additional degree-9 EFPs from \Fig{deg9_subset}.
Since there is a single prior, single curves are shown rather than $\alpha_s$-variation bands.}
\label{fig:herwig_saturation}
\end{figure}

In \Fig{herwig_saturation}, we repeat the information-saturation scan of \Fig{saturation_prime_d5} for the \textsc{Herwig} prior, evaluating the triangular divergence to the target across the same 156 prime EFPs up to degree 6 plus the ten degree-9 trees.
The unweighted prior sits at divergences of a few~$\times\,10^{-3}$, several times closer to the target than the degraded priors of \Fig{saturation_prime_d5}.
The saturation pattern nevertheless mirrors the degraded-\textsc{Sherpa} study.
Training on the single $d\leq1$ EFP already reduces the divergence by more than an order of magnitude across the entire basis, the $d\leq3$ set brings a further improvement, and the gains saturate by $d_{\rm max}=3$--$5$, with the trained and transferred EFPs, including the untrained degree-6 and degree-9 graphs, improved by one to two orders of magnitude overall.
The residual divergences after reweighting reflect cross-generator differences that the four moments per EFP cannot fully remove.
We also observe that the polynomial-moment reweighting at $d\leq5$ is often slightly worse than at $d\leq3$, echoing the same behavior observed for the degraded \textsc{Sherpa} priors.

The reweighting itself is milder than for the degraded priors.
For the $d_{\rm max}=5$ mixed-moment training, the weights remain healthy, with an effective sample fraction of $86\%$, compared to $68\%$ for the worst-case $\alpha_s$ variation at $d_{\rm max}=5$ in \Fig{weight_diagnostics}.
We conclude that, for IRC-safe observables, the information transfer documented in this section does not rely on machinery shared between the prior and the target. A compact set of EFP moments suffices to move an angular-ordered \textsc{Herwig} prior onto the dipole-shower \textsc{Sherpa} target, for trained and untrained EFPs alike.
The boundary of this cross-generator transfer appears for observables sensitive to hadron-level dynamics outside the IRC-safe basis, which we examine in \App{substructure}.


\section{Conclusion and outlook}
\label{sec:conclusion}

In this paper, we studied how maximum-entropy reweighting can be used to systematically improve parton-shower predictions by incorporating precision QCD constraints organized through EFPs.
The framework of~\Reference{Assi:2025ibi} takes a parton-shower prior and produces strictly positive, normalized per-event weights that enforce a set of target moment constraints while minimizing the KL divergence to the prior.
The resulting posterior is fully exclusive: any observable can be computed on the reweighted sample without rebinning or regeneration, and the improvement propagates automatically to observables not included in the training set.

EFPs serve as a natural basis for this program.
They form an over-complete basis for IRC-safe observables, admit a well-defined complexity grading by graph degree and chromatic number, and are closed under multiplication, so that polynomial moments of arbitrary IRC-safe observables can be expressed in terms of EFP moments.
These properties make it possible to specify and systematically enlarge the constraint set without ad hoc choices, while the connections to energy correlators (through shared polynomial moment structures) and to collinear power counting (through the strongly-ordered basis reduction) provide concrete links to precision QCD calculations.

To test these ideas, we performed a proof-of-concept study in $e^+e^- \to \text{hadrons}$, deliberately degrading the parton shower by removing the non-singular parts of the splitting functions and disabling the $g\to q\bar{q}$ channel while preserving the leading soft singularities.
Despite these severe (from a QCD perspective) modifications, reweighting with EFP moments up to degree $d_{\rm max} \sim 3$--$5$ restores quantitative agreement with the target for the large majority of observables.
Several specific findings are worth highlighting, which we summarize below.

First, information saturates rapidly with the number of EFP constraints: training on EFPs up to degree 3 already produces substantial improvement on higher-degree EFPs and on standard hemisphere event shapes (thrust, sphericity, $C$-parameter, broadening) not included in training. This rapid saturation reflects the strong correlations among EFPs induced by the universal soft-collinear structure of QCD, and it means that a compact set of precision inputs can effectively constrain the full observable space. Among the three moment families we tested, mixed moments (combining polynomial and logarithmic terms) prove most effective, as they interpolate between Sudakov-peak sensitivity and bulk coverage. Weight-health diagnostics remain well-behaved throughout, with effective sample fractions above 68\% and moderate tail concentration even for the most extreme prior-target gaps.

Second, transfer degrades to $N$-jettiness with increasing $N$, and transfer to aplanarity and the $D$-parameter is relatively poor.
These provide instructive exceptions to the general success: although hemisphere aplanarity can be nonzero already for three particles ($\chi = 3$), its tails probe multi-particle correlations that are not directly constrained by EFPs up to degree 5.
Adding the single $\chi = 4$ graph $K_4$ does not fully resolve this, indicating that the bottleneck is degree coverage rather than chromatic number.
A complete correction would likely require higher-degree training or an alternative way to identify EFPs that efficiently probe multi-particle correlations.

Third, the basis-comparison tests reveal an interesting interplay between QCD structure and constraint design.
For polynomial moments, the physics-motivated strongly-ordered basis significantly outperforms a maximally redundant set of the same size, confirming that the collinear power counting argument from \Reference{Cal:2022fnm} identifies genuinely distinct angular correlations.
For logarithmic and mixed moments, however, this hierarchy collapses entirely.
This is because logarithmic moments are Sudakov-safe observables that probe the soft/collinear region differently from polynomial moments, rendering the collinear basis reduction ineffective.
Understanding how to organize constraints for Sudakov-safe observables remains an open and important question: a complete basis for Sudakov-safe observables, analogous to the EFP basis for IRC-safe observables, would enable more efficient capture of resummation information and could substantially improve the overall performance of the reweighting.

Fourth, the information transfer survives the change to a prior from a genuinely different event generator.
Reweighting an angular-ordered \textsc{Herwig} prior to the \textsc{Sherpa} target with the same compact constraint sets corrects trained and untrained EFPs alike by one to two orders of magnitude, with healthy weights, even though the differences between \textsc{Herwig} and \textsc{Sherpa} in the evolution variable, the kinematic mappings, and the hadronization model were not engineered for this test.
The same test also marks out the boundary of the method. IRC-safe subjet observables are corrected across generators, while charged multiplicity and the soft part of the fragmentation spectrum, whose cross-generator differences are dominated by the different hadronization models, are not (see \App{substructure}).
This is the expected behavior, since such hadron-level observables lie outside the span of the IRC-safe constraint set, and correcting them would require hadron-level inputs beyond the IRC-safe basis.

The numerical results presented here are encouraging, but several conceptual questions raised in \Sec{it_meets_qcd} remain to be fully addressed.
As discussed in \Sec{accuracy}, the formal accuracy of the posterior is guaranteed only for the specific moments imposed as constraints. Every other improvement documented in this paper is empirical information transfer, demonstrated numerically rather than derived. In the traditional parton-shower literature, accuracy is benchmarked by verifying that a shower reproduces the correct Sudakov exponent at a given logarithmic order (LL, NLL, etc.)\ for specific classes of observables~\cite{Banfi:2004yd}.
It would be valuable to establish an analogous criterion for the posterior: given precision constraints at a certain order, can one prove that the posterior distribution achieves the corresponding logarithmic accuracy for a well-defined class of observables, in the same sense that NLL accuracy is verified for parton showers?
Such a formal connection would place the reweighting on the same theoretical footing as direct shower improvements and clarify precisely what is gained by the precision inputs.

Related to this, the question of whether higher-order constraints can inadvertently degrade the prior's accuracy along unconstrained directions (\Sec{accuracy}) was addressed here only through practical diagnostics: monitoring transfer quality, weight health, and robustness across prior variations.
While we found no evidence of degradation in our proof-of-concept study, and the maximum-entropy principle mitigates this by design, a more rigorous understanding of the conditions under which degradation can or cannot occur would strengthen the theoretical foundations of the approach.
The geometric picture developed in \Sec{accuracy}, where degradation occurs only along projections partially correlated with the constraints, provides a starting point for such an analysis.

Several practical extensions are also natural.
The present study used a degraded shower as prior and the standard shower as target to isolate the information-transfer mechanism, supplemented by the cross-generator test with a \textsc{Herwig} prior in \Sec{herwig}.
A natural next step is to use higher-accuracy targets: NLL or NNLL resummed predictions, fixed-order calculations, or even experimental data.
This would allow one to characterize how formal accuracy propagates to unconstrained observables, and to connect to related frameworks such as the organization of perturbative corrections in SCET factorization and to systematic moment-matching protocols~\cite{Desai:2024yft}.
As discussed in \Sec{Constraints}, the design of optimal constraint functions that balance the resummation, fixed-order, and nonperturbative regions of phase space deserves further study, as does the incorporation of theoretical uncertainties on the precision inputs themselves (see \App{uncertainties}).
At the same time, the training sets used here are based on marginal moments of individual EFPs, and generalizing to multi-dimensional moments involving correlations between EFPs (as explored in \App{2d_transfer}) could provide further improvement, particularly for observables that depend on multi-particle correlations not well captured by marginals alone.

Extension to hadron colliders introduces additional challenges from initial-state radiation, event-by-event variation of the hard process, and color connections between initial and final states, though jet-based EFPs provide a natural generalization and the maximum-entropy framework applies without modification, and the needed precision fixed-order inputs are becoming available through NNLO subtraction methods~\cite{Czakon:2024tjr,Czakon:2021ohs,Czakon:2014oma,Czakon:2010td}.
As emphasized in \Sec{efficient_priors}, the framework works in tandem with efforts to improve parton-shower priors directly: a more accurate prior produces a more efficient reweighting, and both communities benefit from the synergy.
The interplay between perturbative reweighting (correcting parton-shower structure) and nonperturbative reweighting (correcting hadronization) also remains to be explored, as do applications to experimental fiducial corrections where event-level weights can be propagated through detector simulation.

Looking further ahead, we view this work as a step toward building a \emph{foundation model for collider theory}: a unified framework that absorbs all available QCD knowledge, from fixed-order calculations and analytic resummation to nonperturbative modeling and lattice inputs, into a single coherent, fully differential event-level prediction with quantified uncertainties.
The maximum-entropy construction provides the principled backbone for such a program, and the EFP-based constraint organization developed here offers a systematic way to specify and enlarge the space of precision inputs.
The present paper establishes that this approach works in a controlled $e^+e^-$ setting; the challenge now is to scale it to the multi-process precision campaigns demanded by the HL-LHC and future collider programs.

\acknowledgments
We are grateful to Stefan~H\"oche for collaboration in the early stages of this project and for his help in providing the event samples used in this study.
We thank Pedro Cal, Jack Holguin, Bernhard Mistlberger, Rene Poncelet, Gregory Soyez, Frank Tackmann, Melissa van Beekveld, Wouter Waalewijn, and James Whitehead for useful discussions.
B.A. acknowledges support in part by the U.S.\ Department of Energy grants
DE-SC1019775 and DE-SC0026301, and the National Science Foundation grants OAC-2103889, OAC-2411215, and OAC-2417682.
K.L.\ is supported in part by the U.S. Department of Energy under contracts DE-AC02-06CH11357.
J.T.\ was supported by the National Science Foundation under Cooperative Agreement
PHY-2019786 (The NSF AI Institute for Artificial Intelligence and Fundamental Interactions, 
\url{http://iaifi.org/}), by the U.S.\ Department of Energy Office of High Energy Physics under grant
number DE-SC0012567, and by the Simons Foundation through Investigator grant 929241.
J.T.\ also thanks the Institut des Hautes \'Etudes Scientifiques (IHES) and the Institut de Physique Th\'eorique (IPhT) for providing an inspiring sabbatical environment to carry out this research.
This research used resources of the National Energy Research Scientific Computing Center (NERSC), 
a Department of Energy Office of Science User Facility using NERSC award ERCAP0028985.
This work was performed in part at the Aspen Center for Physics, with support for B.A.\ by a grant from the Simons Foundation (1161654, Troyer).
This research was supported in part by grant NSF PHY-2309135 to the Kavli Institute for Theoretical Physics (KITP).

\appendix

\section{Alternative measures of statistical similarity}
\label{app:alternative_measures}

In \Sec{maximum_entropy}, we used the KL divergence as our measure of similarity between the prior $q(\Phi)$ and posterior $p^\star(\Phi)$.
In this appendix, we derive the optimal weight factor $w^\star(\Phi)$ for generic $f$-divergences.
This calculation will explain why the KL divergence is particularly well-suited for our problem setting, since it is guaranteed to yield positive weights.

To generalize the loss function in \eq{totalLoss}, it is convenient to express it in terms of the weight function $w(\Phi) = p(\Phi)/q(\Phi)$ from \Eq{weight_definition}.
Replacing the KL divergence with a generic $f$-divergence, the loss function takes the form:
\begin{align}
L[w] &= \int d\Phi \, q(\Phi) \, f\big(w(\Phi)\big) + (\lambda_0-1) \left(\int d\Phi \, q(\Phi) \, w(\Phi) -1 \right) \nonumber\\
&
~ \qquad +\sum_j \lambda_j \left(\int d\Phi \, q(\Phi) \, w(\Phi) \, m_j(\Phi) - c_j\right),
\end{align}
where $f(w)$ is a convex function with $f(1) = 0$.
For example, $f_{\rm KL}(w) = w \ln w$ corresponds to the KL divergence, which recovers \eq{totalLoss}.
Performing a functional variation of the loss with respect to $w$, we obtain the constraint:
\begin{equation}
    \frac{\delta L}{\delta w} = 0 \quad \Rightarrow \quad  1-f'\big(w(\Phi)\big) = \lambda_0 + \sum_j \lambda_j \, m_j(\Phi) \equiv \Lambda(\Phi),
\end{equation}
such that the optimal weight can be written formally as:
\begin{equation}
    w^\star(\Phi) = (f')^{-1}\big(1-\Lambda^\star(\Phi)\big).
\end{equation}
Using $f_{\rm KL}$, the solution is $w^\star(\Phi) = e^{-\Lambda^\star(\Phi)}$, in agreement with \eq{wweight}.

\setlength{\tabcolsep}{12pt}
\renewcommand{\arraystretch}{1.5}
\begin{table}[t]
\centering
\begin{tabular}{c | c  c  c}
    Name & $f(w)$ & $w(\Lambda)$ & Valid $\Lambda$? \\
    \hline
    KL divergence & $w \ln w$ & $e^{-\Lambda}$ & All\\
    Reverse KL divergence & $-\ln w$ & $\frac{1}{\Lambda-1}$ & $\Lambda > 1$\\
    Neyman $\chi^2$-divergence & $(w-1)^2$ & $\frac{3 - \Lambda}{2}$ & $\Lambda < 3$\\
        Pearson $\chi^2$-divergence & $\frac{(w-1)^2}{w}$ & $\frac{1}{\sqrt{\Lambda}}$ & $\Lambda > 0$\\
        Triangular divergence & $\frac{1}{2} \frac{(w-1)^2}{w+1}$ & $-1 + \frac{2}{\sqrt{2 \Lambda - 1}}$ & $\Lambda \in \left(\frac{1}{2},\frac{5}{2}\right)$
\end{tabular}
\caption{Alternative $f$-divergences to quantify the difference between $q(\Phi)$ and $p(\Phi)$.  For any convex function $f$ with $f(1) = 0$, we can in principle derive weight factors $w(\Lambda)$.
Among these examples, the KL divergence is special since it yields positive weights for any $\Lambda$.}
\label{tab:alternative_f}
\end{table}

While it is possible to use alternative functions for $f$, it is highly inconvenient.
In Tab.~\ref{tab:alternative_f}, we compare the KL divergence to other popular $f$-divergences and derive the corresponding weight functions $w(\Lambda)$.
Because the sign and range of $\Lambda(\Phi)$ is not known a priori, we would prefer to work with weight functions that are manifestly positive.
Of the examples shown in Tab.~\ref{tab:alternative_f}, only the KL divergence satisfies this property.
In other cases, the weight factor can be negative, or even ill-defined depending on the value of the Lagrange multipliers.
It is possible to find $f$-divergences that yield positive weights and can be solved in closed form, e.g.~$f(w) = \frac{w^2 - 1}{2} - \log w$, but the functional forms for the resulting weights are much messier and more difficult to optimize.
This is the reason we focus on the KL divergence in our framework.

\section{Aspects of incorporating uncertainties}
\label{app:uncertainties}

In the main text, we assumed that the moments were perfectly known.
In reality, moment constraints will come with (correlated) uncertainties, which have to be transferred to the Lagrange multipliers.
In this appendix, we discuss how to incorporate such uncertainties in the Gaussian approximation, leaving a full treatment of uncertainties to future work.

To start, recall that the moment constraint in \Eq{constraints} implicitly determines the optimal Lagrange multipliers $\boldsymbol{\lambda^\star}$ as a function of the constraint values $c_i$:
\begin{equation}
    \label{eq:alt_constraint}
    \langle m_i \rangle_{p^\star} = \int d\Phi \, q(\Phi) \, w(\Phi; \boldsymbol{\lambda^\star}) \, m_i(\Phi) = c_i\,,
\end{equation}
where $p^\star(\Phi) = q(\Phi) \, w(\Phi; \boldsymbol{\lambda^\star})$.
We want to understand how $\boldsymbol{\lambda}^\star$ responds when the constraints have uncertainties:
\begin{equation}
    c_i \to c_i + \delta c_i\,, \qquad \boldsymbol{\lambda^\star} \to \boldsymbol{\lambda^\star} + \boldsymbol{\delta \lambda^\star}\,.
\end{equation}
Expanding the right-hand side of \Eq{alt_constraint} to linear order in $\delta \lambda^\star$ is straightforward, since the moment constraint is related to the gradient of the dual objective from \Eq{gradient_of_dual}, whose derivative we already computed in \Eq{cov_hessian}.
This leads to:
\begin{equation}
    \label{eq:constraint_expanded}
    c_i + \delta c_i = \langle m_i\rangle_{p^\star} + \sum_j {\rm Cov}_{p^\star}(m_i,m_j) \, \delta \lambda^\star_j + \cdots\,.
\end{equation}
For notational convenience, we define:
\begin{equation}
    C_{ij} = {\rm Cov}_{p^\star}(m_i,m_j)\,.
\end{equation}

Since $\langle m_i\rangle_{p^\star} = c_i$ by construction, \Eq{constraint_expanded} reduces to the relation $\delta c_i = C_{ij}\, \delta \lambda^\star_j$ at linear order.
In the case that $c_i$ is perfectly known, $\delta c_i = 0$, and the equation is just the statement that $\delta \lambda^\star_j$ vanishes.
If $c_i$ has uncertainties, though, this relation implies that the covariance matrix for $c_i$ should match (to linear order) with the covariance matrix for $\lambda^\star_j$:

\begin{equation}
    {\rm Cov}(c_i, c_j) = C_{ik} \, {\rm Cov}(\lambda_k, \lambda_\ell) \, C^T_{\ell j}.
\end{equation}
As long as the moment constraints are not degenerate, then $C_{ij}$ is invertible, leading to
\begin{equation}
    \label{eq:sol_cov_lambda}
    {\rm Cov}(\lambda_i, \lambda_j) = (C^{-1})_{ik} \, {\rm Cov}(c_k, c_\ell) \, (C^{-1})^T_{\ell j}.
\end{equation}
Therefore, transferring the uncertainties on the moments to uncertainties on the Lagrange multipliers only involves computing and inverting the $C_{ij}$ matrix.

This analysis assumed that there was just one prior distribution $q(\Phi)$.
If one has a distribution of priors $q(\Phi;\boldsymbol{\nu})$ characterized by nuisance parameters $\boldsymbol{\nu}$, then we can extend this Gaussian analysis to include uncertainties on the prior.
Expanding around some baseline value $\boldsymbol{\nu}_0$:
\begin{equation}
    \boldsymbol{\nu} = \boldsymbol{\nu}_0 + \boldsymbol{\delta\nu},
\end{equation}
it is straightforward to compute how the moments shift with $\boldsymbol{\delta\nu}$.
To counteract this shift, the Lagrange multipliers $\boldsymbol{\lambda}$ have to shift in a correlated way to keep the moment constraints satisfied.
This leads to a joint covariance matrix over $\boldsymbol{\nu}$ and $\boldsymbol{\lambda}$, where the $\text{Cov}[\nu_i,\lambda_j]$ entries enforce this correlated moment constraint, the $\text{Cov}[\nu_i,\nu_j]$ entries correspond to the ``prior on the prior'', and the $\text{Cov}[\lambda_i,\lambda_j]$ entries correspond to the moment uncertainties from \Eq{sol_cov_lambda}.
The specific form of this matrix is not particularly enlightening, so we leave a detailed study of it to future work, but it is satisfying to know that one can derive analytic expressions for the uncertainties in the Gaussian limit and avoid costly retrainings to handle uncertainties.

\section{Two-dimensional transfer diagnostics}
\label{app:2d_transfer}
 
The most direct way to assess reweighting quality is the distance between posterior and target at the full phase-space level.
The one-dimensional marginal distributions studied in \Sec{results} provide strong indirect evidence that this distance is small, since close agreement across many projections simultaneously constrains the underlying phase-space distribution.
In this appendix, we probe this more directly by examining two-dimensional joint distributions, which test correlations between pairs of observables that marginal projections alone cannot access.
 
Before presenting numerical results, it is useful to understand why we expect marginal constraints to capture much of the joint distribution structure.
In the strongly-ordered limit of \Sec{EFPbasis}, many prime EFPs at a given degree become proportional to one another, so their joint distribution collapses to a lower-dimensional subspace determined by the marginals of the independent directions.
More concretely, composite EFPs factorize as $\mathrm{EFP}_G = \mathrm{EFP}_{G_a} \mathrm{EFP}_{G_b}$, so that constraining the polynomial moments $\langle \mathrm{EFP}_G^m \rangle$ of the composite automatically imposes correlated constraints of the form $\langle \mathrm{EFP}_{G_a}^m \mathrm{EFP}_{G_b}^m \rangle$.
Similarly, logarithmic moments of composites decompose into sums of cross-moments via $\ln(\mathrm{EFP}_{G_a} \mathrm{EFP}_{G_b}) = \ln \mathrm{EFP}_{G_a} + \ln \mathrm{EFP}_{G_b}$.
As the marginal basis grows and more composite EFPs enter the training set, an increasing number of cross-correlations are constrained implicitly, and the residual joint-distribution information not captured by marginals shrinks.
In the kinematic regions where these power-counting approximations apply, marginal constraints along a sufficiently rich set of EFPs should therefore determine the joint distributions to good approximation.
Away from these limits, however, EFPs that are degenerate under strong ordering become genuinely independent, and correlated moments may carry new information that marginals alone cannot capture.

We probe two-dimensional distributions using bin-by-bin ratio plots.
Each panel shows the ratio of either the prior or the reweighted distribution to the target, where a ratio of unity (shown in white) indicates perfect agreement.
We use the $+20\%$ $\alpha_s$ variation as the prior throughout, since it has the largest prior-to-target gap, and we outline bins with fewer than 10 target counts in black.
In all cases, the prior ratio plot reflects the global $\alpha_s$-driven shift in the radiation pattern: overproduction at large observable values (red, ratios $\gtrsim 1.5$) and underproduction at small values (blue, ratios $\lesssim 0.7$).

\begin{figure}[t]
\centering
\includegraphics[width=0.95\textwidth]{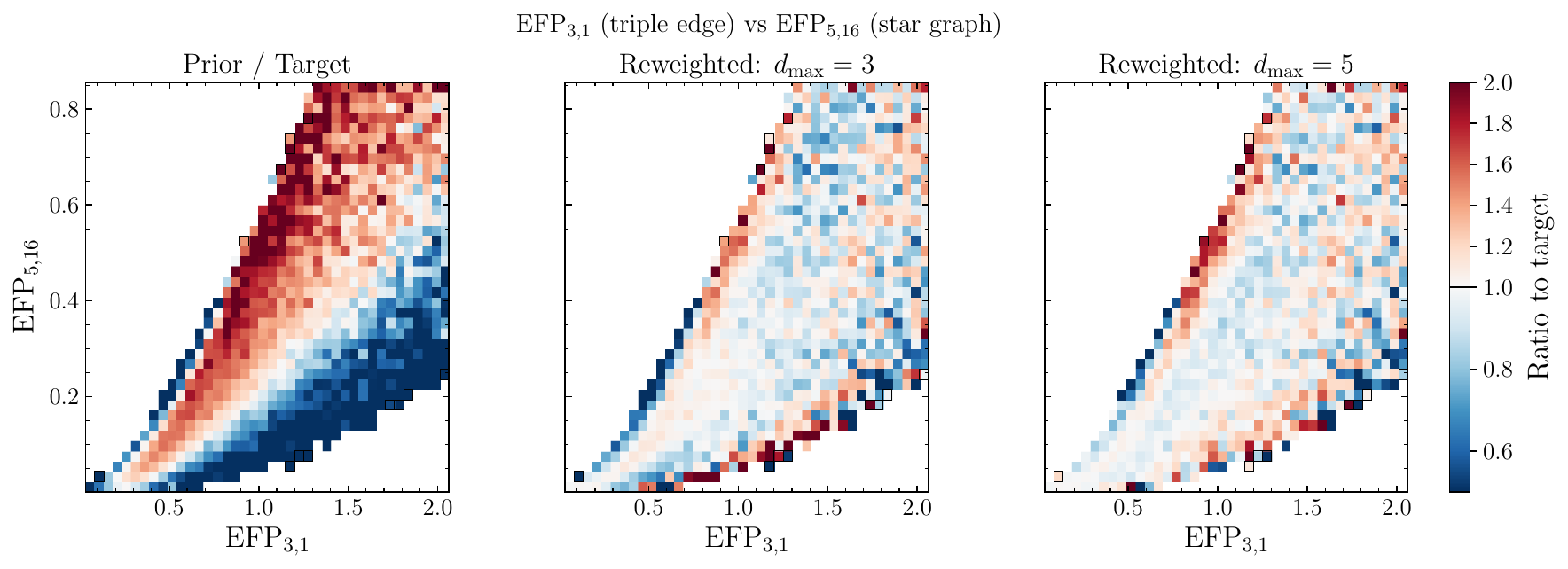}
\caption{Two-dimensional ratio plot for the triple edge $\mathrm{EFP}_{3,1}$ versus the degree-5 star graph $\mathrm{EFP}_{5,16}$.
The left panel shows the prior, the center panel shows the reweighted result at $d_{\rm max} = 3$ where only $\mathrm{EFP}_{3,1}$ is in the basis, and the right panel shows $d_{\rm max} = 5$ where both are in the basis.
The two reweighted panels are nearly indistinguishable, confirming that tree-like EFPs transfer well from lower-degree training.}
\label{fig:2d_efp_pair}
\end{figure}
 
We begin by selecting two EFPs that probe complementary aspects of the radiation pattern: $\mathrm{EFP}_{3,1}$, the triple multi-edge $(\theta_{12})^3$, a degree-3 observable that is purely sensitive to collinear splittings, and $\mathrm{EFP}_{5,16}$, a degree-5 star graph with chromatic number $\chi=2$ that probes multi-prong radiation patterns.
These two have a relatively low Pearson correlation of $|r|=0.90$, making them a useful choice for probing complementary directions in the EFP space.

The joint plane of $\mathrm{EFP}_{3,1}$ and $\mathrm{EFP}_{5,16}$ is shown in \Fig{2d_efp_pair}. The prior (left panel) displays the expected $\alpha_s$-driven mismatch, with overproduction at large values and underproduction at small values.
The center panel shows the reweighted result at $d_{\rm max}=3$, where only the triple edge is directly constrained: the mismatch is largely eliminated, and the ratio is close to unity throughout the populated region.
The right panel uses $d_{\rm max}=5$ weights, where both observables are now in the basis.
The two reweighted panels are nearly indistinguishable, confirming that the joint distribution is already well-corrected by marginals alone at $d_{\rm max}=3$.
This is consistent with the strong-ordering expectation: both EFPs are tree graphs ($\chi=2$) whose strongly-ordered scalings are related, so constraining a few low-degree marginals effectively determines their correlations.
 
\label{app:cross_moments}
 
\begin{figure}[t]
\centering
\includegraphics[width=0.95\textwidth]{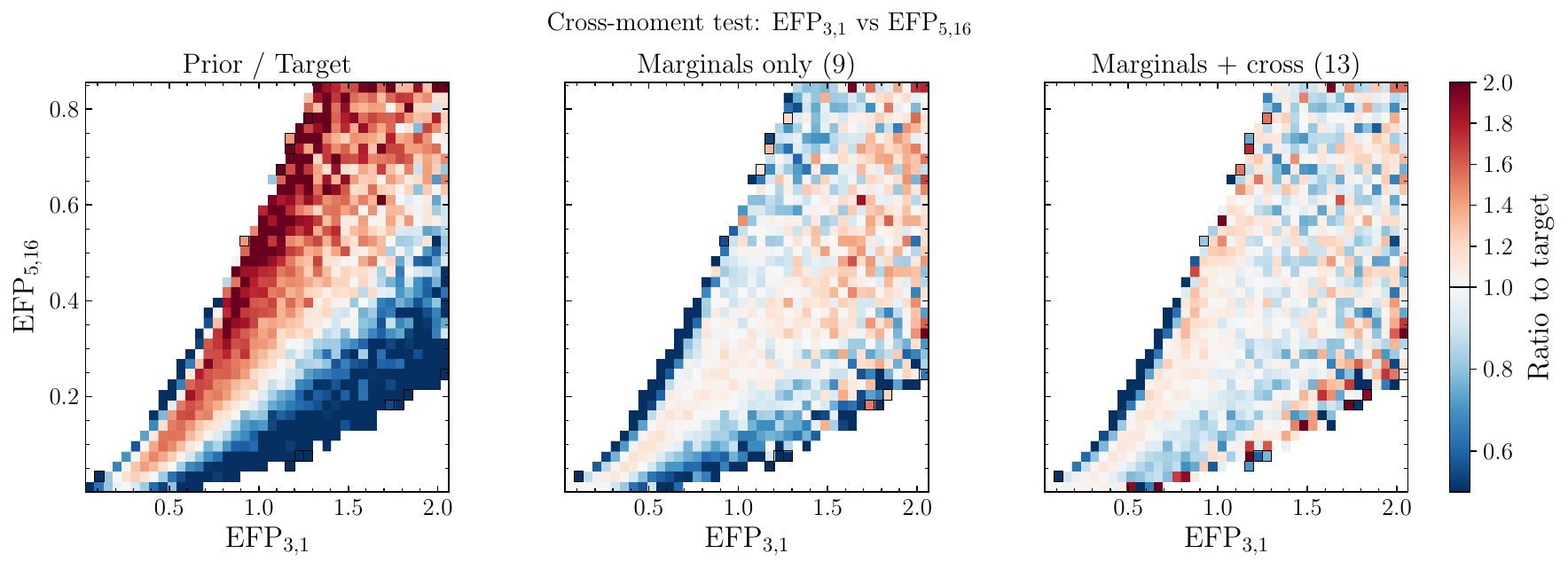}
\caption{Cross-moment test for a two-EFP basis containing $\mathrm{EFP}_{3,1}$ and $\mathrm{EFP}_{5,16}$. The left panel shows the prior, the center panel shows marginals-only reweighting with 9 constraints, and the right panel adds four bilinear cross-moments for a total of 13 constraints. Adding cross-moments produces a modest improvement in this sparse basis.}
\label{fig:2d_2efp_cross_test}
\end{figure}
 
The $d_{\rm max}=3$ result above used 12 EFPs, so the good joint-distribution agreement could simply reflect the large number of marginal constraints (49 total).
To test this, we reduce the basis to just $\mathrm{EFP}_{3,1}$ and $\mathrm{EFP}_{5,16}$ themselves, with only 9 marginal constraints (one normalization constraint plus four mixed moments per EFP).
We can also ask whether cross-moment constraints of the form in \Eq{cross_constraints} can be used to directly capture the joint distribution.
Adding four bilinear cross-moments brings the total to 13 constraints, a $44\%$ increase, making this a sensitive test.
All three panels are shown side by side in \Fig{2d_2efp_cross_test}.
The marginals-only reweighted panel corrects most of the prior mismatch, but a band of blue overcorrection remains at low $\mathrm{EFP}_{5,16}$ values.
Adding cross-moments reduces this overcorrection modestly, demonstrating that with only two EFPs the marginal constraints do not fully capture the correlated structure.

\begin{figure}[t]
\centering
\includegraphics[width=0.95\textwidth]{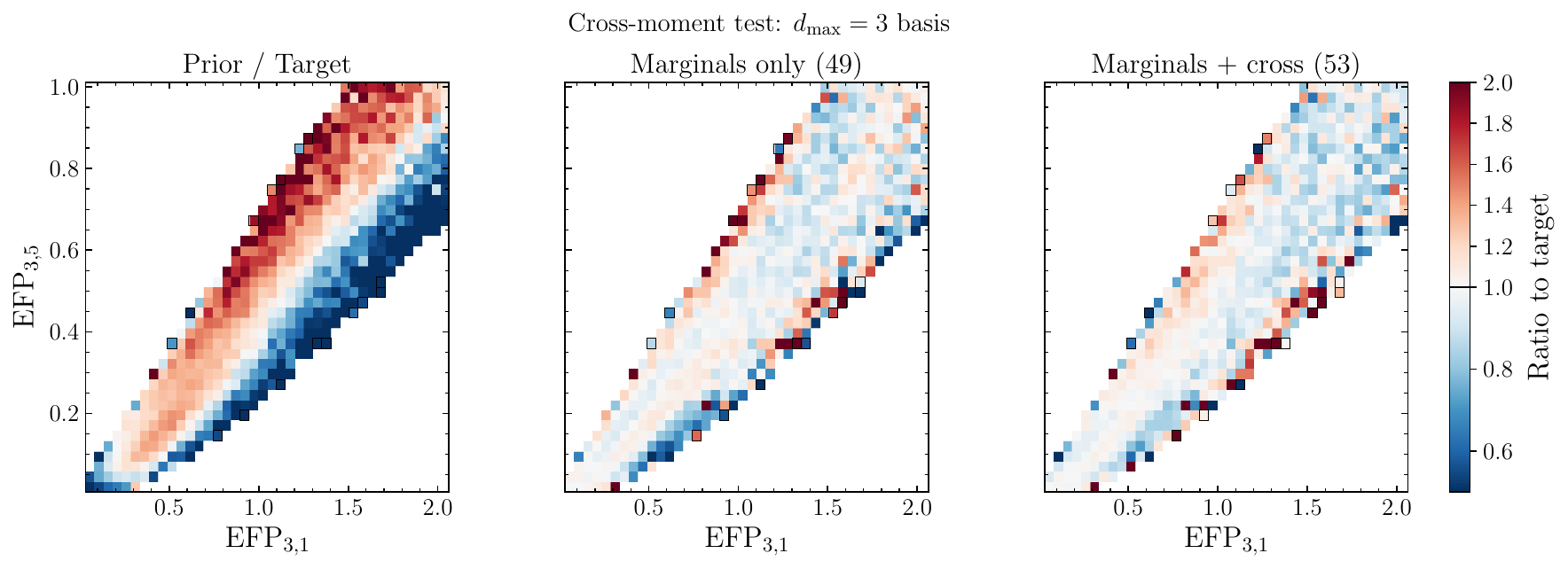}
\caption{Cross-moment test for the full $d_{\rm max}=3$ basis of 12 EFPs, projected onto $\mathrm{EFP}_{3,1}$ versus $\mathrm{EFP}_{3,5}$, a path graph on four vertices. The left panel shows the prior, the center panel shows marginals-only reweighting with 49 constraints, and the right panel adds four cross-moments for a total of 53 constraints. The two reweighted panels are nearly identical, confirming that cross-moments become redundant as the marginal basis grows.}
\label{fig:2d_deg3_cross_test}
\end{figure}
 
We then repeat the test with the full $d_{\rm max}=3$ training set of 12 EFPs (49 marginal constraints), projecting onto the plane of $\mathrm{EFP}_{3,1}$ versus $\mathrm{EFP}_{3,5}$, a degree-3 path graph with Pearson correlation $|r|=0.97$ to the triple edge.
Adding four cross-moments between these two observables brings the total to 53 constraints, only an $8\%$ increase.
The comparison is shown in \Fig{2d_deg3_cross_test}.
With 12 EFPs already constraining the marginals, the cross-moments produce only a slight further improvement at small observable values.
This confirms the expectation from the strong-ordering argument: as the marginal basis grows, the composite EFPs in the training set implicitly constrain more and more cross-correlations, and explicit cross-moment constraints become redundant.

\begin{figure}[t]
\centering
\includegraphics[width=0.95\textwidth]{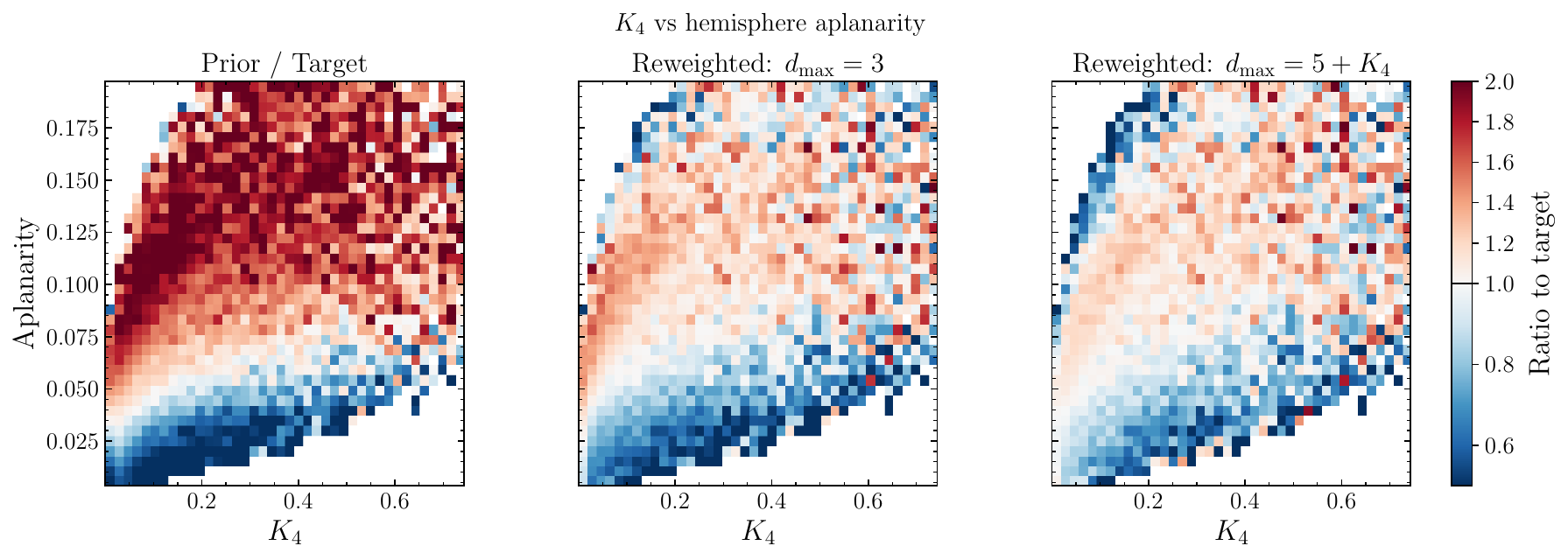}
\caption{Two-dimensional ratio plot for $K_4$ versus hemisphere aplanarity. The left panel shows the prior, the center panel shows the reweighted result at $d_{\rm max} = 3$ where neither observable is in the basis, and the right panel shows $d_{\rm max} = 5 + K_4$ where $K_4$ is in the basis.
Including $K_4$ explicitly corrects the bulk of the joint plane, but residual disagreement persists in the anti-correlated region where one observable is large while the other is small.}
\label{fig:2d_K4_aplanarity}
\end{figure}
 
Finally, we use the two-dimensional plane to examine the multi-particle region where the marginal EFP reweighting of \Sec{aplanarity} showed incomplete transfer.
Since $K_4$ (chromatic number $\chi=4$) requires at least four particles to be nonzero, being differential in $K_4$ effectively selects the $\geq 4$-particle multiplicity region and provides a more targeted diagnostic than the one-dimensional aplanarity distribution, which mixes contributions from three-particle and higher-multiplicity events. 
Their joint plane is shown in 
\Fig{2d_K4_aplanarity}.
The right panel uses $d_{\rm max}=5+K_4$ weights, so that $K_4$ is in the basis. Including $K_4$ corrects the bulk of the plane, but residual disagreement concentrates in the anti-correlated region: overcorrection (blue) at large $K_4$ with small aplanarity, and residual underproduction (red) at large aplanarity with small $K_4$.
The middle panel uses $d_{\rm max}=3$ weights, where neither observable is in the basis, and retains substantially more mismatch throughout, confirming that $K_4$ benefits from explicit inclusion in a way that the tree-like star graph does not.
This is consistent with the one-dimensional results of \Sec{aplanarity}.
 
The residual mismatch pattern has an instructive physical interpretation.
Events with large $K_4$ but small aplanarity correspond to configurations where four particles are well-separated in angle yet remain roughly coplanar.
Conversely, events with large aplanarity but small $K_4$ correspond to configurations with substantial out-of-plane momentum spread but smaller mutual pairwise angles, which can arise already from three non-coplanar particles without requiring the four well-separated particles that $K_4$ probes.
These two sectors of multi-particle phase space are controlled by genuinely different angular correlations that a single graph like $K_4$ cannot span.
Correcting the remaining mismatch would require EFPs (or combinations thereof) that are specifically sensitive to the planarity of multi-particle configurations.
Finding an explicit decomposition of aplanarity in terms of EFPs would clarify quantitatively which graphs and at what degree the information saturates, but this is beyond the scope of the present work.

\section{\texorpdfstring{Transfer to jet substructure and fragmentation observables}{Transfer to jet substructure and fragmentation observables}}
\label{app:substructure}

The event shapes and correlators studied in \Sec{transfer_eventshapes} are built from global momentum sums over the hemisphere.
In this appendix, we consider a complementary class of observables that probe the shower evolution and the splitting kernels more directly, through the clustering history of the event and through the individual hadrons.
Specifically, we study Durham subjet rates, the primary Lund plane, the charged-particle multiplicity, and the charged fragmentation spectrum.
Two of these observables, the charged multiplicity and the fragmentation spectrum, are not even IRC safe and therefore lie outside the EFP basis of \Sec{efps} entirely, directly probing the limits of information transfer.
We study them in two settings, first reweighting the degraded \textsc{Sherpa} priors to the \textsc{Sherpa} target as in the main text, and then reweighting the \textsc{Herwig} prior of \Sec{herwig} to the same \textsc{Sherpa} target.
All observables are computed on the heavy hemisphere, consistent with the hemisphere EFP moments used as constraints, and none of them enter the training in any form.

The observables are defined as follows.
We cluster the heavy-hemisphere particles with the Durham ($e^+e^-$ $k_T$) algorithm~\cite{Catani:1991hj}, with the resolution variable defined relative to the hemisphere energy, and record the scale $-\ln y_{23}^{(H)}$ at which the hemisphere changes from two to three subjets, along with the exclusive subjet multiplicity $N_{\rm subjet}^{(H)}$ at $y_{\rm cut}=0.01$.
The primary Lund plane~\cite{Dreyer:2018nbf} is obtained by reclustering the hemisphere with the Cambridge--Aachen algorithm~\cite{Dokshitzer:1997in} and recording the coordinates $(\ln 1/\theta, \ln k_T)$ of each emission along the hard branch of the declustering sequence.
For hadron-level dynamics, we use the charged-particle multiplicity $n_{\rm ch}^{(H)}$ and the charged fragmentation spectrum $\xi^{(H)}=\ln(1/x_p)$ with $x_p = 2|\vec{p}\,|/Q$.
The subjet resolution, subjet multiplicity, and Lund-plane density are IRC safe, but unlike the event shapes of \Sec{transfer_eventshapes}, which sum over all particles at once, they resolve the emission pattern through the sequential clustering structure of the radiation.
The multiplicity and fragmentation spectrum are not IRC safe and cannot be expanded in the EFP basis via \Eq{O_expand_EFP}.

\subsection{\texorpdfstring{Transfer from the degraded \textsc{Sherpa} priors}{Transfer from the degraded Sherpa priors}}

\begin{figure}[t]
\centering
\subfloat[][]{
    \includegraphics[width=0.32\textwidth]{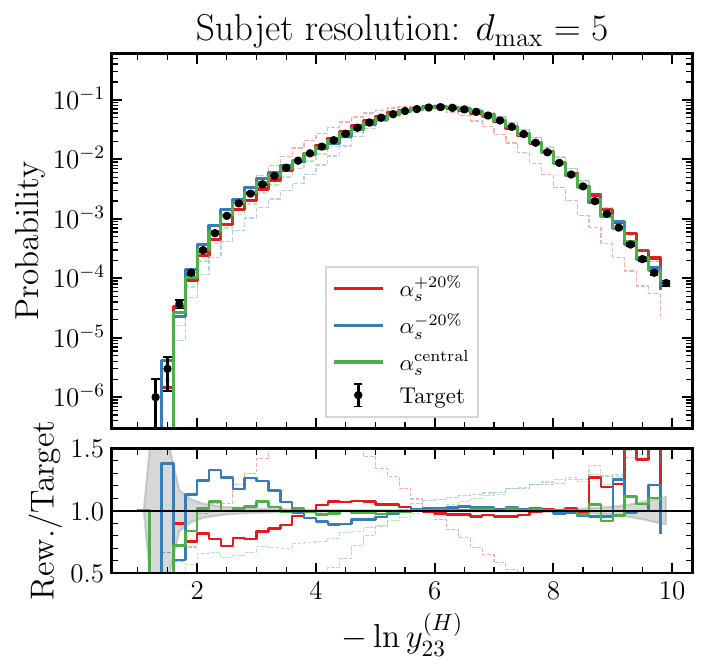}
    }
\subfloat[][]{
    \includegraphics[width=0.32\textwidth]{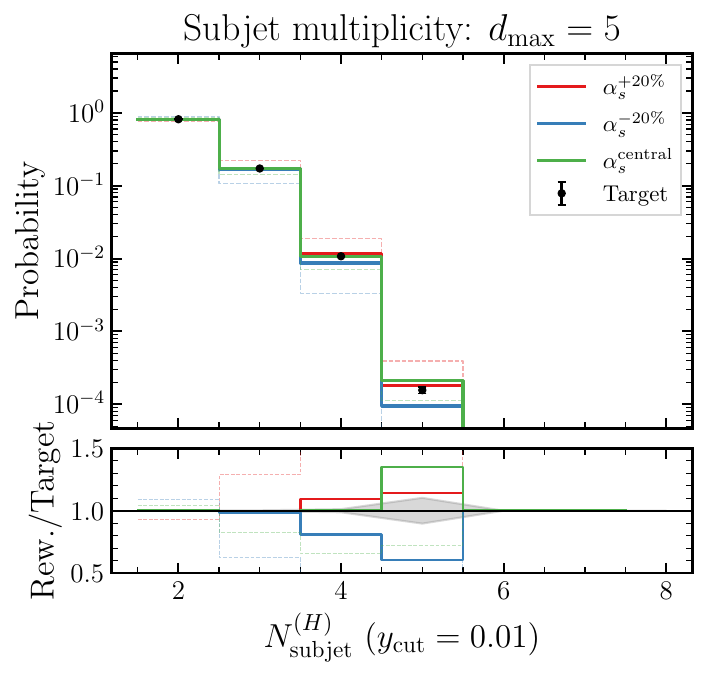}
    }
\subfloat[][]{
    \includegraphics[width=0.32\textwidth]{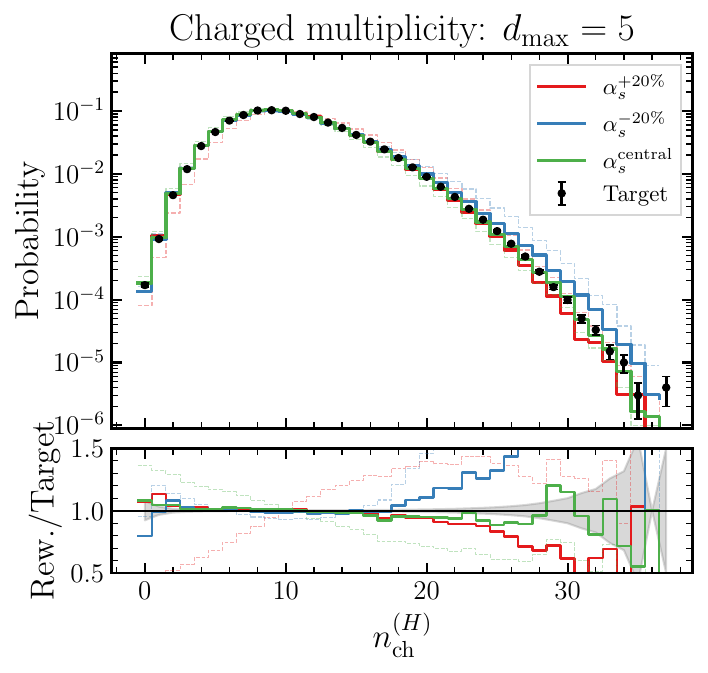}
    }
\caption{Transfer to heavy-hemisphere jet observables for $d_{\rm max}=5$ mixed-moment training with the degraded \textsc{Sherpa} priors: (a) Durham $2\to3$ subjet resolution $-\ln y_{23}^{(H)}$, (b) exclusive subjet multiplicity at $y_{\rm cut}=0.01$, and (c) charged-particle multiplicity $n_{\rm ch}^{(H)}$.
The plot formatting follows \Fig{thrust_saturation}.}
\label{fig:substructure_transfer}
\end{figure}

\begin{figure}[t]
\centering
\subfloat[][]{
    \includegraphics[width=0.4\textwidth]{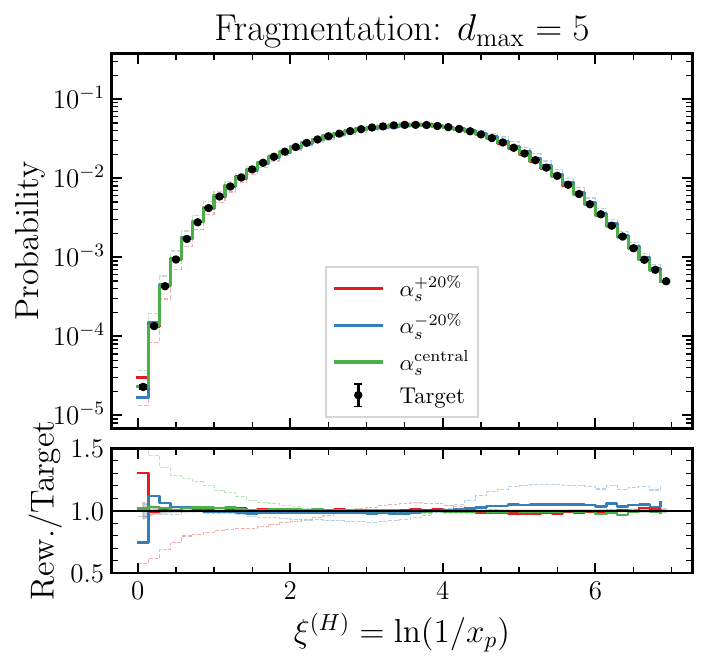}
    }
\subfloat[][]{
    \includegraphics[width=0.4\textwidth]{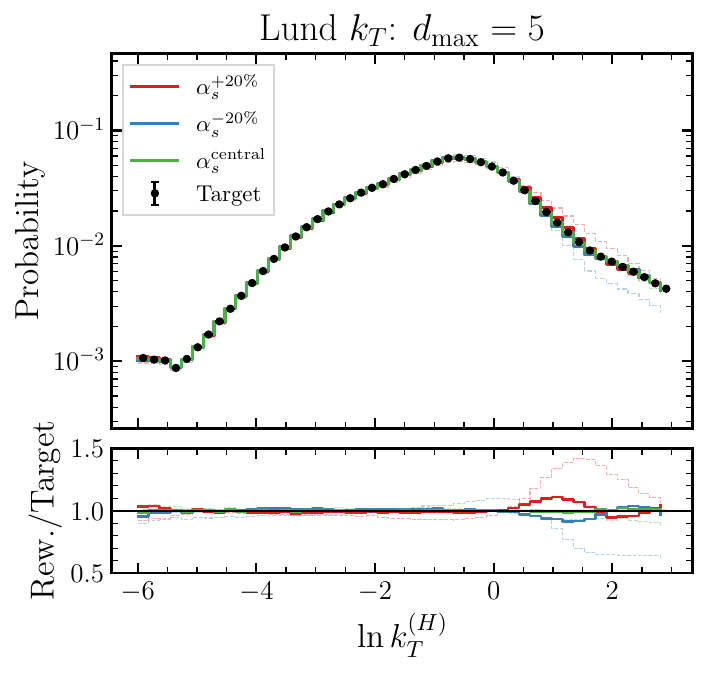}
    }
\caption{Transfer to heavy-hemisphere fragmentation and Lund-plane observables for $d_{\rm max}=5$ mixed-moment training with the degraded \textsc{Sherpa} priors: (a) charged fragmentation spectrum $\xi^{(H)}=\ln(1/x_p)$ and (b) the $\ln k_T$ marginal of the primary Lund plane.
The plot formatting follows \Fig{thrust_saturation}.}
\label{fig:fragmentation_transfer}
\end{figure}

\begin{figure}[t]
\centering
\includegraphics[width=0.95\textwidth]{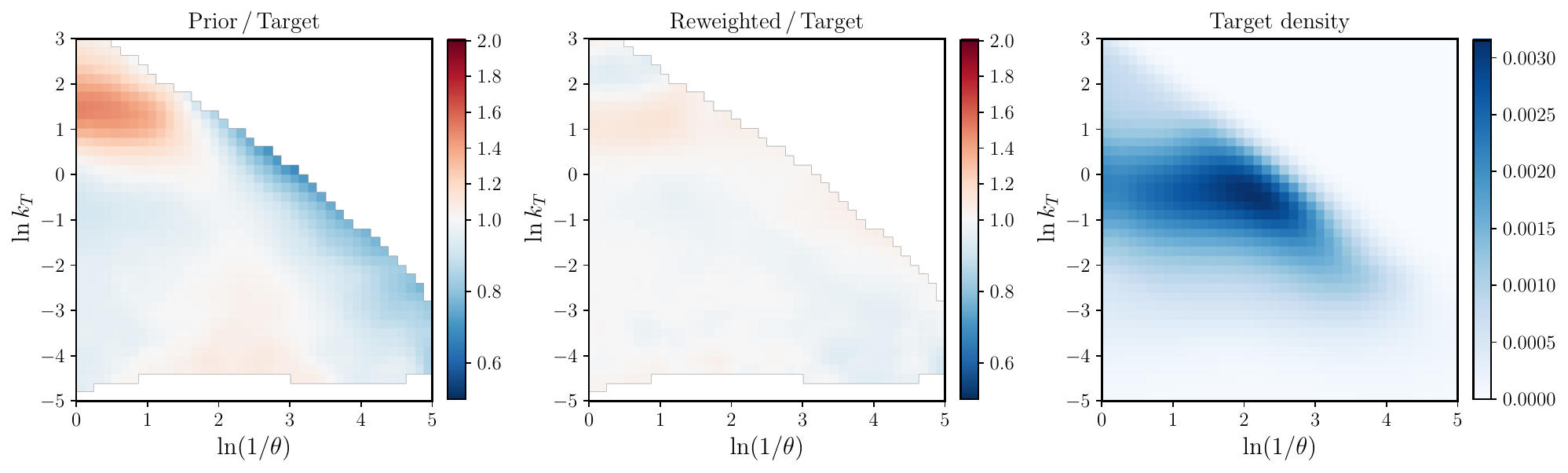}
\caption{Heavy-hemisphere primary Lund-plane density in $(\ln 1/\theta, \ln k_T)$ for the $+20\%$ $\alpha_s$ degraded \textsc{Sherpa} prior with $d_{\rm max}=5$ mixed-moment training.
The left panel shows the ratio of the prior to the target, the center panel shows the ratio of the reweighted prior to the target, and the right panel shows the target density.
The prior overpopulates the wide-angle, large-$k_T$ region and underpopulates the hard boundary of the plane, and the reweighting flattens the ratio toward unity across the populated plane.}
\label{fig:lund2d}
\end{figure}

In \Fig{substructure_transfer}, we show the subjet resolution, the exclusive subjet multiplicity, and the charged multiplicity for the three degraded \textsc{Sherpa} priors, reweighted with the $d_{\rm max}=5$ mixed moments of the main text.
The subjet resolution $-\ln y_{23}^{(H)}$ is corrected to near-target agreement across the Sudakov peak, with residuals at the 10--30\% level surviving at the resolved (fixed-order) end of the distribution for the $\pm20\%$ priors, and comparable or larger excursions in the sparsely populated deepest bins of the resummation tail.
The subjet multiplicity is corrected to within a few percent in the dominant two- and three-subjet bins, while the rare four- and five-subjet bins retain residuals at the 15--40\% level (improving or unchanged for the $\pm20\%$ priors but overshooting for the central prior in the five-subjet bin), consistent with the multi-particle-correlation pattern observed for $\tau_3$ and aplanarity in \Sec{transfer_eventshapes}.
The charged multiplicity, despite not being IRC safe, is corrected to the few-percent level throughout the bulk of the distribution, with the $\pm 20\%$ priors retaining residuals that grow to $\mathcal{O}(50\%)$ in the sparsely populated high-multiplicity tail.

In \Fig{fragmentation_transfer}, we show the charged fragmentation spectrum and the $\ln k_T$ marginal of the primary Lund plane.
Both are corrected to near-target agreement across their range, with mild residuals at large $k_T$ for the $\pm20\%$ priors and 25--30\% residuals confined to the rare hardest-fragmentation bin ($x_p \to 1$).
The two-dimensional Lund-plane density itself is examined in \Fig{lund2d} for the $+20\%$ prior, which has the largest prior-target gap.
The ratio of prior to target reveals the anatomy of the degradation.
The prior overpopulates the wide-angle, large-$k_T$ region, reflecting the enhanced coupling, and underpopulates the hard boundary of the plane, where the removed non-singular kernel terms and the disabled $g \to q\bar{q}$ channel control the emission rate.
After reweighting, the ratio is flattened to near unity across the populated plane, confirming that the EFP moment constraints reshape the emission pattern differentially in both angle and transverse momentum, not merely at the level of one-dimensional projections.

The success on the hadron-level observables deserves comment.
The multiplicity and fragmentation spectrum cannot be expanded in the EFP basis, so there is no a priori guarantee of transfer.
Here, however, the degraded \textsc{Sherpa} priors and the \textsc{Sherpa} target share an identical hadronization model, so their differences in any hadron-level observable are driven entirely by the upstream perturbative cascade, which is exactly what the EFP constraints correct.
The cross-generator study below probes what happens when this is no longer true.

\subsection{\texorpdfstring{Transfer from the \textsc{Herwig} prior}{Transfer from the Herwig prior}}

\begin{figure}[t]
\centering
\subfloat[][]{
    \includegraphics[width=0.32\textwidth]{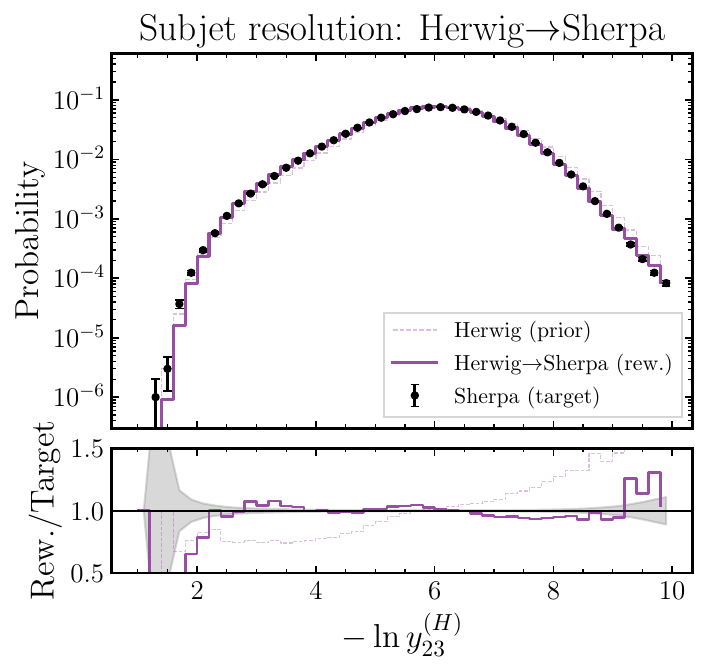}
    }
\subfloat[][]{
    \includegraphics[width=0.32\textwidth]{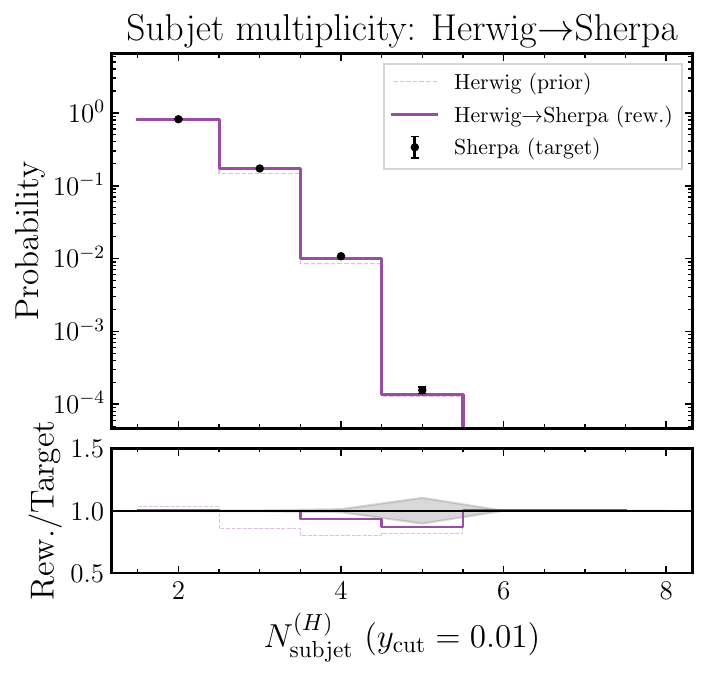}
    }
\subfloat[][]{
    \includegraphics[width=0.32\textwidth]{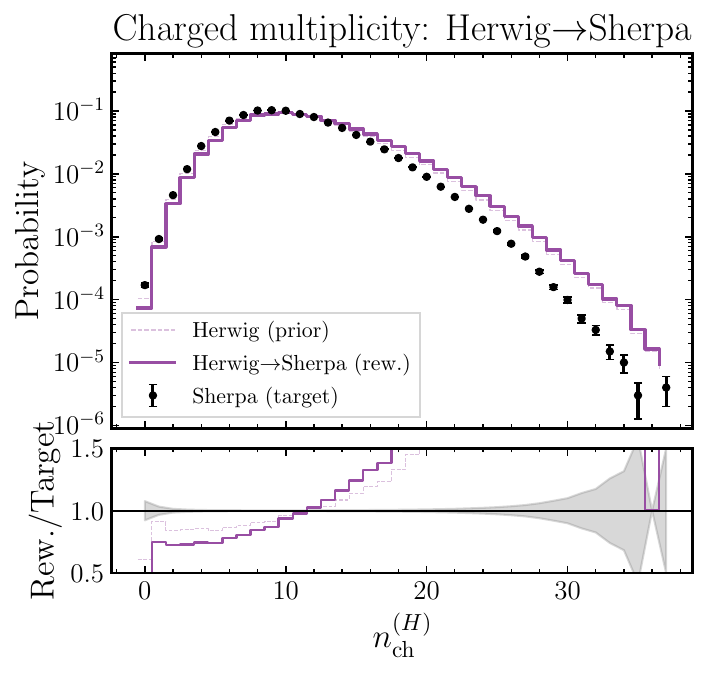}
    }
\caption{Cross-generator reweighting of the \textsc{Herwig} prior for heavy-hemisphere jet observables: (a) Durham $2\to3$ subjet resolution, (b) exclusive subjet multiplicity at $y_{\rm cut}=0.01$, and (c) charged-particle multiplicity.
The plot formatting follows \Fig{herwig_efp_dists}.
The IRC-safe subjet observables are largely corrected, while the charged multiplicity is essentially unchanged by the reweighting.}
\label{fig:herwig_substructure}
\end{figure}

\begin{figure}[t]
\centering
\subfloat[][]{
    \includegraphics[width=0.4\textwidth]{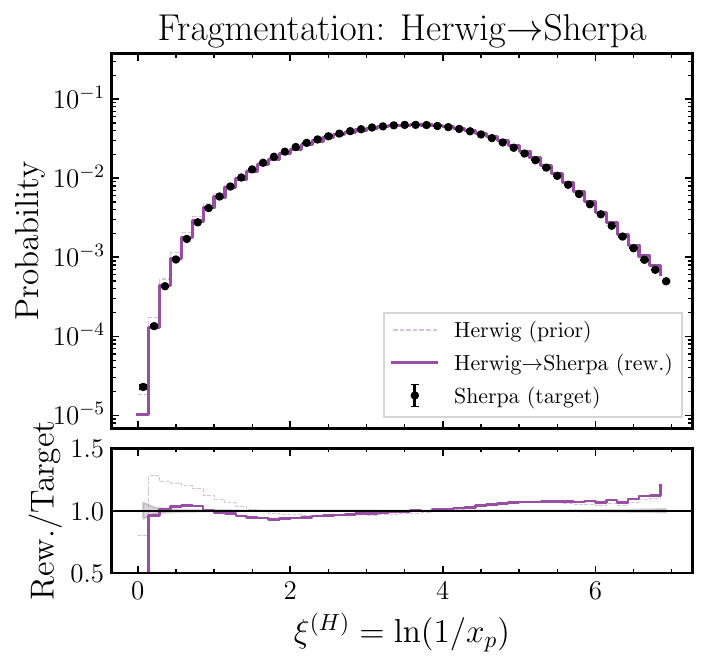}
    }
\subfloat[][]{
    \includegraphics[width=0.4\textwidth]{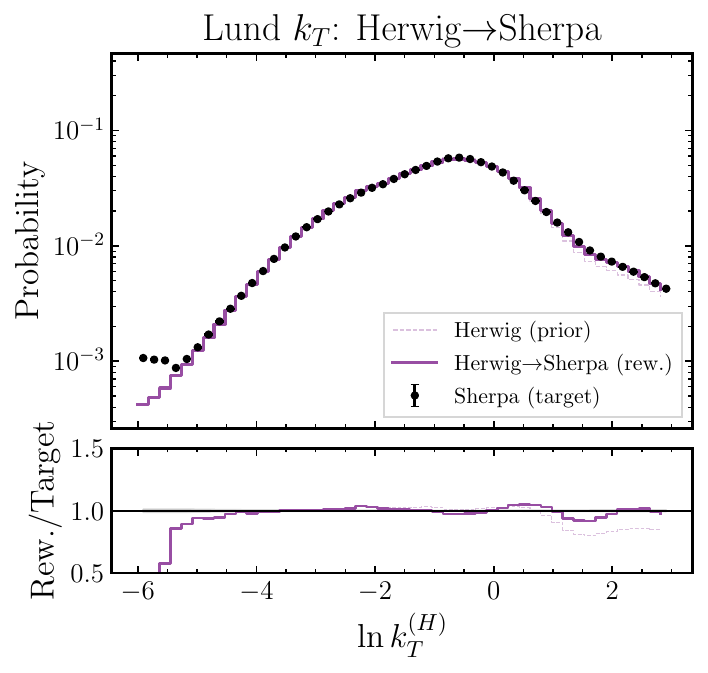}
    }
\caption{Cross-generator reweighting of the \textsc{Herwig} prior for (a) the charged fragmentation spectrum and (b) the $\ln k_T$ marginal of the primary Lund plane.
The plot formatting follows \Fig{herwig_efp_dists}.
The hard end of the fragmentation spectrum and the perturbative region of the Lund plane are improved, while the soft, hadronization-dominated regions are not.}
\label{fig:herwig_fragmentation}
\end{figure}

We now repeat these studies for the cross-generator reweighting of \Sec{herwig}, where the \textsc{Herwig} prior and the \textsc{Sherpa} target no longer share a hadronization model.
In \Fig{herwig_substructure}, the IRC-safe subjet observables behave much as they do for the degraded \textsc{Sherpa} priors.
The subjet resolution is corrected from prior deviations of up to $30\%$ (larger still in the deepest bins) to few-percent agreement across the bulk of the distribution, though an excess remains in the sparsely populated deep tail, and the dominant subjet-multiplicity bins are corrected with mild residuals remaining in the rare high-multiplicity bins.
The charged multiplicity, by contrast, is essentially unchanged by the reweighting, with the reweighted distribution tracking the prior, underpopulating low multiplicities and overpopulating the high tail.

The fragmentation and Lund-plane observables in \Fig{herwig_fragmentation} complete this picture.
The hard end of the fragmentation spectrum (small $\xi$), which is strongly correlated with the IRC-safe energy flow, is markedly improved, apart from the sparsely populated $x_p \to 1$ endpoint bin, while the soft region retains deviations at the $5$--$15\%$ level that the reweighting does not remove.
Similarly, the $\ln k_T$ marginal of the Lund plane is corrected throughout the perturbative region, including the large-$k_T$ tail where the prior undershoots the target by up to $20\%$, but the deficit in the deeply nonperturbative region $\ln k_T^{(H)} \lesssim -5$ persists.

\begin{figure}[t]
\centering
\includegraphics[width=0.95\textwidth]{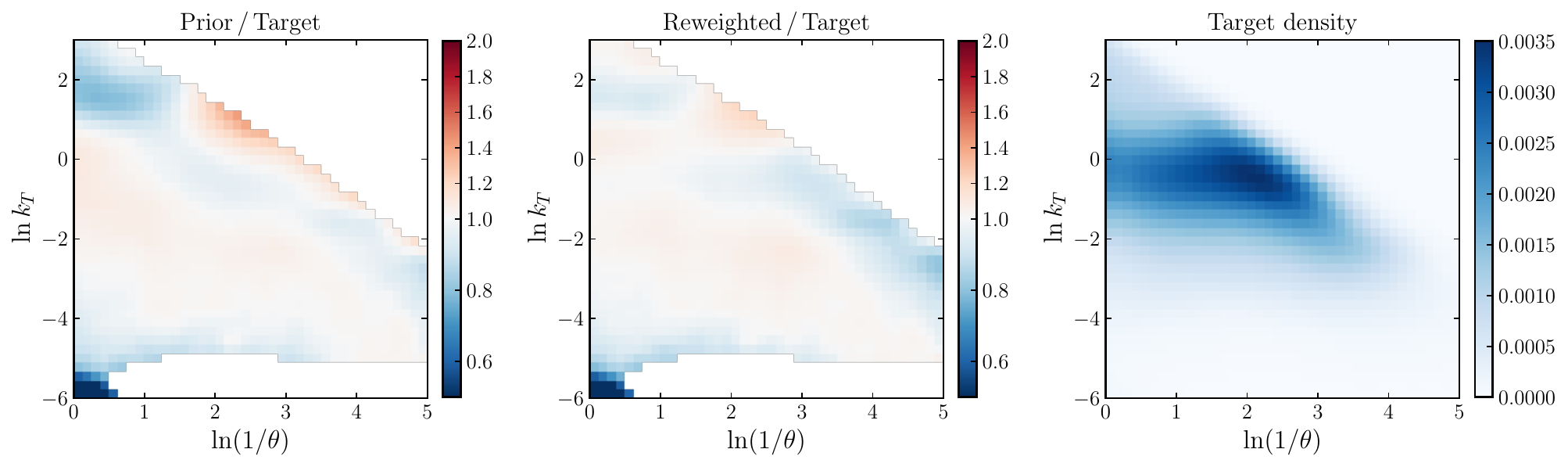}
\caption{Cross-generator analog of \Fig{lund2d}: the heavy-hemisphere primary Lund-plane density in $(\ln 1/\theta, \ln k_T)$ for the angular-ordered \textsc{Herwig} prior reweighted to the \textsc{Sherpa} target with $d_{\rm max}=5$ mixed moments.
Left: ratio of the \textsc{Herwig} prior to the target.
Center: ratio of the reweighted prior to the target.
Right: target density.
The reweighting reduces the excess of hard, collinear emission along the upper edge of the plane, while residual structure driven by the different hadronization models of the two generators remains.}
\label{fig:lund2d_herwig}
\end{figure}

These results delineate the boundary of cross-generator information transfer in a physically transparent way.
The residual differences between \textsc{Herwig} and \textsc{Sherpa} for multiplicity, soft fragmentation, and deeply nonperturbative emissions are dominated by their different hadronization models, and this information lies outside the span of the IRC-safe constraint set, in the sense of \Sec{accuracy}.
The contrast with the degraded-\textsc{Sherpa} study, where the identical observables were corrected because hadronization was common to prior and target, is the cleanest illustration in this paper of which information the constraints do and do not carry.
This is seen directly in the two-dimensional primary Lund-plane density of \Fig{lund2d_herwig}, the cross-generator analog of \Fig{lund2d}, where the reweighting removes the \textsc{Herwig} excess of hard, collinear emission along the upper edge of the plane, while the deeply nonperturbative region set by hadronization is left largely unchanged.
Correcting such observables across generators would require supplementing the framework with hadron-level precision inputs, such as moments of fragmentation observables, which the general construction of \Sec{it_meets_qcd} accommodates but which we leave to future work.

\bibliographystyle{JHEP}
\bibliography{refs}

\end{document}